\documentstyle[11pt]{article}
\input epsf

\begin{document}

\renewcommand{\thesection}{\Roman{section}}
\renewcommand{\thesubsection}{\Alph{subsection}}
\renewcommand{\thesubsubsection}{\arabic{subsubsection}} 

\begin{titlepage}
\begin{flushright}
SUSSEX-AST 96/7-5\\
(July 1996)\\
\end{flushright}
\begin{center}
\Large
{\bf The Behaviour Of Cosmological Models With Varying-G}\\  
\vspace{.3in}
\normalsize
\large{John D. Barrow and Paul Parsons} \\
\normalsize
\vspace{.6 cm}
{\em Astronomy Centre, \\ CPES,
University of Sussex, \\ Brighton BN1 9QH, U.~K.}\\
\vspace{.6 cm}
\end{center}
\parskip=18pt
\begin{abstract}
\noindent
We provide a detailed analysis of Friedmann-Robertson-Walker 
universes in a wide range of scalar-tensor theories of
gravity. We apply solution-generating methods to three
parametrised classes of scalar-tensor theory which lead
naturally to general relativity in the weak-field limit. We
restrict the parameters which specify these theories by the
requirements imposed by the weak-field tests of gravitation
theories in the solar system and by the requirement that
viable cosmological solutions be obtained. We construct a
range of exact solutions for open, closed, and flat
isotropic universes containing matter with equation of state
$P\leq \frac{1}{3}\rho$ and in vacuum. We study the range of
early and late-time behaviours displayed, examine when there
is a `bounce' at early times, and expansion maxima in closed
models. 

\end{abstract}

\begin{center}
\vspace{1cm}
PACS number(s)~~~98.80.Cq, 98.80.Hw, 04.50.+h, 04.60.-m, 12.10.-g 
\end{center}

\end{titlepage}


{\center \section{Introduction}}

Cosmological models arising from theories of gravity in which the
Newtonian gravitational `constant', $G$, varies with time have a long
history. They
were first studied in detail in response to Dirac's claims that a
coincidence between the values of `large numbers' arising in
dimensionless
combinations of physical and cosmological constants could emerge
naturally
if one of the constants involved possessed a time-variation that was 
significant over cosmological timescales \cite{DIR}. Dirac ascribed
that
time variation to $G$ and simply wrote the time-variation into the
Newtonian
expressions which held for constant $G$. Subsequently, mathematically
well-posed gravitation theories were developed in which Einstein's
theory of
general relativity (GR) was generalised to include a varying-$G$ by
deriving
it from a scalar field satisfying a conservation equation. These
scalar-tensor gravity theories, first formulated by Jordan \cite{JOR},
were
most fully exploited by Brans and Dicke in 1961 \cite{BD}. Motivated
by
claims that the observations of light-bending by the sun were in
significant
disagreement with the predictions of GR, Brans and Dicke explored the
possibility that the simplest scalar-tensor theory could provide
predictions
of the weak-field solar-system tests in agreement with light-bending
and
other geological and paleontological observations \cite{DICKE}.
Cosmological
models could be found in Brans-Dicke (BD) theory, but astronomical
observations were unable to impose stronger limits upon them than had
been
found from solar system experiments. Subsequently, the doubts
regarding the
compatibility between observations of solar light-bending and the
predictions of GR were removed by a fuller understanding of the
uncertainties surrounding measurements of the solar diameter at times
of
high solar surface activity \cite{WILL}. This removed the one
observation
that called for the replacement of GR by a scalar-tensor theory
displaying
varying-$G$ in the weak-field limit. Since that time there have been
two
observations which have led to renewed interest in gravity theories
with
non-Newtonian variation in $G$: one was the claim that a 'fifth force'
of
Nature existed on laboratory length scales and influenced
E\"{o}tv\"{o}s
experiments as if there existed a deviation from the inverse-square
form of
the law of gravitation in the non-relativistic Newtonian regime
\cite{FISH};
the other has been the occasional claim that the flatness of the
rotation
curves displayed by spiral galaxies might be a signal of non-Newtonian
gravitational attraction \cite{MIL}. The evidence for the 'fifth
force'
variations has not been supported by other experiments and flat
rotation
curves appear as natural outcomes of protogalaxy formation in
conventional
gravitation theories. In both cases, the deviations from conventional
gravitation theory, with constant $G$, would introduce a new
characteristic
length scale into the law of gravity with no fundamental basis other
than to
explain a particular set of observations. Some attempts were made by
Gibbons
and Whiting \cite{GIB} to see how simple scaling arguments might
relate
preferred high-energy physics scales to those observed in fifth force
experiments. More recently, the study of scalar-tensor gravity
theories has
been rejuvenated by theoretical developments in the study of the
evolutionary possibilities open to the early universe.

In a metric scalar-tensor theory of gravity the gravitational coupling
is
derived from some scalar field, $\phi$, so $G=G(\phi )$. Historically,
most
interest has been focused upon the first and simplest theory of this
type, 
presented by Brans and Dicke \cite{BD}, in which the coupling function
$
\omega (\phi )$ is a constant. In general, the consideration of
scalar-tensor theories with non-constant $\omega (\phi )$, \cite{WILL,
BERG}
greatly enlarges the range of possible $G$-variations and weakens the
impact
of observational limits accordingly. In the weak-field limit Nordtvedt
found
an expression for the observed value of the gravitation `constant' in
these theories, to leading order, as \cite{NORD1}.

\[
G(t)=\phi ^{-1}\left( \frac{4+2\omega (\phi )}{3+2\omega (\phi
)}\right)\,, 
\]
so that

\[
\frac{\dot G}G=-\dot{\phi} \left( \frac{3+2\omega
}{4+2\omega 
}\right) \left( G+\frac{2\omega ^{\prime }(\phi )}{(3+2\omega
)^2}\right)\,.  
\]
We see that in Brans-Dicke theory, with constant $\omega $ the
variation of $
G(t)\ $is just inversely proportional to $\phi (t).\ $Moreover, for
one particular choice of $\omega (\phi ),$

\[
\omega (\phi )=\frac{4-3\phi }{2(\phi -1)}\,, 
\]
it is possible to have $\dot G=0$ to first order in the weak-field
limit \cite{BARK}.

Our observational limits are a mixture of cosmological limits, studies
of
astrophysical objects, and weak-field tests of gravitation in the
solar
system. In Brans-Dicke theory the scope for significant deviations
from
constant $G$ is very small because of the constancy of $\omega $.
However,
if $\omega $ varies then it can increase with cosmic time in such a
way that 
$\omega \rightarrow \infty $ and $\omega ^{\prime }\omega
^{-3}\rightarrow 0$
as $t\rightarrow \infty $ so that weak-field observations at the
present
time accord well with the predictions of GR even though the theory may
possess significant deviations from GR predictions at very early
cosmological times \cite{BAR1}.

The principal observational upper limits on $\dot G/G$ come from many
different observations. Passive radar data on the distances to Mercury
and
Venus \cite{SHAP1, REAS1} give the limit $\dot G/G<4\times
10^{-10}yr^{-1}$.
This was improved using the Mariner 9 Mars orbiter by Anderson {\em et
al.} 
\cite{AND1} to $\dot G/G<1.5\times 10^{-10}yr^{-1}.$ Anderson {\em et
al.} 
\cite{AND2} then used Mariner 10 data and radar ranging to Mercury and
Venus
to obtain $\dot G/G<0.0\pm 2.0\times 10^{-12}yr^{-1}.$ Viking landers,
Mars
orbiters and transponders gave upper limits of $\dot G/G<3\times
10^{-11}yr^{-1},$ \cite{HELL1}, $\dot G/G<2\pm 4\times
10^{-12}yr^{-1}, $ 
\cite{HELL2}, and $\dot G/G<-2\pm 10\times 10^{-12}yr^{-1},$
\cite{JJD}.
These are made uncertain by our incomplete knowledge of the asteroid
distribution. Studies of the Binary pulsar PSR 1913+16 by Damour,
Gibbons
and Taylor \cite{DAM} give a limit of $\dot G/G=-(1.10\pm 1.07)\times
10^{-11}yr^{-1}$ with uncertainties due to the pulsar's proper motion.
Gravitational lensing promises to yield new tests, but depends upon
other
uncertain cosmological parameters \cite{KRAUSS}. Krauss and White
argue that
a variation in $G$ of $\Delta G/G\leq 20$ would influence lenses at $z
\sim 1.5 $ and these studies might eventually achieve constraints of
order $\dot{G%
}/G\leq 10^{-11}$ in Brans-Dicke theories. White dwarf cooling will be
affected by variations in $G$, \cite{VILA}, and recently,
Garc\'ia-Berro 
{\em et al.} \cite{BERRO} gave limits, depending upon the chemical
composition of the white dwarf of $\dot G/G\leq -(1\pm 1)\times
10^{-11}yr^{-1}$ for chemically stratified models, and $\dot{G}/G\leq
-(3_{-3}^{+1})\times 10^{-11}yr^{-1}$ for models with constant C/O
composition. However, these studies have theoretical uncertainties
introduced by the quasi-Newtonian modelling of the variation of $G$,
\cite
{BAR2}. This is also true of attempts to constrain $\dot{G}/G$ by
measurements of the time-variation of neutron star masses \cite{THOR}.
Cosmological nucleosynthesis gives limits that are of order $\dot
G/G\leq
0.01H$ in Brans-Dicke theory. Limits have also been obtained in
theories
with $2\omega +3\propto \phi ^\alpha$, which display a slow
logarithmic
decrease in $G(t), $ \cite{TORRES, BAR3}. Other limits include those
from
solar evolution, \cite{CHIN}, of $\dot G/G\leq 10^{-10}yr^{-1};$ lunar
occultations and eclipses, \cite{MORR}, of $\dot G/G\leq 0.4\times
10^{-10}yr^{-1}$; lunar laser-ranging studies of the Moon's orbit
around the 
Earth, \cite{JGW}, of $\dot G/G\leq 0.3\times 10^{-10}yr^{-1}$, and
M\"uller 
{\em et al.} \cite{MULL1} derive $\dot G/G<(0.1\pm 10)\times
10^{-12}yr^{-1}. $ There are proposals to measure via radar the
position of
a transponder in orbit around Mercury, \cite{BEND}. The ranging
sensitivity
is claimed to be able to reach just a few centimetres which would
translate 
into limits of order $\dot G/G<3\times 10^{-13}yr^{-1}$. Most of these
limits (especially cosmological ones from nucleosynthesis) are
uncertain if
varying $G$ is coupled to the variation of other constants
\cite{BAR4}. 

The coupling function of scalar tensor theories, $\omega (\phi ),$ is
related to the PPN parameters by, \cite{NORD1}, 
\begin{eqnarray*}
\beta &=&1+\frac{\omega^{\prime}}{(3+2\omega )^2(4+2\omega
)}\rightarrow
1+O\left( \frac{\omega^{\prime}}{8\omega ^3}\right)\; {\rm
as\;}\;\omega
\rightarrow \infty \\
&&
\end{eqnarray*}

\[
\gamma =1-\frac 1{\omega +2}\rightarrow 1\;{\rm as\;}\;\omega
\rightarrow 
\infty 
\]
Observational limits on the weak field PPN parameter $\gamma $ are
$\gamma
\simeq 1\pm 0.002$ from radio timing delays \cite{REAS2} ; $\gamma
\simeq
1.0002\pm 0.002$ from light deflection using VLBI observations of
quasars
\cite{ROB} ; $\gamma \simeq 1\pm 0.02$ from lunar laser ranging,
\cite{MULL1} 
. Future experiments, GPB, POINTS and Mercury Relativity Satellite,
hope to
reach sensitivities of $\left| \gamma -1\right| \sim 3\times 10^{-7},$
\cite
{CIUO1}. For combinations of two PPN parameters determining $\omega ,$
limits of $\left| 4\beta -\gamma -3\right| <5\times 10^{-3}(1\sigma )$
have
been found \cite{JJD}. If we take the observational limits as $\left|
\gamma 
-1\right| <0.002$ and $\left| 4\beta -\gamma -3\right| <0.001$ then we
have $ \omega >498$ and

$\ $
\[
\left| \frac{\omega ^{\prime }}{(3+2\omega )^2(4+2\omega )}\right|
<0.001 
\]
\[
\]
where $\omega ^{\prime }$ is evaluated at the asymptotic $\phi$ value $
\phi_0 $ where $\omega \rightarrow \omega (\phi _0).$ Hence, we have
only a
rather weak limit of $\left| \omega ^{\prime }\right| <10^{-6}\times
999^2\sim O(1). $

Planetary data \cite{TALM} also provides a limit on the spatial
gradient of $
G$ over solar system scales of $\nabla G/G<3\times 10^{-10}AU^{-1}.$ A
limit
of $\delta G/G<10^{-13}$ on possible spatial anisotropy of $G$ in the
solar
system has been derived by studying the alignment of the Sun's
rotation axis
with the direction of the solar system's angular momentum vector
\cite{NORD2}
, and of $\ \delta G/G<2\times 10^{-12}$ from satellite and LAGEOS
laser ranging data \cite{CIUO2}.

A brief period of interest in extended Kaluza-Klein cosmological
theories 
\cite{FREU}, which culminated in their replacement by superstring
theories 
\cite{GREEN}, revealed how time-variations in any extra ($>$ 3)
dimensions of
space would manifest itself through the time-variation of the
`constants'
defined in the three dimensions we observe. Over the same period,
cosmologists showed growing interest in the behaviour of all scalar
fields
during the early stages of the universe. The time variation of a field
energy source in the early universe at a rate slower than the universe
is
expanding is a general possibility only for {\em scalar} fields and
has
become known as `slow-rolling' of the field. Typically, it produces an
acceleration of the expansion scale factor of the universe with time:
a
phenomenon known as `inflation' \cite{GUTH}. This has led to the
investigation of general-relativistic cosmological models containing a
wide
range of self-interacting scalar-field sources \cite{BAR5}, the
classification of different varieties of inflation that can result
from
their slow-rolling evolution, and the extraction of detailed
predictions
concerning the fluctuations imprinted in the cosmic microwave
background
radiation by the spectra of gravitational waves and density
perturbations
that emerge from a period of primordial inflation. However, it has
also been
recognised that natural scalar fields might be provided by the scalar
component of a scalar-tensor theory of gravity \cite{MATH}. Such
theories
possess close conformal relationships with GR plus explicit scalar
fields,
and although the scalar field determining the strength of the
gravitational
coupling does not easily drive inflation it does have a significant
effect
upon the pace of inflation that arises when self-interacting scalar
fields
are included in the universal energy-momentum tensor. Following the
long
initial study of BD cosmological models, Barrow \cite{BAR1} showed how
to
generate cosmological solutions of vacuum or radiation-dominated
isotropic
cosmological models in any scalar-tensor gravity theory. The method
used
works for any isotropic cosmological model with an energy momentum
tensor
possessing a vanishing trace. Barrow and Mimoso \cite{BAR3} then
devised a
more complicated procedure which allows solutions to be generated for
isotropic, zero-curvature universes with the pressure $p$ and density
$\rho $
related by a perfect-fluid equation of state $p=(\gamma -1)\rho ,$
with the
constant $\gamma $ lying in the range $0\leq \gamma \leq 4/3$. In
particular, they found the first dust solutions for scalar-tensor
theories
more general than BD together with a wide range of new inflationary
solutions. Recently, the techniques introduced by Barrow and Mimoso
have
been used to study the asymptotic behaviour of isotropic and
anisotropic
cosmological models in scalar-tensor theories \cite{SERNA, DERU}. The
anisotropic case has also been studied by Mimoso and Wands
\cite{WANDS}. In
this paper we are going to extend these studies to arrive at a more
general
and systematic understanding of the behaviour of isotropic
cosmological 
models containing matter with equation of state $p\leq \rho /3.$ We
will be
interested in studying scalar-tensor gravity theories which can
approach
general relativity in the weak-field limit at late cosmic epochs. By
studying the classes of gravity theory which allow this approach to
occur we
shall use our solution-generating techniques to build up a detailed
picture
of the behaviour of scalar-tensor cosmological models. We shall be
particularly interested in models containing radiation ($p=\rho /3$),
dust ($
p=0$), and inflationary stresses ($p=-\rho $).

The plan of the paper is as follows. In section \ref{st} we introduce
the
Lagrangians and field equations which define scalar-tensor gravity
theories
in terms of their coupling function $\omega (\phi)$ and dictate their
evolution. We shall specialise our study to the cases of Friedmann
universes containing perfect fluid sources. In section \ref{methods}
we describe the two
techniques for finding complete and asymptotic solutions of these
equations
for arbitrary choices of $\omega (\phi).$ The `direct' method works
for
universes of all curvatures but is restricted to the vacuum and
radiation-dominated cases. The `indirect' method works only for flat
universes but for any equation of state. We shall be especially
interested
in the vacuum, dust ($p=0$), radiation and inflationary ($p=-\rho $)
cases.
These techniques allow us to draw important conclusions about the
small and
late-time evolution of cosmological models in scalar-tensor theories.
In
section \ref{coupling} we introduce three broad classes of gravity
theory, defined by the
form of $\omega (\phi)$, which allow distinct forms of cosmological
evolution. The free parameters can be restricted to allow the theories
to
reproduce the successful weak-field predictions of general relativity
and to
give cosmological solutions. The cosmological consequences of these
three
classes of theory are explored systematically in sections 
\ref{th1}, 
\ref{th2}, and \ref{th3}. In
each case we are interested in determining the early and late time
behaviours, and finding exact solutions which describe the dust,
radiation,
and inflationary phases of expansion. The results are discussed in
section \ref{disc}. 

\begin{center}
\section{Scalar-tensor cosmologies}
\label{st}

\subsection{Field equations}
\end{center}

We shall consider scalar-tensor theories of gravity defined by the
action,
\begin{equation}  \label{lag}
S_{G} = \int d^4 x \sqrt{-g} \left[ -\phi {\cal {R}} +
\frac{\omega(\phi)}{ 
\phi}g^{{\rm ab}} \partial_{{\rm a}}\phi \partial_{{\rm
b}}\phi\right]\,.  \\
\end{equation}
where ${\cal {R}}$ is the curvature scalar arising from the spacetime
metric 
$g_{{\rm ab}}$, $g$ is the determinant of $g_{{\rm ab}}$, $\phi$ is a
scalar
field, and $\omega(\phi)$ is a function determining the strength of
the
coupling between the scalar field and gravity. We are working in units
such
that Newton's constant, $G_{N}$, is equal to unity. The field $\phi$ is
the
analogue of $G_{N}^{-1}$ in GR except that here, in contrast to
Einstein's
theory, $\phi$ is a dynamical quantity. Scalar-tensor gravitational
theories
therefore permit histories in which the value of the gravitational
``constant'' varies. The simplest such case is that explored by Brans
and
Dicke, where $\omega$ is a constant. The action in Eq.~(\ref{lag})
offers
more general $\omega(\phi)$ theories as natural extensions to BD
gravity. It
is these theories that will be of primary interest in this paper.
Demanding
that the first-order variations of Eq.~(\ref{lag}) with respect to
$\phi$
and $g_{{\rm ab}}$ vanish, we derive the field equations:

\begin{eqnarray}  \label{field2}
{\cal R}_{{\rm ab}}-\frac{1}{2}g_{{\rm ab}}{\cal R} & = & -8\pi
\frac{T_{ 
{\rm ab}}} {\phi} - \frac{\omega(\phi)}{\phi^2}\left(\phi_{{\rm a}}
\phi_{ 
{\rm b}} - \frac{1} {2}g_{{\rm ab}} \phi_{{\rm c}} \phi^{{\rm
c}}\right) - 
\frac{1}{\phi}\left( \phi_{{\rm a;b}} - g_{{\rm ab}} \Box \phi
\right)\,, \\
\Box \phi & = & \frac{1}{2\omega(\phi)+3}\left(8 \pi T - \omega
^{\prime}(\phi) \phi_{{\rm c}} \phi^{{\rm c}}\right)\,,
\end{eqnarray}
and the conservation law, 
\begin{equation}  \label{cons}
{T^{{\rm ab}}}_{{\rm ;b}} = 0\,,
\end{equation}
where ${\cal R}_{{\rm ab}}$ is the Ricci curvature tensor, $T_{{\rm
ab}}$ is 
the energy-momentum tensor specifying the properties of the matter
occupying
the universe. Primes indicate derivatives with respect to $\phi$. The
first
of these three equations is the scalar-tensor analogue of Einstein's
equations, the second is the wave equation for $\phi$, and the final
expression is the energy-momentum conservation law for the matter,
which
ensures that each theory is consistent with the equivalence principle.

{\center \subsection{Friedmann universes}}

We shall examine solutions to these equations that describe
homogeneous,
isotropic\linebreak
Friedmann-Robertson-Walker (FRW) cosmological models, with
time-varying-$G$.
The FRW metric line element in spherical polar co-ordinates
($t,\;r,\;\theta
,\;\psi $) is given by: 
\begin{equation}
ds^2=-dt^2+a^2(t)\left[ \frac{dr^2}{1-kr^2}+r^2\left( d\theta
^2+sin^2\theta
d\psi ^2\right) \right] \,,
\end{equation}
where the curvature parameter is $k=-1,:0,:+1$ for open, flat or
closed
cosmologies respectively, and the scale-factor, $a(t)$, characterises
the
expansion history of the universe. We shall assume the universe
contains a
simple perfect-fluid which may be accurately described by a
perfect-fluid
equation of state, 
\begin{equation}
p=(\gamma -1)\rho \,;\;\;\;0\leq \gamma \leq 2,:\;\;\gamma {\rm 
constant}\,;  \label{state}
\end{equation}
with these prescriptions, the equations of motion become, 
\begin{equation}
H^2+H\frac{\dot \phi }\phi -\frac{\omega (\phi )}6\frac{{\dot \phi
}^2}{\phi
^2}+\frac k{a^2}=\frac{8\pi }3\frac \rho \phi \,,  \label{cos1}
\end{equation}

\begin{equation}  \label{cos2}
\ddot{\phi} + \left[3H +
\frac{\dot{\omega(\phi)}}{2\omega(\phi)+3}\right] 
\dot{\phi} = \frac{8\pi \rho}{2\omega(\phi)+3}(4-3\gamma)\,,
\end{equation}

\begin{equation}  \label{cos3}
\dot{H} + H^2 +\frac{\omega(\phi)}{3}\frac{{\dot{\phi}}^2}{\phi^2} -
H\frac{
\dot{\phi}}{\phi} = -\frac{8 \pi \rho}{3\phi}
\frac{(3\gamma-2)\omega(\phi) 
+3}{2\omega(\phi)+3} +\frac{1}{2}
\frac{\dot{\omega(\phi)}}{2\omega(\phi)+3} 
\frac{\dot{\phi}}{\phi}\,,
\end{equation}

\begin{equation}  \label{cos4}
\dot{\rho} +3\gamma H \rho = 0\,,
\end{equation}
where $H \equiv \dot{a}/a$ is the Hubble expansion parameter and
overdots 
denote derivatives with respect to comoving proper time, $t$.

The structure of the solutions to these equations is sensitive to the
form
of the coupling function $\omega (\phi )$, which defines the
scalar-tensor
theory of gravity. As Nordvedt first showed \cite{NORD1}, it is
possible to
place bounds on the parameter-space of many models, prior to
extracting a
full solution from the field equations, simply by inspecting the
explicit
form of $\omega (\phi )$. Nordvedt's constraints demand that the
theories
tend to GR in the weak-field limit, so that they concur with the
observational limits on light-bending and perihelion precession. This
requirement manifests itself explicitly in the conditions $\omega
\rightarrow \infty $ and $\omega ^{\prime }\omega ^{-3}\rightarrow 0$
as $
t\rightarrow \infty $. Typically, scalar tensor theories add a term
proportional to $\omega ^{\prime }\omega ^{-3}$ to the weak-field
predictions of general relativity. Whilst the first condition ($\omega
\rightarrow \infty $) is well known, the second ($\omega ^{\prime
}\omega
^{-3}\rightarrow 0$) is not. As we shall see in Section \ref{coupling}
there
are many theories which satisfy the first condition but not the second:
for
example those defined by $\omega (\phi )\sim \left( 1-\phi /\phi
_0\right)
^{-\alpha }$ , with $0<\alpha <1/2,$ as $\phi \rightarrow \phi _0$
from below.

\begin{center}
\section{Methods of solution}

\label{methods} 
\end{center}

Solutions to Eqs.~(\ref{cos1}) - (\ref{cos4}) for Brans-Dicke FRW
models
have existed for some years \cite{NARIAI, LORENZ}. The Brans-Dicke
case has
also been studied qualitatively by Kolitch and Eardley, \cite{KOL}.
Recently, Barrow \cite{BAR1, BAR6} and later Barrow and Mimoso
\cite{BAR3}
have extended these treatments to general $\omega (\phi )$ theories,
providing a method for obtaining zero-curvature cosmological solutions
when $
\gamma <4/3$, and solutions for vacuum and radiation-dominated
universes of
any curvature.

We now recapitulate the two methods of solution introduced in refs.
\cite
{BAR1} and \cite{BAR3} to solve the cosmological field equations for
theories specified by any $\omega (\phi )$.

\begin{center}
\subsection{Vacuum and radiation models}

\subsubsection{Exact solutions}
\end{center}

The general solutions to Eqs.~(\ref{cos1}) - (\ref{cos4}) contain four
arbitrary integration constants, one more than their GR counterparts,
the
extra degree of freedom being attached to the value of $\dot{\phi}$.
When
the energy-momentum tensor is trace-free there exists a conformal
equivalence between the theory and GR, the right-hand side of
Eq.~(\ref
{field2}) vanishes and $\dot{\phi}=0$ is always a particular solution,
corresponding to a special choice of the additional constant possessed
by
the model over GR. Consequently, the exact general solution of
Einstein's
equations when $T_{{\rm ab}}$ is trace-free is also a particular
solution to
Eqs.~(\ref{cos1}) -- (\ref{cos4}) with $\phi$, and hence
$\omega(\phi)$,
constant.

It will seldom be the case that the particular solution obtained in
this way
will form the general solution for that particular choice of $\omega
(\phi )$
. Usually, however, it will be the late or early time attractor of the
general solution. For example, in the case of Brans-Dicke theory the
special
GR solution is the late-time attractor for flat and open universes but
not
the early-time attractor. However, one of these authors \cite{BAR1}
developed a method for integrating the field equations for models
containing
trace-free matter. The procedure is as follows.

Eq.~(\ref{cos4}) integrates immediately to yield 
\begin{equation}
8\pi \rho =3\Gamma a^{-3\gamma }\,,  \label{cos5}
\end{equation}
where $\Gamma \geq 0$ is a constant of integration; $\Gamma =0$
describes
vacuum models. Making the choice $\gamma =4/3$, corresponding to
blackbody
radiation, and introducing the conformal time co-ordinate, $\eta $,
defined
by 
\begin{equation}
ad\eta =dt\,,  \label{conf}
\end{equation}
Eq.~(\ref{cos2}) becomes 
\begin{equation}
\phi _{\eta \eta }+\frac 2aa_\eta \phi _\eta =-\frac{\omega ^{\prime
}(\phi )
}{2\omega (\phi )+3}\left( \phi _\eta \right) ^2\,,
\end{equation}
where subscript $\eta $ denotes a derivative with respect to conformal
time.
This integrates exactly to give, 
\begin{equation}
\phi _\eta a^2=3^{1/2}A(2\omega (\phi )+3)^{-1/2}\,;\;\;\;A\;\;{\rm
const.}
\label{phip}
\end{equation}
We now employ the variable used by Lorenz-Petzold to study Brans-Dicke
models
\cite{LORENZ}, 
\begin{equation}
y=\phi a^2\,,  \label{ydef}
\end{equation}
to re-write the scalar-tensor version of the Friedmann equation,
Eq.~(\ref
{cos1}), as 
\begin{equation}
\left( y_\eta \right) ^2=-4ky^2+4\Gamma y+\frac 13\left( \phi _\eta
\right) 
^2a^4(2\omega (\phi )+3)\,.  \label{ycom}
\end{equation}
Dividing Eq.~(\ref{phip}) by Eq.~(\ref{ydef}), and using
Eq.~(\ref{phip}),
we obtain the coupled pair of differential equations 
\begin{eqnarray}
\label{trace1}
\frac{\phi _\eta }\phi  &=&3^{1/2}Ay^{-1}(2\omega (\phi
)+3)^{-1/2}\,,\\
\label{trace2}
\left( y_\eta \right) ^2 &=&-4ky^2+4\Gamma y+A^2\,.
\end{eqnarray}
We may now obtain the general solution for a particular choice of
$\omega
(\phi )$, given $k$. Integrating Eq.~(\ref{trace2}) yields $y(\eta )$
which,
in conjunction with $\omega (\phi )$, implies $\phi (\eta )$ from
Eq.~(\ref
{trace1}) and, without further integration, $a(\eta )$ from
Eq.~(\ref{ydef} 
). If Eq.~(\ref{conf}) is both integratable and invertible we may
compute $
\phi (t)$ and $a(t)$, so completing the solution.

The vacuum models are obtained by setting $\Gamma = 0$ and in this
case Eq.~(
\ref{trace2}) has three possible solutions, according to the value of
$k$:  
\begin{equation}  \label{vacy}
y(\eta) = \left\{ 
\begin{array}{ll}
A(\eta + \eta_0)\,, & k=0\,, \\ 
&  \\ 
\frac{1}{2}A \sinh [2(\eta + \eta_0)]\,, & k=-1\,, \\ 
&  \\ 
\frac{1}{2}A \sin[2(\eta + \eta_0)]\,, & k=+1\,,
\end{array}
\right.
\end{equation}
where $\eta_0$ is an arbitrary constant fixing the origin of
$\eta$-time.
Combining these results with Eq.~(\ref{trace1}) yields the set of
integral
relations: 
\begin{equation}  \label{vacint}
\int \frac{(2\omega(\phi) +3)^{1/2}}{\phi} d\phi = \left\{ 
\begin{array}{ll}
\sqrt{3}\ln |\eta + \eta_0|\,, & k=0\,, \\ 
&  \\ 
\sqrt{3}\ln \left| \tanh (\eta + \eta_0) \right|\,, & k=-1\,, \\ 
&  \\ 
\sqrt{3}\ln \left| \tan (\eta + \eta_0) \right|\,, & k=+1\,.
\end{array}
\right.
\end{equation}
Specification of $\omega(\phi)$ allows the full solutions to be
completed.
When $k=-1$ the negativity of the right-hand side, arising because $
0<|\tanh(\eta+\eta_0)|<1$ $\forall \; \eta$, places strong constraints
on
the allowed form of the integral on the left.

Similarly, one may perform this treatment on the radiation models, ie
those
cases for which $\Gamma > 0$. Again, the results are classified by the
value
of $k$: 
\begin{equation}  \label{rady}
y(\eta) = \left\{ 
\begin{array}{ll}
\Gamma(\eta+\eta_0)^2 - A^2 / 4\Gamma\,, & k=0\,, \\ 
&  \\ 
-\frac{1}{2}\Gamma + \frac{1}{2}(A^2 - \Gamma^2)^{1/2} \sinh[2(\eta +
\eta_0) 
]\,, & k=-1\, \\ 
&  \\ 
\frac{1}{2}\Gamma + \frac{1}{2}(\Gamma^2 + A^2)^{1/2} \sin[2(\eta +
\eta_0) 
]\,, & k=+1\,.
\end{array}
\right.
\end{equation}
Integrating Eq.~(\ref{trace1}) with the solutions above leads to 
\begin{equation}  \label{radint}
\int \frac{(2\omega(\phi) +3)^{1/2}}{\phi} d\phi = \left\{ 
\begin{array}{ll}
\sqrt{3} \ln \left| \frac{2\Gamma \eta + 2\Gamma \eta_0 -A} {2\Gamma
\eta +
2\Gamma \eta_0 +A} \right|\,, & k=0\,, \\ 
&  \\ 
\sqrt{3} \ln \left| \frac{\Gamma \tanh(\eta + \eta_0) +
(A^2-\Gamma^2)^{1/2} 
-A}{\Gamma\tanh(\eta+\eta_0) + (A^2-\Gamma^2)^{1/2} +A}\right| \,, &
k=-1\,, 
\\ 
&  \\ 
\sqrt{3} \ln \left| \frac{\Gamma \tan(\eta + \eta_0) + (\Gamma^2 +
A^2)^{1/2} -A}{\Gamma \tan(\eta + \eta_0) + (\Gamma^2 + A^2)^{1/2} +A}
\right|\,, & k=+1\,.
\end{array}
\right.
\end{equation}
These expressions can be simplified by choosing the arbitrary
integration
constant $\eta_0$ such that $2\Gamma\eta_0 =A$.

The domain on which each right-hand side exists strongly constrains
the
integral on the left. For instance, if we require $\phi \in (0,\phi
_0)$
then the corresponding range of the function of $\phi $ resulting from
the
integral on the left must be compatible with the allowed range on the
right.
In general, $\phi $ must tend to its general relativistic form, ie a
constant, at large $\eta $. We shall find in Sections \ref{th1} --
\ref{th3}
that this behaviour is not generic and often requires the integration
constant associated with the left-hand integral to assume a particular
value. This integration constant can be interpreted as an initial
boundary
condition on $\phi $ or $\omega $ at, say, $\eta =0.$ Further
restrictions
can be found by studying the evolution at early times.

In Sections \ref{th1} -- \ref{th3} we shall exploit these relations to
derive general $\omega(\phi)$ solutions for all values of the
curvature parameter.

{\center \subsubsection{Approximation techniques}}

Many of the models we shall present are insoluble in terms of $t$. The
reason for this is the non-invertibility of $t(\eta)$, arising from
integrating $a(\eta)$. In these cases we shall invert
$t(\eta)$
approximately at early and late times to obtain series solutions for
the
behaviour of $\phi(t)$ and $a(t)$, indicating their limiting forms and
their
approach to these forms to leading order. We shall use the inversion
technique of Olver \cite{OLV}. If we have an expression 
\begin{equation}  \label{examp}
y(x) = f(x) g(x)\,,
\end{equation}
and we require an approximation to $x(y)$, valid in a region where $g$
dominates over $f$, we neglect $f$ and write 
\begin{equation}
x(y) \simeq g^{-1}(y)\,.
\end{equation}
The next-order approximation is easily obtained by substituting this
result
into $f$, yielding $f \simeq f(g^{-1}(y))$. Using this in
Eq.~(\ref
{examp}) we have 
\begin{equation}  \label{it2}
x(y) \simeq g^{-1}\left(\frac{y}{f\left(g^{-1}(y)\right)} \right)\,.
\end{equation}
This procedure can be iterated indefinitely but we shall use the
$2^{nd}$
iteration form as given in Eq.~(\ref{it2}).

To analyse the asymptotic behaviour we need to establish the value of
$\eta$
as $t\rightarrow \infty $ and our primary interest is in models which
display GR, ie $\phi \rightarrow \phi _0$, in this limit.
Radiation-dominated, $k=0$, universes in GR evolve like $a\propto
t^{1/2}$ at
late times and hence $t\propto \eta ^2$ and $\eta \rightarrow \infty $
as $
t\rightarrow \infty $. There does not exist a spatially flat vacuum
model in
GR, however, the negatively curved general relativistic models are
asymptotically vacuum-dominated, (as the matter density redshifts to
zero)
tending to the Milne solution, $a\propto t$, as $t\rightarrow \infty
$, and $
\eta \propto \ln t$ which diverges with $t$. We therefore study the
asymptotic behaviour of vacuum and radiation models in the limit $\eta
\rightarrow \infty $ when $k\leq 0$. We will not consider the late-time
limit of the $k>0$ models. Typically, they recollapse to a final
singularity
if $\rho >0$ and $\rho +3p>0,\ $although a bounce occurs for many
choices of 
$\omega (\phi )$. The behaviour in their recollapse phase is similar
to the
time-reverse of the early expansion of $k=0$ models. 

We now define the early-time limit. We examine first models with
$0<\phi
<\phi _0$, in this range $\sqrt{2\omega +3}$ does not change sign, as
guaranteed by the choices of the coupling function, $\omega (\phi )$,
detailed in Section \ref{coupling} and Eq.~(\ref{phip}) ensures that
$\phi$
is monotonically increasing. We thus extrapolate the evolution back to
$\phi
=0$ and treat this as the early-time limit in vacuum and radiation
models.
In more general perfect-fluid models we find that we are easily able
to
examine the behaviour when $\phi >\phi _0$ by exploiting the conformal
invariance of the theory (this procedure is explained in detail in the
Appendix) and we so examine early-time behaviour in the neighbourhood
of the
last zero, or non-zero minimum, of $a$.

{\center \subsubsection{Maxima and minima}}

It is easy to study the structure of non-singular vacuum and
radiation-dominated cosmological models within the framework presented in
this section. Differentiating Eq.~(\ref{ydef}) and substituting from
Eqs.~(\ref{phip}) and (\ref{ydef}) leads to 
\begin{equation}
(a^2)_\eta =\frac{y_\eta }\phi -\frac{\sqrt{3}A}{\phi \sqrt{2\omega
+3}}\,.
\end{equation}
The condition for the scale-factor to contain a stationary point, $a_\eta
=0$
, is equivalent to $(a^2)_\eta =0$ when $a\neq 0$, which leads to the
simple
relation 
\begin{equation}
y_\eta =\frac{\sqrt{3}A}{\sqrt{2\omega +3}}\,,  \label{mintest}
\end{equation}
when $\phi $ is finite. This relation will prove useful for locating
minima
when we lack an exact solution for $a(\eta )$.

Eq.~(\ref{mintest}) is a simple test for the existence of stationary
points
in the evolution of $a$.  Stationary points of the type $_{{\rm
maximum}}^{
{\rm minimum}}$ will occur when $a_{\eta \eta }{}_{<}^{>}0$,
respectively
but, because $a>0$, it is sufficient to replace $a_{\eta \eta }$ with
$
(a^2)_{\eta \eta }$ in this condition. Differentiating
Eq.~(\ref{ydef})
twice and evaluating at a stationary point, ie where $(a^2)_\eta $
vanishes,
we obtain 
\begin{equation}
\left( a^2\right) _{\eta \eta }=\frac{y_{\eta \eta }}\phi -\frac{\phi
_{\eta
\eta }a^2}\phi \,.
\end{equation}
A condition for the stationary point to be a $_{{\rm maximum}}^{{\rm
minimum} 
}$ is then 
\begin{equation}
\frac{y_{\eta \eta }}\phi {}_{<}^{>}\frac{\phi _{\eta \eta }a^2}\phi
\,.
\label{minnat}
\end{equation}
Differentiating Eq.~(\ref{phip}) and discarding $(a^2)_\eta $ we have 
\begin{equation}
\phi _{\eta \eta }a^2=-\frac{3A^2\omega ^{\prime }(\phi )}{a^2\left(
2\omega
+3\right) ^2}\,.
\end{equation}
Substituting this into Eq.~(\ref{minnat}), and remembering $\phi >0$
yields 
\begin{equation}
y_{\eta \eta }{}_{<}^{>}-\frac{3A^2\omega ^{\prime }(\phi )}{a^2\left(
2\omega +3\right) ^2}\,,  \label{minnatmod}
\end{equation}
as the condition for the stationary point to be a $_{{\rm
maximum}}^{{\rm
minimum}}$. When $0<\phi <\phi _0$ we have $\omega ^{\prime }(\phi
)>0$ for
the choices in Section \ref{coupling} and the right-hand side of
Eq.~(\ref
{minnatmod}) is negative definite. Thus whenever we can prove $y_{\eta
\eta
}\geq 0$ we may exclude the possibility of $a(\eta )$ containing
maxima, and
then by continuity we can limit the number of minima to one. We test
this
condition for the forms of $y(\eta )$ given in Eqs.~(\ref{vacy}) and
(\ref
{rady}), the results are summarised in Table \ref{dder}. From the table
it
can be seen that all spatially flat and negatively curved models may
only
contain a single minimum. The $k=+1$ models are not bounded in this
way and
may contain an undetermined number of minima and maxima.

\begin{table}[tbp]
\begin{center}
\begin{tabular}{||c||c|c|c||c|c|c||}
\hline
Matter source & \multicolumn{3}{c||}{Vacuum} &
\multicolumn{3}{c||}{Radiation
} \\ \hline
$k$ & $-1$ & $0$ & $+1$ & $-1$ & $0$ & $+1$ \\ \hline\hline
$y_{\eta \eta}$ & $\,\;\;\;>0\;\;\;\,$ & $\,\;\;\;=0\;\;\;\,$ & $%
\,\;\;\;<0\;\;\;\,$ & $>2\Gamma>0$ & $=2\Gamma>0$ & $\;\;<2\Gamma\;\;$
\\ 
\hline
\end{tabular}
\end{center}
\caption{$y_{\eta \eta}$ for vacuum and radiation models of all
curvatures.}
\label{dder}
\end{table}

{\center \subsection{General perfect-fluid cosmologies}}

When $T$ is non-vanishing the situation is substantially more
complicated.
In this instance, $\dot \phi =0$ is no longer a particular solution of
the
field equations, forcing us to resort to more elaborate methods to
obtain
solutions. Barrow and Mimoso \cite{BAR3} have done this, for the $k=0$
models, by generalising the method of Gurevich {\em et al.}
\cite{NARIAI}
for BD models to the case of varying $\omega $. We now outline this
procedure.

Introducing the new time co-ordinate $\xi$, and the two new variables
$x$
and $v$ such that 
\begin{equation}  \label{timeperf}
dt = a^{3(\gamma-1)}\;\sqrt{\frac{2\omega+3}{3}} \; d\xi \quad,
\label{edeftimevar}
\end{equation}
\begin{equation}
x\equiv \left[\phi a^{3(1-\gamma)}\left( a^3\right)_{\xi} \right] \; ,
\label{edefx}
\end{equation}
\begin{equation}
v\equiv \left[a^{3(2-\gamma)}\phi_{\xi}\right]\quad,  \label{edefy}
\end{equation}
and confining attention to the $k=0$ models, Eqs.~(\ref{cos1}),
(\ref{cos2})
and (\ref{cos3}) transform to 
\begin{equation}
\left( \frac{2}{3} x+v \right)^2 =\left(\frac{2\omega+3}{3}\right)\,
\left[
v^2 +4 \Gamma\,\phi\,a^{3(2-\gamma)}\right] \; ,  \label{eFriedm}
\end{equation}
\begin{equation}
v_{\xi} = \Gamma\;(4-3\gamma) \; ,  \label{edy}
\end{equation}
and 
\begin{equation}
x_{\xi} = 3\Gamma\, \left[(2-\gamma)\omega+1\right] +
\frac{3}{2}\left(\frac{ 
2}{3(2\omega+3)}x+v\right)\, \omega_{\xi} \; ,  \label{ex'}
\end{equation}
where subscript-$\xi$ represents a derivative with respect to
$\xi$-time. 
Eqs.~(\ref{edy}) and (\ref{ex'}) integrate easily to yield 
\begin{eqnarray}
v &=& \Gamma (4-3\gamma)\,(\xi-\xi_1) \quad ,  \label{eyeta} \\
x &=&\frac{3}{2}\;\left[-v + \sqrt{2\omega+3} \left( C +
\Gamma(2-\gamma) \,
\int_{\xi_1}^{\xi}\,\sqrt{2\omega+3}\, d \bar{\xi}\right) \right]
\quad ,
\label{exeta}
\end{eqnarray}
$C$ is an integration constant and $\xi_1$ fixes the origin of
$\xi$-time.
Noting the relation 
\begin{equation}
\frac{3}{a\phi} a_{\xi} \phi_{\xi} = \frac{1}{\phi^2}\left(\phi_{\xi} 
\right)^2\; 3\frac{\phi}{a}\frac{a_{\xi}}{\phi_{\xi}} =
\frac{1}{\phi^2}\,
\left(\phi_{\xi}\right)^2 \frac{x}{v} \; ,  \label{erelxy}
\end{equation}
and differentiating $y$, with respect to $\xi$, yields 
\begin{equation}
\left(\frac{\phi_{\xi}}{\phi}\right)_{\xi} + \left[
\frac{3\gamma-4}{2}+ 
\frac{1}{\xi-\xi_1}f_{\xi}(\xi) \right]\;
\left(\frac{\phi_{\xi}}{\phi} \right)^2 = \frac{1}{\xi-\xi_1}\,
\frac{\phi_{\xi}}{\phi}\quad ,
\label{ephi''}
\end{equation}
where a new function $f(\xi)$, is defined by 
\begin{equation}
f(\xi) \equiv \int_{\xi_1}^{\xi}\,
\frac{3(2-\gamma)}{2\Gamma(4-3\gamma)}\, 
\sqrt{2\omega(\phi)+3}\; \left[
C+\Gamma(2-\gamma)\,\int_{\xi_1}^{\xi}\, 
\sqrt{2\omega(\phi)+3}\, d {\tilde{\xi}}\right] \, d \bar{\xi} \; .
\label{edeff}
\end{equation}
Solving Eq.~(\ref{ephi''}), we have the solution 
\begin{equation}  \label{gdef}
\ln{\left(\frac{\phi}{\phi_0}\right)} =\int_{\xi_1}^{\xi} \,
\frac{\xi-\xi_1 
}{g(\xi)} d \xi \,,  \label{elnphi}
\end{equation}
with $g(\xi)$ simply related to $f(\xi)$ by 
\begin{equation}
g(\xi) \equiv f(\xi) +\frac{3\gamma-4}{4}\, (\xi-\xi_1)^2 + D\,, \quad 
\label{edefg}
\end{equation}
where $D$ is a constant of integration. Eq.~(\ref{erelxy}) immediately
reveals a simple formula for the scale-factor: 
\begin{equation}  \label{aofg}
a^3 = a_0^3 \; \left(\frac{g}{\phi}\right)^{\frac{1}{2-\gamma}}\; ;
a_0 \; 
{\rm constant}\,.  \label{eV}
\end{equation}
Finally, the scalar-tensor coupling function $\omega(\phi)$ is given
as a
function of $f$ by 
\begin{equation}
2\omega\left(\phi(\xi)\right) + 3 = \frac{4-3\gamma}{3(2-\gamma)^2}\;
\frac{ 
(f^{\prime})^2}{\left[f+\frac{4-3\gamma}{3(2-\gamma)^2}\,f_0\right]}\quad
,
\label{edefomega}
\end{equation}
where $f_0$ is another arbitrary constant.

In the calculations to follow, we shall exploit the arbitrariness of
$\xi_1$ 
and set it to zero. In addition, we note some relations between the
other 
constants in the solution, arising from the scalar-tensor Friedmann
constraint, Eq.~(\ref{eFriedm}) 
\begin{eqnarray}  \label{phiconstr}
D &=& \frac{4-3\gamma}{3(2-\gamma)^2}\, f_0 \; ,  \label{e5cons1} \\
\phi_0\,a_0^{3(2-\gamma)} &=& \Gamma\,(4-3\gamma) \; , 
\label{e5cons2} 
\end{eqnarray}
and from the requirement that the BD theory be recovered when $%
\omega=\omega_0$ is a constant we obtain the further condition, 
\begin{equation}
C^2 = \left( \frac{2\Gamma(4-3\gamma)}{3(2-\gamma)}\right)^2\;f_0 \,.
\end{equation}
We shall require $\phi_0$, ie $G^{-1}$ today, to be positive and from
Eq.~(\ref{phiconstr}) we see that this requires $\gamma < 4/3$. We
shall also
confine our attention to theories with $2\omega + 3 >0$. Using
Eq.~(\ref{elnphi}) we can re-write Eq.~(\ref{edefomega}), as a
function of
$\phi$ and
its derivatives, as 
\begin{equation}  \label{omphid}
2\omega(\xi) + 3 =
\frac{4-3\gamma}{3(2-\gamma)^2}\frac{\left(\frac{\phi}
{\phi_{\xi}} - \xi \frac{\phi \phi_{\xi \xi}}{(\phi_{\xi})^2} + \frac{
6-3\gamma}{2} \xi \right)^2}{\left(\xi \frac{\phi}{\phi_{\xi}} -
\frac{ 
3\gamma -4}{4} \xi^2 \right)}\,.
\end{equation}
Asymptotically, $2\omega +3 \rightarrow (g_{\xi})^2/g$ as $\xi
\rightarrow 
\infty$ and the above relation becomes 
\begin{equation}
2\omega + 3 \rightarrow \frac{\phi}{\xi \phi_{\xi}} \left( 1 + \xi
\frac{ 
\phi_{\xi}}{\phi} - \xi \frac{\phi_{\xi \xi}}{\phi_{\xi}} \right)^2\,.
\end{equation}
The choice of $\phi(\xi)$ is thus equally fundamental as that of
$g(\xi)$,
and amounts to specifying $G(\xi)$. We shall use the former. Once
$\phi(\xi)$
has been specified, we may infer $g(\xi)$ from Eq.~(\ref{elnphi}),
$a(\xi)$
from Eq.~(\ref{eV}) and $\omega(\xi)$ from Eq.~(\ref{omphid}). If
Eq.~(\ref
{edeftimevar}) can be integrated and inverted to yield $\xi(t)$ these
variables may be expressed in terms of cosmic time, $t$.

An important benchmark is provided by the behaviour of the BD theory,
where $
\omega(\phi) = \omega_0 =$ constant. In this case, the generating
function, $
f(\xi)$, is given by a quadratic in $\xi$: 
\begin{equation}
f_{{\rm BD}}(\xi) = \frac{3(2-\gamma)}{2\Gamma(4-3\gamma)}\, \sqrt{
2\omega_0+3}\; \left[ C\; \xi+\frac{\Gamma(2-\gamma)}{2}\,\xi^2\;
\sqrt{
2\omega_0+3}\right]\,.  \label{eg(eta)JBD}
\end{equation}
Hence, in general $(C\neq 0 \neq \Gamma)$ when $\gamma \neq 4/3\,, \;
2$, we
see that $f_{{\rm BD}} \propto \xi^2$ as $\xi \rightarrow \infty$ and
$f_{
{\rm BD}} \propto \xi$ as $\xi \rightarrow 0$, where $dt \propto
a^{3(\gamma-1)}d\xi$. If we choose $C=0$ then $f_{{\rm BD}} \propto
\Gamma
\xi^2$ as $\xi \rightarrow 0$. The choice $C=0$ restricts the solution
to
the special `matter-dominated' solutions (termed `Machian' by Dicke
\cite
{BD, DICKE}, see also Weinberg \cite{WEIN}) which were first found for
all
perfect-fluids by Nariai \cite{NARIAI}. If $C \neq 0$ then the
early-time
behaviour is dominated by the dynamics of the $\phi$-field; such
solutions
are termed `$\phi$-dominated' (or `non-Machian' by Dicke).

Therefore if we choose a generating function $g(\xi)$ that grows
slower than 
$\xi^2$ as $\xi \rightarrow \infty$ it will produce a theory that
approaches
BD at late times ($\phi \rightarrow$constant, $\omega(\phi)
\rightarrow$
constant), whilst if $g(\xi)$ decreases slower than $\xi$ as $\xi
\rightarrow 0$ then the theory will approach the behaviour of
$\phi$-dominated BD theory at early-times. This means that we
will find new
(non-BD) late-time behaviours by studying generating functions which
increase faster than $g(\xi)=\xi^2$ as $\xi \rightarrow \infty$ and
new
(non-BD) early-time behaviour by picking generating functions which
decrease
slower than $g(\xi)=\xi$ as $\xi \rightarrow 0$ or $\xi \rightarrow
\xi_{ 
{\rm min}}$ (if there is no zero of $\xi$ at the minimum of $a(t)$).

\begin{center}
\section{The coupling function}

\label{coupling}
\end{center}

We are interested in ascertaining the general behaviours displayed by
cosmological models in the range of scalar-tensor gravity theories
that
approach GR in the weak-field, late-time limit. This requires $2\omega
+3\rightarrow \infty $ as $t\rightarrow \infty $ and also $\omega
^{\prime
}\omega^{-3}\rightarrow 0$ if the solar system tests are to accord
with
observation. The specific form of the leading-order corrections to the
general relativistic predictions of light-bending, perihelion
precession,
and radar echo delay are all almost equal to $\omega^{\prime
}\omega^{-3}$
in the large $\omega $ limit. However, the rate at which $\omega (\phi
)$
tends to infinity will determine the form of the cosmological models.
In an
earlier paper \cite{BAR3} we explored the behaviour of simple
power-law
forms for $\omega (\phi )$ which, although growing with time, only
attain
the GR limit when $\phi =\infty $, although at any finite time
$\omega$ can
be made as large as we wish by the choice of the constants defining
$\omega
(\phi ).$ Here, we turn our attention to a potentially more
interesting
situation in which $\omega \rightarrow \infty $ as $\phi \rightarrow
\phi_0$
where $\phi _0$ may be taken as the present value of $\phi (t)$, which
determines the observed value of the Newtonian gravitation
constant, $G=\phi_0^{-1}$.

\begin{figure}[tbp]
\begin{center}
{\large Theory 1: $2\omega(\phi)+3 = 2B_{1}|1-\phi /\phi _0| ^{-\alpha
}$}\\ 
\leavevmode
{\bf a)} \epsfxsize=6cm \epsfbox{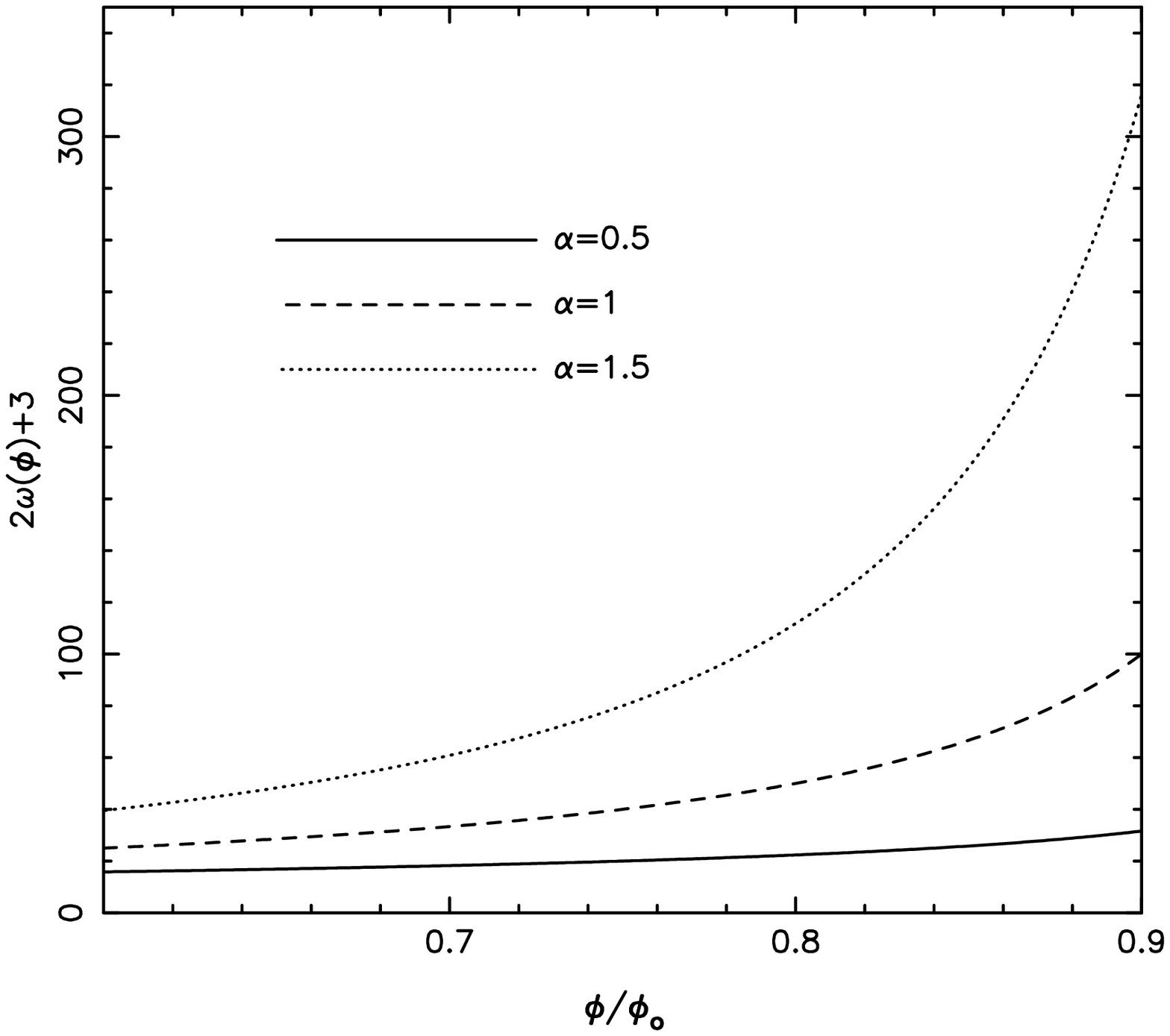} \hspace{0.5in} {\bf b)}
\epsfxsize
=6cm \epsfbox{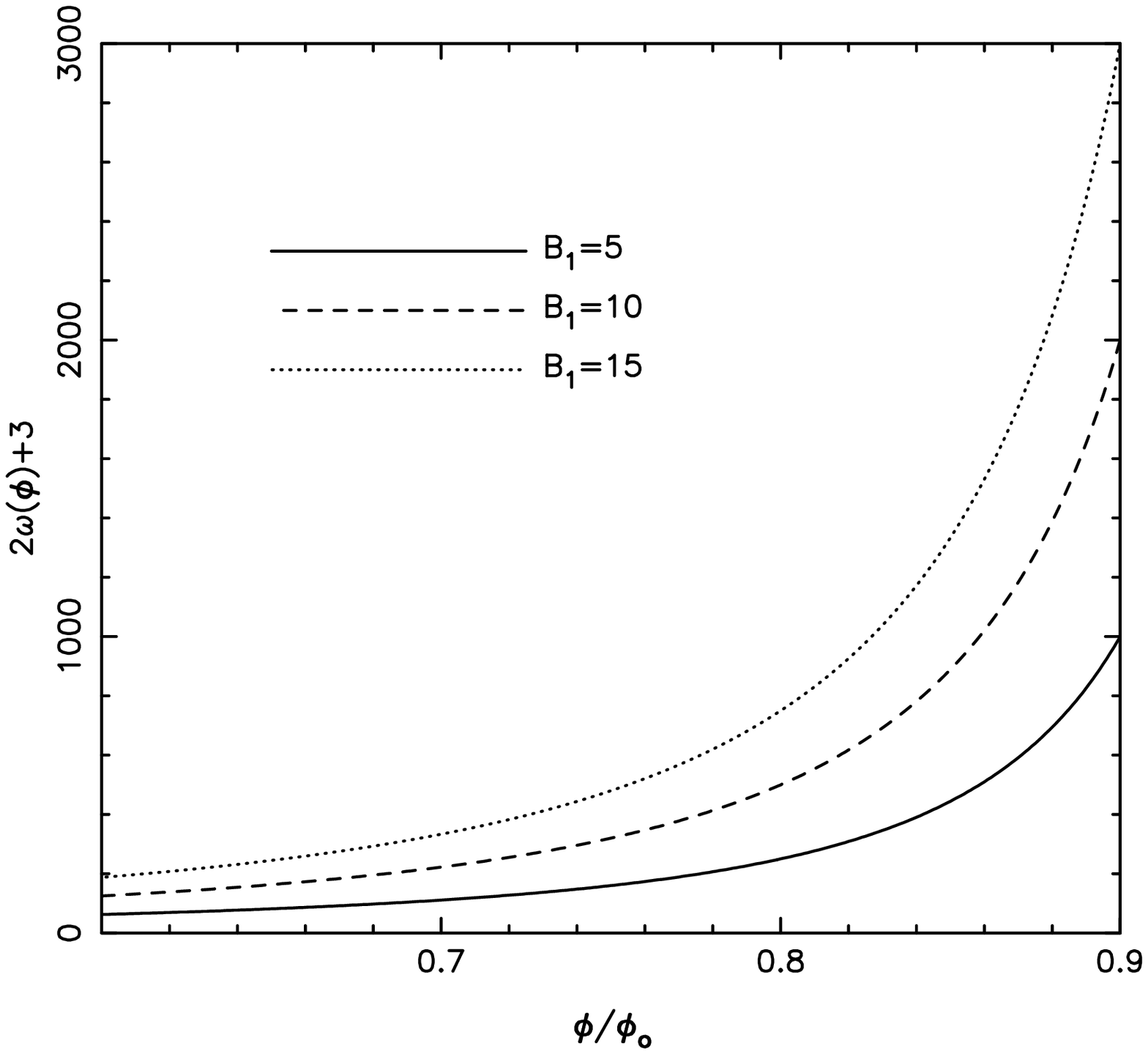}\\
\end{center}
\caption{Theory 1 with {\bf a)} $B_1 = 5$ and $\alpha = 0.5$, $1$,
$1.5$;
and {\bf b)} $\alpha=2$ and $B_1 = 5$, $10$, $15$.}
\label{coup1}
\end{figure}

\begin{figure}[tbp]
\begin{center}
{\large Theory 2: $2\omega (\phi )+3=B_{2}\left| \ln (\phi /\phi
_0)\right|^{-2\delta}$}\\\leavevmode
{\bf a)} \epsfxsize=6cm \epsfbox{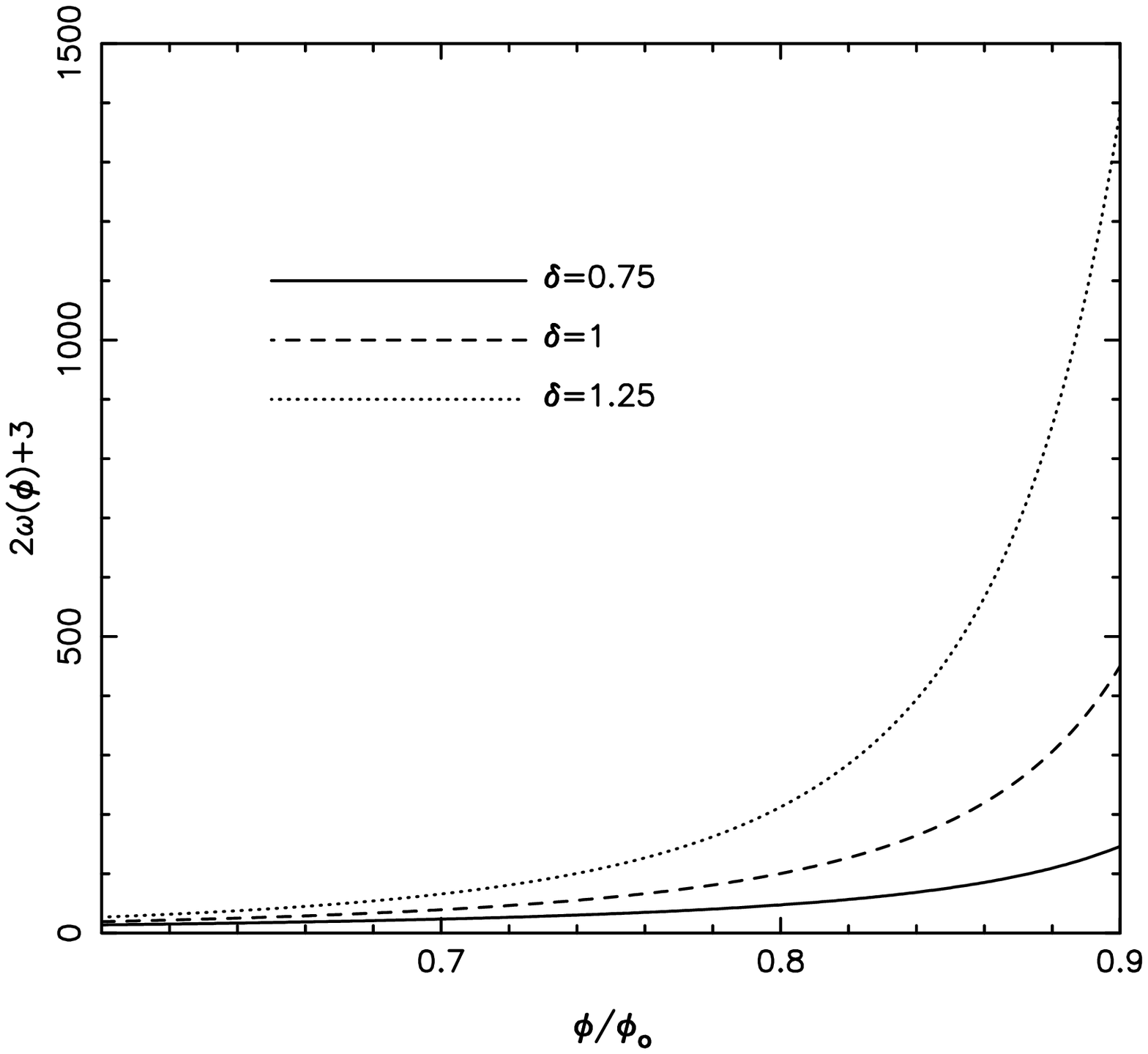} \hspace{0.5in} {\bf b)}
\epsfxsize 
=6cm \epsfbox{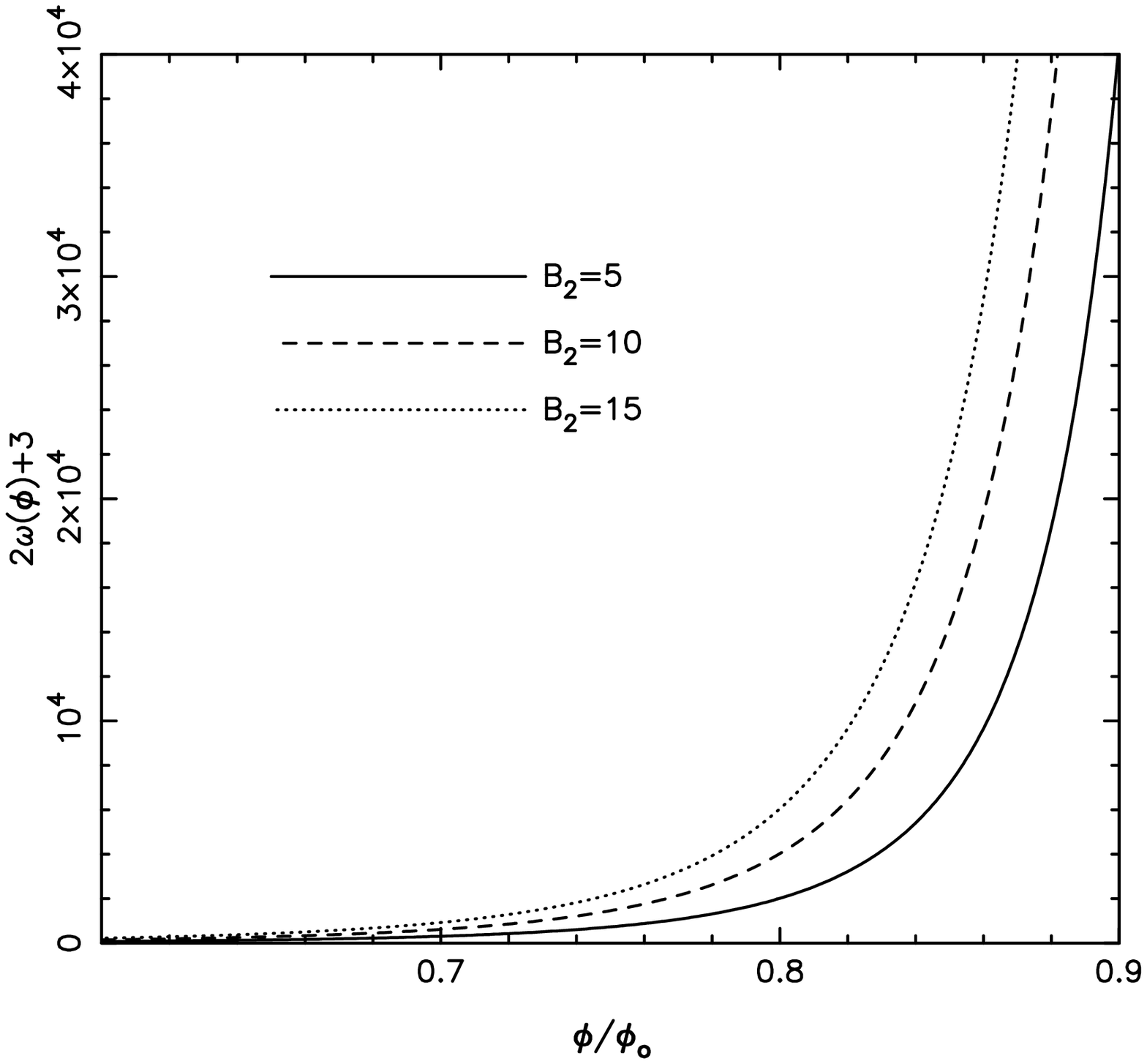}\\
\end{center}
\caption{Theory 2 with {\bf a)} $B_2 = 5$ and $\delta = 0.75$, $1$,
$1.25$;
and {\bf b)} $\delta=2$ and $B_2 = 5$, $10$, $15$.}
\label{coup2}
\end{figure}

\begin{figure}[tbp]
\begin{center}
{\large Theory 3: $2\omega (\phi )+3=B_{3}\left| 1-(\phi /\phi _0)
^\beta\right| ^{-1}$}\\{\bf a)} \epsfxsize=6cm \epsfbox{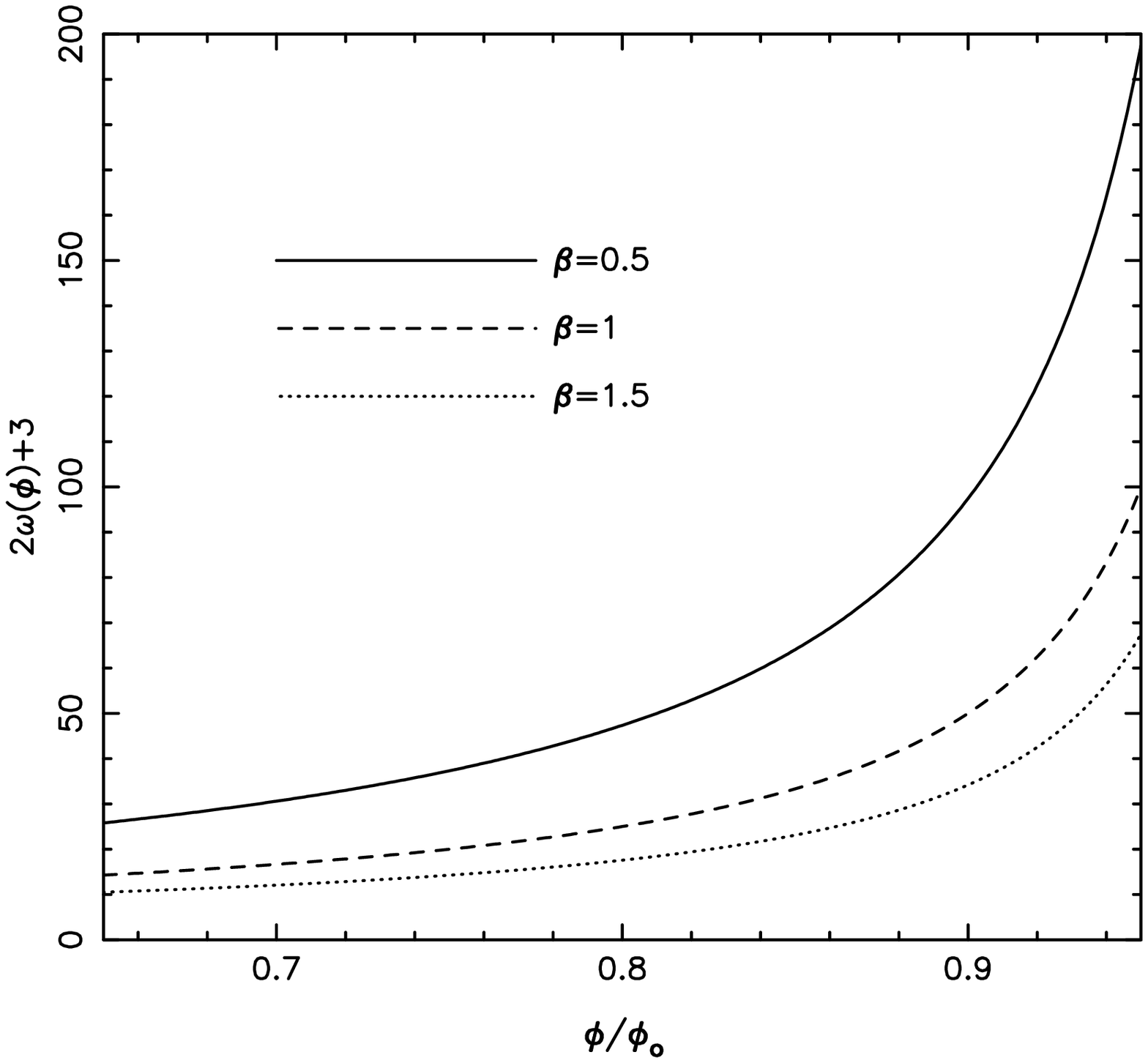} 
\hspace{0.5in} {\bf b)} \epsfxsize=6cm \epsfbox{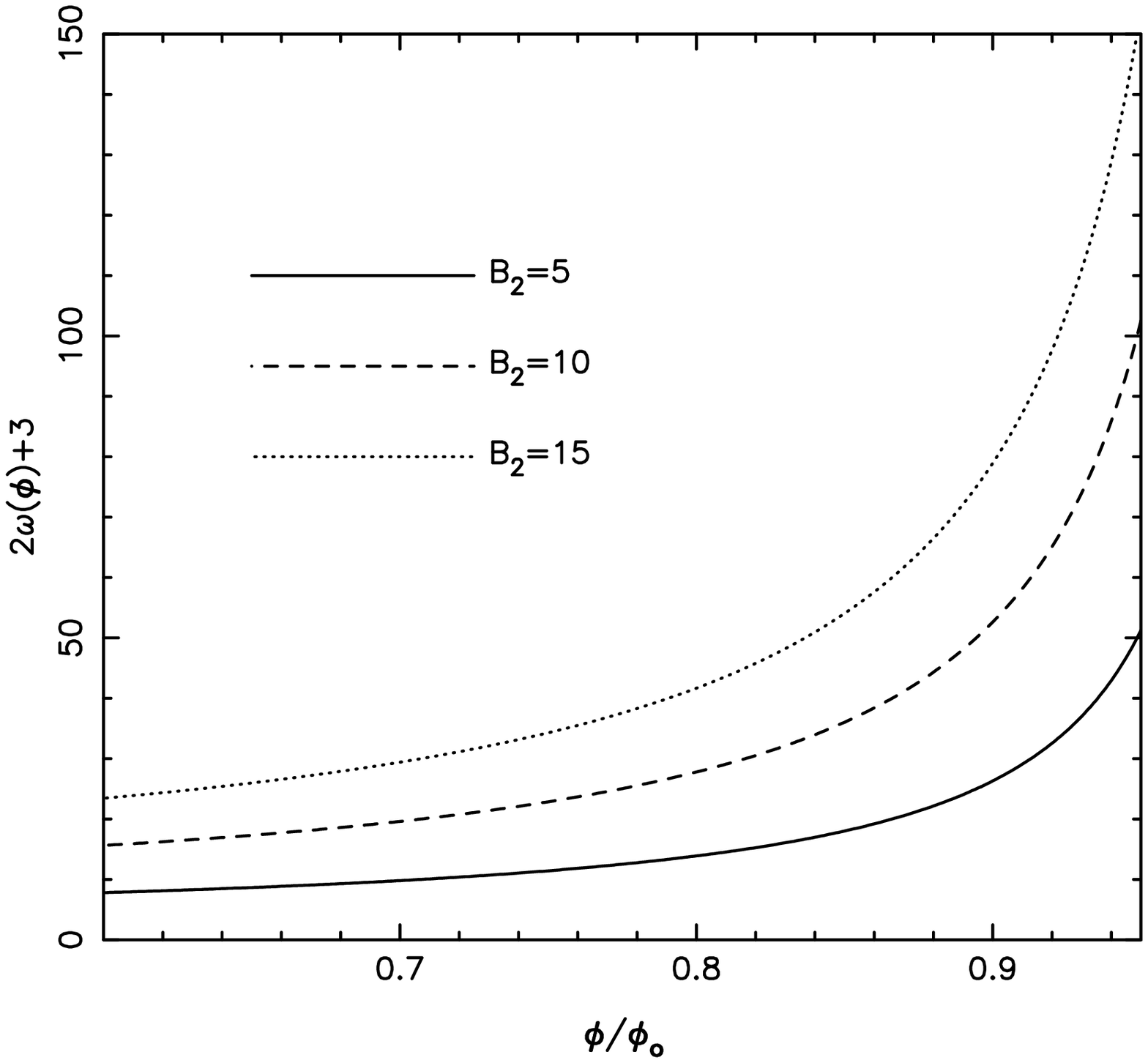}
\end{center}
\caption{Theory 3 with {\bf a)} $B_3=5$ and $\beta = 0.5$, $1$, $1.5$;
and 
{\bf b)} $\beta=2$ and $B_3=5$, $10$, $15$.}
\label{coup3}
\end{figure}

We shall study the three general classes of theory, some examples of
which are displayed in Figs.~\ref{coup1} -- \ref{coup3}:

\begin{itemize}
\item  {\bf Theory 1.} $2\omega (\phi )+3=2B_1|1-\phi /\phi
_0|^{-\alpha}$; $\alpha >0$, $B_1>0$ constants.

\item  {\bf Theory 2.} $2\omega (\phi )+3=B_2\left| \ln (\phi /\phi
_0)\right| ^{-2\delta }$, $\delta >0$, $B_2>0$ constants.

\item  {\bf Theory 3.} $2\omega (\phi )+3=B_3\left| 1-(\phi /\phi
_0)^\beta
\right| ^{-1}$, $\beta >0$, $B_3>0$ constants.
\end{itemize}

Theory 1 has been studied previously by Serna and Alimi \cite{SERNA},
Comer 
{\em et al.} \cite{DERU}, Barrow \cite{BAR6}, and Garcia-Bellido and
Quiros
\cite{JUAN}. Serna and Alimi \cite{SERNA} paid particular attention to
the
radiation eras of these models, extending their treatment to allow
$2\omega
+ 3<0$, leading to $\phi$-dominated initial conditions. Their analysis
dealt
mainly with the early-time behaviour of the models, we shall examine
in
detail the late-time approach to GR. Theories 1--3 will all permit
$\omega
\rightarrow \infty $ if $\phi \rightarrow \phi _0$ at late times and
span a
wide range of different rates of approach to GR in the weak-field
limit.
They can all be reduced to Barker's constant-$G$ theory \cite{BARK} as
$\phi
\rightarrow \phi_0$ for special parameter choices. Theories 1 and 3
approach
Brans-Dicke theory \cite{BD} as $\phi \rightarrow 0$ and the power law
form
studied in \cite{BAR1} as $\phi \rightarrow \infty$. General
functional
forms for $2\omega (\phi )+3$ can be expanded in a series of functions
of
this form and their asymptotic behaviours at small and large times
will be
dominated by one term of the above type. In fact, further restrictions
can
be placed upon the allowed theories within these three classes by the
weak-field limit requirements, as follows:

\noindent
{\bf Theory 1:} We see that $\omega \rightarrow \infty$ as $\phi
\rightarrow
\phi_0$ is guaranteed if $\alpha > 0$ and that 
\begin{equation}  \label{weak1}
\frac{\omega^{\prime}}{\omega^3} \propto \left( 1- \frac{\phi}{\phi_0}
\right)^{2\alpha -1}\,.
\end{equation}
Hence $\omega^{\prime}\omega^{-3} \rightarrow 0$ and the weak field
limit
will be compatible with observation as $\phi \rightarrow \phi_0$ so
long as $
\alpha > 1/2$.

\noindent
{\bf Theory 2:} Here, $\omega \rightarrow \infty$ as $\phi \rightarrow
\phi_0 $ for $\delta >0$, but 
\begin{equation}  \label{weak2}
\frac{\omega^{\prime}}{\omega^3} \propto
\left(\frac{\phi_0}{\phi}\right)
\ln^{4\delta -1} \left( \frac{\phi}{\phi_0} \right) \,,
\end{equation}
and this tends to zero as $\phi \rightarrow \phi_0$ so long as $\delta
> 1/4$.

\noindent
{\bf Theory 3:} Here, $\omega \rightarrow \infty$ as $\phi \rightarrow 
\phi_0 $, but 
\begin{equation}  \label{weak3}
\frac{\omega^{\prime}}{\omega^3} \propto \left[1 -
\left(\frac{\phi}{\phi_0}
\right)^{\beta}\right]\left(\frac{\phi}{\phi_0}\right)^{\beta-1}\,,
\end{equation}
and this tends to zero for all $\beta$ as $\phi \rightarrow \phi_0$.

These constraints on the parameter ranges of theories 1-3 are
independent of
the form of the cosmological solutions so long as $\phi \rightarrow \phi
_0$
at late cosmological times. The latter will introduce further
restrictions
on the allowed values of $\alpha $, $\beta $ and $\delta $. With the
exception of the $k\neq -1$ vacuum solutions, where there exists no
FRW
model within GR, we shall focus our attention on those particular
parameter
choices which reproduce Einstein's theory at late times and which we
have
just delineated in Eqs.~(\ref{weak1}) -- (\ref{weak3}). When $k=-1$
the
Milne model, $a\propto t$, supplies the general relativistic solution
at
late times.

\begin{center}
\section{Theory 1: $2\omega (\phi )+3=2B_{1}|1-\phi /\phi
_0| ^{-\alpha }$; $
\alpha>1/2$, $B_{1} >0$ constants.}
\label{th1} 
\end{center}

We study the evolution in the interval $\phi \in (0,\phi_0)$,
allowing us to
drop the modulus signs from $2\omega +3$. Making the
substitution $u= (1-
\phi/ \phi_{0})$, this bound becomes $u \in (0,1)$ and
Eq.~(\ref{vacint}) is 
\begin{equation}  \label{rad1}
\int \frac{\sqrt{2\omega+3}}{\phi} d\phi = \sqrt{2B_{1}} \int
\frac{du}{
u^{\alpha /2}(1-u)}-\sqrt{3}\ln{K}\,, \quad \alpha \neq
1,\quad 2\,,
\end{equation}
where $K$ is an integration constant. In the models which
asymptote to GR, $%
\omega \rightarrow \infty$ as $\phi \rightarrow \phi_0$.

\begin{center}
\subsection{Vacuum solutions ($k=0$)}

\subsubsection{Late-time behaviour}
\end{center}

For the theory to reduce to GR at late times we require $u
\rightarrow 0$ as 
$\eta \rightarrow \infty$. Choosing $\eta_0 = 0$ with $k=0$
in vacuum we
then have
\begin{equation}  \label{vacap1}
-\frac{1}{\lambda} \frac{u^{1-\alpha/2}}{1-\alpha/2} = \ln\left(
K\eta 
\right) \,,
\end{equation}
as $u \rightarrow 0$. $\lambda = \sqrt{3/2B_1}$ and its sign
determines the
sign of $(2\omega+3)^{1/2}$.
The requirement that 
$u 
\rightarrow 
0$ as $\eta
\rightarrow \infty$ demands we must have $\alpha >2$ and hence
$\lambda > 0$
to ensure the right-hand side of Eq.~(\ref{vacap1}) is
positive. Picking the
appropriate right-hand side from Eq.~(\ref{vacy}) we thus
obtain 
\begin{eqnarray}
\phi(\eta) & \rightarrow & \phi_0 \left\{ 1 - \left[ \lambda
\left(\frac{ 
\alpha-2} {\alpha}\right)
\right]^{\frac{2}{2-\alpha}}\ln^{\frac{2}{2-\alpha} 
} \left(K\eta\right) \right\}\,, \\
a^2(\eta) & \rightarrow & \frac{A}{\phi_0}\eta \left\{ 1 + \left[
\lambda 
\left(\frac{\alpha-2} {\alpha}\right)
\right]^{\frac{2}{2-\alpha}}\ln^{\frac{ 
2}{2-\alpha}} \left(K\eta\right) \right\}\,,
\end{eqnarray}
at large $\eta$. This leads to the asymptotic form for $t(\eta)$,  
\begin{equation}  \label{conf10}
t(\eta) \simeq \frac{2}{3} \sqrt{\frac{A}{\phi_0}}\eta^{3/2} \left\{1+
\frac{ 
1}{2} \left[ \lambda \left(\frac{\alpha-2}{2}\right)\right]
^{\frac{2}{ 
2-\alpha}}\ln^{\frac{2}{2-\alpha}}(K\eta)\right\}\,.
\end{equation}
The first-order inversion of this at large $\eta$ is 
\begin{equation}
\eta(t) \simeq
\left(\frac{3}{2}\right)^{2/3}\left(\frac{\phi_0}{A}\right) 
^{1/3}t^{2/3}\,.
\end{equation}
This result is obtained by ignoring the factor in curly brackets on
the
right-hand side of Eq.~(\ref{conf10}). To obtain the next-order
expression
we substitute this first-order result into the weakly-varying
$\ln(K\eta)$
term on the right-hand side of Eq.~(\ref{conf10}), take the curly
brackets
onto the left-hand side and solve for $\eta$. This yields 
\begin{equation}
\eta(t) \simeq
\left(\frac{3}{2}\right)^{2/3}\left(\frac{\phi_0}{A}\right)
^{1/3}t^{2/3} \left\{ 1-
\frac{1}{3}\left(\frac{2}{3}\right)^{\frac{2}{2
-\alpha}}\left[\lambda\left(\frac{\alpha-2}{2}
\right)\right]^{\frac{2}{
2-\alpha}}\ln^{\frac{2}{2-\alpha}} t\right\}\,,
\end{equation}
and $\eta \rightarrow \infty$ as $t \rightarrow \infty$. At late times
we
then obtain the solution 
\begin{eqnarray}  \label{a70}
\phi(t) & \rightarrow & \phi_0\left\{1 - \left[ \lambda \left(\frac{
\alpha-2 
}{3}\right)\right]^{\frac{2}{2-\alpha}}\ln^{\frac{2}{2-\alpha}} t
\right\}\,,\\
a(t) & \rightarrow & \left( \frac{3A}{2\phi_0} \right)^{1/3}
t^{1/3}\left\{1
+ \frac{1}{3} \left[\lambda\left(\frac{\alpha-2}{3}
\right)\right]^{\frac{2}{ 
2-\alpha}}\ln^{\frac{2}{2-\alpha}} t\right\}\,.
\end{eqnarray}

{\center \subsubsection{Early-time behaviour}}

At early times we see from Eqs.~(\ref{vacint}) and (\ref{rad1}) that
$u
\rightarrow 1$ as $\eta \rightarrow 0$ is the only possibility
consistent
with $u \in (0,1)$ and $\lambda>0$. Treating $u^{\alpha/2} \simeq 1$
in Eq.~(
\ref{rad1}) leads to the approximate relations 
\begin{equation}
\ln(1-u) = \lambda\ln(K\eta)\,,
\end{equation}
and 
\begin{eqnarray}  \label{a47}
\phi(\eta) & \simeq & \frac{\phi_0}{K^{-\lambda}} \eta^{\lambda}\,, \\
a^2(\eta) & \simeq & \frac{AK^{-\lambda}}{\phi_0} \eta^{1-\lambda}\,.
\end{eqnarray}
Taking the square root of the latter, integrating and inverting leads
to 
\begin{equation}  \label{conf11}
\eta(t) \simeq \left(\frac{\phi_0}{AK^{-\lambda}}\right)^{\frac{1}{3-
\lambda
}}\left(\frac{3-\lambda}{2}\right)^{\frac{2}{3-\lambda}} t^{\frac
{2}{3-\lambda}}\,, \quad \lambda \neq 3\,.
\end{equation}
When $0<\lambda<3$ we see from the above that $t \rightarrow 0$ as
$\eta
\rightarrow 0$ and when $\lambda>3$ we have $t \rightarrow -\infty$ as
$\eta
\rightarrow 0$. In both cases we obtain the early-time behaviour 
\begin{eqnarray}  \label{a9}
\phi(t) & \rightarrow & A^{\frac{\lambda}{\lambda-3}} \left(
\frac{\phi_0}{ 
K^{-\lambda}}\right)^{\frac{3}{3-\lambda}} \left( \frac{2}{3-\lambda} 
\right)^{\frac{2\lambda}{\lambda-3}} t^{\frac{2\lambda}{\lambda-3}}\,,
\\
a(t) & \rightarrow & \left(\frac{A K^{-\lambda}}{\phi_0}\right)^{\frac
{1}{3-\lambda}}\left(\frac{3-\lambda}{2}\right)^{\frac{1-\lambda}
{3-\lambda}} t^{\frac{1-\lambda}{3-\lambda}}\,.
\end{eqnarray}
When $\lambda = 3$, Eq.~(\ref{conf11}) becomes 
\begin{equation}
\eta(t) = \exp \left( \sqrt{\frac{\phi_0 K^{3}}{A}} t \right)\,,
\end{equation}
and $t \rightarrow -\infty$ as $\eta \rightarrow 0$. The early-time
evolution is 
\begin{eqnarray}
\label{phit10} 
\phi(t) & \rightarrow & \phi_0 K^3 \exp\left(3 \sqrt{\frac{\phi_0
K^{3}}{A}} 
t \right)\,, \\
\label{at10}
a(t) & \rightarrow & \sqrt{\frac{A}{\phi_0 K^{3}}} \exp \left( -
\sqrt{\frac{
\phi_0 K^{3}}{A}}t \right)\,,
\end{eqnarray}
as $t \rightarrow -\infty$. Differentiating Eq.~(\ref{a9}), we deduce
that
the early-time models will be expanding as long as $\lambda>3$ or
$\lambda<1$.

{\center \subsubsection{Minima}}

We now probe the $k=0$ vacuum models generated by this choice of
coupling
for the existence of expansion minima. Eq.~(\ref{mintest}) for these
universes becomes 
\begin{equation}
u_{*} = \left( \frac{1}{\lambda}\right)^{2/\alpha}\,,
\end{equation}
at a minimum. For $u_{*}$ to be real for all $\alpha$ requires
$\lambda>0$
which is guaranteed when $u \in (0, \; 1)$. In general, however, we
find 
\begin{eqnarray}
0 < \lambda < 1 & \leftrightarrow & u_{*}>1\,, \\
\lambda = 1 & \leftrightarrow & u_{*} = 1\,, \\
\lambda > 1 & \leftrightarrow & u_{*}<1\,,
\end{eqnarray}
and when $u \in (0, 1)$, minima are only present when $\lambda>1$. We
have
the early-time solution for this model (Eqs.~(\ref{phit10}) and
(\ref{at10}
)) in which $a(0)=0$ for $\lambda>1$ and the universe experiences a
phase of
contraction before bouncing and approaching late time general
relativistic
expansion. When $\lambda=1$, $a(0)$ itself becomes a non-zero minimum
($a^2 \rightarrow A/K\phi_0$).

{\center \subsubsection{Exact solution}}

The models with $\alpha =1$ are not described by the solutions
presented
above. These models tend (as $\phi \rightarrow \phi _0$) to the theory
of
gravity proposed by Barker \cite{BARK}, for their evolution we find
the
exact results 
\begin{eqnarray}
\label{km1et1vp}
\phi (\eta ) &=&\frac{4\phi _0K^{-\lambda }\eta ^{-\lambda }}{\left(
K^{-\lambda }\eta ^{-\lambda }+1\right) ^2}\,,\\  
\label{km1et1va}
a^2(\eta ) &=&\frac A{4\phi _0K^{-\lambda }}\eta ^{1+\lambda }\left(
K^{-\lambda }\eta ^{-\lambda }+1\right) ^2\,.
\end{eqnarray}
When $\eta \rightarrow \infty $ however, we observe that there is no
combination of parameters allowing $\phi \rightarrow \phi _0$.
Consequently,
we shall not pursue this model any further. Theories with $\alpha =2$
have
been solved exactly and studied earlier by Barrow \cite{BAR1},
\cite{BAR6}
and in ref. \cite{SERNA}.

\begin{center}
\subsection{Vacuum solutions ($k=-1$)}

\subsubsection{Late-time behaviour}
\end{center}

When $k=-1$ the integral equation for the field evolution becomes
(setting
$ \eta _0=0$) 
\begin{equation}
-\frac 1\lambda \int \frac{du}{u^{\alpha /2}(1-u)}=\ln \left[ K\tanh
\eta 
\right] \,.  \label{negint}
\end{equation}
at late times. As $\eta \rightarrow \infty $, $\tanh \eta \rightarrow
1-2e^{-2\eta }$. Demanding $u\rightarrow 0$ in this extreme, Eq.~(\ref
{negint}) may be approximated by 
\begin{equation}
-\frac 1\lambda \frac{u^{1-\alpha /2}}{1-\alpha /2}\simeq \ln K-2\exp
\left(
-2\eta \right) \,.
\end{equation}
As $\eta \rightarrow \infty $ the right-hand side is finite and we
require $
1-\alpha /2>0$ (ie, $\alpha <2$) in order that the left-hand side does
not
diverge as $u\rightarrow 0$. We also require $K=1$ to ensure that
$\eta
\rightarrow \infty $ and $u\rightarrow 0$ correspond to the same
limit. The
right-hand side approaches zero from below as $\eta $ increases and
$u>0$;
thus to keep the left-hand side negative we need $\lambda >0$. The
form of $
y(\eta )$ for the negatively curved vacuum models, selected from
Eq.~(\ref
{vacy}), approaches $Ae^{2\eta }/4$ as $\eta \rightarrow \infty $ and
the
late-time solutions are 
\begin{eqnarray}
\label{phi45}
\phi (\eta ) &\rightarrow &\phi _0\left[ 1-\left\{ \lambda (2-\alpha
)\right\} ^{\frac 2{2-\alpha }}\exp \left( \frac{4\eta }{\alpha
-2}\right)
\right] \,, \\
\label{a45}
a^2(\eta ) &\rightarrow &\frac A{4\phi _0}e^{2\eta }\left[ 1+\left\{
\lambda 
(2-\alpha )\right\} ^{\frac 2{2-\alpha }}\exp \left( \frac{4\eta
}{\alpha -2}
\right) \right] \,,
\end{eqnarray}
as $\eta \rightarrow \infty $. Integrating and asymptotically
inverting the
second of these expressions we obtain the $\eta (t)$ relation 
\begin{equation}
\eta (t)\rightarrow \ln \left\{ 2\sqrt{\frac{\phi _0}A}t\left[ 1-\frac
12\left\{ \lambda (2-\alpha )\right\} ^{\frac 2{2-\alpha }}\left(
\frac{ 
\alpha -2}{\alpha +2}\right) \left( 2\sqrt{\frac{\phi _0}A}\right)
^{\frac
4{\alpha -2}}t^{\frac 4{\alpha -2}}\right] \right\} \,. 
\label{etakm1}
\end{equation}
Substituting this back into Eqs.~(\ref{phi45}) and (\ref{a45}) yields
the
asymptotic forms 
\begin{eqnarray}
\phi (t) &\rightarrow &\phi _0\left[ 1-\left\{ \lambda (2-\alpha
)\right\}
^{\frac 2{2-\alpha }}\left( 2\sqrt{\frac{\phi _0}A}\right) ^{\frac
4{\alpha 
-2}}t^{\frac 4{\alpha -2}}\right] \,,  \label{akm1} \\
a(t) &\rightarrow &t\left[ 1+\frac 2{\alpha +2}\left\{ \lambda
(2-\alpha 
)\right\} ^{\frac 2{\alpha -2}}\left( 2\sqrt{\frac{\phi _0}A}\right)
^{\frac
4{\alpha -2}}t^{\frac 4{\alpha -2}}\right] \,,
\end{eqnarray}
as $t\rightarrow \infty $. We note the asymptotic approach of this
model to
the general relativistic Milne universe at late times.

{\center \subsubsection{Early-time behaviour}}

At early times $u \rightarrow 1$ as $\eta \rightarrow 0$ and 
\begin{equation}
\ln (1-u) \simeq \lambda \ln\left[\tanh \eta \right]\,,
\end{equation}
which leads to 
\begin{eqnarray}
\phi(\eta) & \rightarrow & \phi_0 \eta^{\lambda}\,, \\
a^2(\eta) & \rightarrow & \frac{A}{\phi_0} \eta^{1-\lambda}\,.
\end{eqnarray}
The functional form of this early-time behaviour is identical to that
detailed in Eqs.~(\ref{conf11}) -- (\ref{at10}), after enforcing the
choice $ K=1$.

{\center \subsubsection{Minima}}

Searching for stationary points in the scale-factor evolution for this
theory, Eq.~(\ref{mintest}) identifies 
\begin{equation}
\cosh (2\eta _{*})=\lambda u_{*}^{\alpha /2}\,.
\end{equation}
The $\cosh $ function is bounded below by unity, as is the value of $
u^{-\alpha /2}$ when $u\in (0,1)$, which gives $\lambda \geq 1$ as the
condition for the existence of a stationary point. When $0<\lambda <1$
the
universe expands monotonically away from an initial singularity at
$\eta =0$ 
. When $1\leq \lambda \leq 3$, $a\rightarrow \infty $ as $\eta
\rightarrow 0$
, the subsequent evolution will contain a minimum, allowing the
initial 
contraction to `bounce' back into late-time expansion. When $\lambda
>3$ the
evolution is expanding at all times. The special case $\alpha =1$ is
treated
by transforming $\eta +\eta _0\rightarrow \tanh (\eta +\eta _0)$ in
Eqs.~( 
\ref{km1et1vp}) and (\ref{km1et1va}). Solutions for the $\alpha =2$
case may
be obtained from the solutions in ref.\cite{BAR6} after applying the
same transformation.

\begin{center}
\subsection{Vacuum solutions ($k=+1$)}

\subsubsection{Late-time behaviour}
\end{center}

The positively curved ($k=+1$) vacuum solutions are governed by 
\begin{equation}
-\frac{1}{\lambda}\int\frac{du}{u^{\alpha/2}(1-u)} = \ln \left[ K \tan
\eta \right]\,,
\end{equation}
taking $\eta_0 = 0$. In the GR limit ($u \rightarrow 0$) this is
approximated by 
\begin{equation}
-\frac{1}{\lambda}\frac{u^{1-\alpha/2}}{1-\alpha/2} \simeq \ln \left[
K \tan
\eta \right]\,.
\end{equation}
In conjunction with Eq.~(\ref{vacy}) we obtain 
\begin{eqnarray}
\phi(\eta) & \rightarrow & \phi_0 \left[1 - \left\{\lambda\left(\frac
{\alpha-2}{2}\right)\right\}^{\frac{2}{2-\alpha}}\ln^{\frac{2}
{2-\alpha}}
\left(K \tan \eta \right)\right]\,, \\
a^2(\eta) & \rightarrow & \frac{A}{2\phi_0} \frac{\sin 2\eta}{\left[1-
\left\{ \lambda \left(\frac{\alpha-2}{2}\right)\right\}^{\frac{2}
{2-\alpha}}\ln^{\frac{2}{2-\alpha}}\left(K\tan\eta\right)\right]}\,.
\end{eqnarray}
These expressions approach GR when $\alpha<2$ as $\eta \rightarrow
\tan^{-1}(K^{-1})$ and when $\alpha>2$ as $\eta \rightarrow n\pi/2$
where $n$
is an integer. The behaviour when $\alpha=2$ can be extracted from the
$k=0$
treatment in ref. \cite{BAR6} by substituting $2(\eta + \eta_0)
\rightarrow
\tanh(\eta + \eta_0)$.

\begin{center}
\subsection{Radiation solutions ($k=0$)}

\subsubsection{Late-time behaviour}
\end{center}

When $k=0$ Eq.~(\ref{radint}) leads to the asymptotic behaviour 
\begin{equation}
-\frac 1\lambda \frac{u^{1-\alpha /2}}{1-\alpha /2}=\ln
{K}-\frac{2\eta_0} 
\eta \,,  \label{radap1}
\end{equation}
where we have picked $2\Gamma \eta _0=A$ to fix the arbitrary
integration $
\eta _0$. The requirement that $u\rightarrow 0$ as $\eta \rightarrow
\infty $
then demands $K=1$ and $\alpha <2$. This upper bound on $\alpha $, in
conjunction with the lower limit implied by the requirement that
$\omega
^{\prime }\omega ^{-3}\rightarrow 0$, yields the powerful constraint $
1/2<\alpha <2$. We examine models for which $u\in (0,1)$,
corresponding to $
\phi \in (0,\phi _0)$. In this case we have the constraint $\lambda
\eta _0>0 
$. Using $y(\eta )$ from Eq.~(\ref{rady}), the late-time evolution of
$\phi $
and $a$ as functions of $\eta $ is 
\begin{eqnarray}
\phi (\eta ) &\rightarrow &\phi _0\left[ 1-\left( \frac{(2-\alpha
)\lambda
\eta _0}\eta \right) ^{\frac 2{2-\alpha }}\right] \,,  \label{a4} \\
a^2(\eta ) &\rightarrow &\frac{\Gamma \eta ^2}{\phi _0}\left[ 1+\left(
\frac{
(2-\alpha )\lambda \eta _0}\eta \right) ^{\frac 2{2-\alpha }}\right]
\,.
\end{eqnarray}
The latter of these allows us to deduce 
\begin{equation}
\eta (t)\rightarrow \left( \frac{4\phi _0}\Gamma \right)
^{1/4}t^{1/2}\left[
1-\frac 14\left( \frac{\alpha -2}{\alpha -1}\right) \left\{ \lambda
\left(
2-\alpha \right) \eta _0\right\} ^{\frac 2{2-\alpha }}\left(
\frac{4\phi_0} 
\Gamma \right) ^{\frac 1{2(\alpha -2)}}t^{\frac 1{\alpha -2}}\right]
\,, 
\label{etarkzt1}
\end{equation}
and hence, the evolution as a function of cosmic time: 
\begin{eqnarray}
\phi (t) &\rightarrow &\phi _0\left[ 1-\left( \lambda ^2{\eta
_0}^2(\alpha
-2)^2\sqrt{\frac \Gamma {4\phi _0}}\right) ^{\frac 1{2-\alpha
}}t^{\frac
1{\alpha -2}}\right] \,,  \label{arkzt1} \\
a(t) &\rightarrow &\left( \frac{4\Gamma }{\phi _0}\right)
t^{1/2}\left[
1+\frac \alpha {4(\alpha -1)}\left\{ \lambda ^2\eta _0^2(\alpha
-2)^2\sqrt{
\frac \Gamma {4\phi _0}}\right\} ^{\frac 1{2-\alpha }}t^{\frac
1{\alpha
-2}}\right] \,,
\end{eqnarray}
as $t\rightarrow \infty ,$ so there is a power-law approach to the GR
solution at late times.

{\center \subsubsection{Early-time behaviour}}

At early times, we analyse the behaviour in the neighbourhood of $u
\rightarrow 1$, ie 
\begin{equation}  \label{app10}
\ln(1-u) \simeq \lambda \ln \left[ \frac{\eta}{\eta + 2\eta_0}
\right]\,. 
\end{equation}
The left-hand side of Eq.~(\ref{app10}) tends to $-\infty$ as GR is
approached. In order that the right-hand side approach the same limit
we see
that $\eta \rightarrow 0$ when $\lambda>0$ and $\eta \rightarrow
-2\eta_0$
when $\lambda<0$. Analysing the former case leads to 
\begin{eqnarray}
\phi(\eta) & \rightarrow & (2\eta_0)^{-\lambda}\phi_0
\eta^{\lambda}\,, \\
a^2(\eta) & \rightarrow & \frac{(2\eta_0)^{1+\lambda} \Gamma}{\phi_0}
\eta^{1-\lambda}\,,
\end{eqnarray}
after setting $2\Gamma \eta_0 =A$. The behaviour of this system as a
function of cosmic time $t$ mirrors that described in
Eqs.~(\ref{conf11}) --
(\ref{at10}) with $K^{\lambda}$ replaced by $(2\eta_0)^{-\lambda}$ and
$A$
replaced by $2\Gamma \eta_0$. The bound $\lambda \eta_0>0$ implies
$\eta_0
>0 $ when $\lambda>0$. When $\lambda<0$, $\eta \rightarrow -2\eta_0$
at
early times. The inequality $\lambda \eta_0>0$ implies $\eta_0<0$ and
hence $
-2\eta_0 >0$. the evolution is approximately 
\begin{eqnarray}
\phi(\eta) & \simeq & \phi_0 (-2\eta_0)^{\lambda}(\eta + 2\eta_0)
^{-\lambda}\,, \\
a^2(\eta) & \simeq & \frac{\Gamma}{\phi_0}(-2\eta_0)^{1-\lambda}(\eta
+
2\eta_0)^{1+\lambda}\,.
\end{eqnarray}
The $t$-parametrised behaviour of these equations is given by
Eqs.~(\ref
{conf11}) -- (\ref{at10}), after applying the transformations 
\begin{eqnarray}
K^{\lambda} & \rightarrow & (-2\eta_0)^{\lambda}\,, \\
\lambda & \rightarrow & -\lambda\,, \\
A & \rightarrow & -2\Gamma \eta_0\,.
\end{eqnarray}

{\center \subsubsection{Minima}}

As with the vacuum solutions, we can search for turning points of
$a^2$
when 
$\lambda >0$. Eq.~(\ref{mintest}) is 
\begin{equation}
\frac{\eta _{*}}{\eta _0}=-\left( 1-\lambda u_{*}^{\alpha /2}\right)
\,,
\end{equation}
where subscript $*$ denotes the value of a quantity at the stationary
point.
When $0<\lambda <1$, the scale-factor expands away from $a=0$ at $\eta
=0$
by Eq.~(\ref{a47}). Monotonicity of $\eta $ implies $\eta _{*}\geq 0$
and we
know $\eta _0>0$ from the sign of $\lambda $, which together give rise
to
the inequality 
\begin{equation}
u_{*}^{\alpha /2}\geq \frac 1\lambda \,.  \label{mincond}
\end{equation}
The range of $0\leq u\leq 1$ together with the condition $\alpha >0$
confirm 
$\lambda \geq 1$ as a necessary condition for the existence of minima.
When $
\lambda =1$, $\eta _{*}=0$, and the universe expands from a
non-singular
state of size $a\simeq (2\eta _0)^{1+\lambda }\Gamma /\phi _0$. When $ 
\lambda >1$ the solution is initially contracting, bounces at $\eta
_{*}$
and tends to general relativistic behaviour at late times.
Eq.~(\ref{mincond}
) still applies when $\lambda <0$, and asserts that none of them can
contain
stationary points in the evolution of the scale-factor. This does not
effect
the models in which $-1<\lambda <0$, which begin from $a=0$ at $\eta
=-2\eta
_0$ and monotonically expand. When $\lambda \leq -1$ they are
initially
contracting and, due to the absence of minima, will always contract
and
never approach late-time general relativistic expansion.

{\center \subsubsection{Exact solution}}

The case $\alpha =1$ possesses a simple exact form, which is
instructive.
Solving Eqs.~(\ref{radint}) and (\ref{rady}) leads to 
\begin{eqnarray}
\phi (\eta ) &=&\frac{4\phi _0K^\lambda \eta ^\lambda \left( \eta
+2\eta
_0\right) ^\lambda }{\left[ \left( \eta +2\eta _0\right) ^\lambda
+K^\lambda
\eta ^\lambda \right] ^2}\,, \\
a^2(\eta ) &=&\frac{\Gamma \eta \left( \eta +\eta _0\right) \left[
\left(
\eta +2\eta _0\right) ^\lambda +K^\lambda \eta ^\lambda \right]
^2}{4\phi
_0K^\lambda \eta ^\lambda \left( \eta +2\eta _0\right) ^\lambda }\,,
\end{eqnarray}
where in these expressions we have again made the special choice
$2\Gamma
\eta _0=A$. Examining the large $\eta $ limit reveals that $\phi
\rightarrow
\phi _0$ if and only if $K=1$. In this case we integrate the
asymptotic $
a(\eta )$ to obtain 
\begin{equation}
\eta (t)\rightarrow \left( \frac{4\phi _0}\Gamma \right)
^{1/4}t^{1/2}\left[
1-\frac{\eta _0}2\left( \frac \Gamma {4\phi _0}\right)
^{1/4}t^{-1/2}+\frac
14\left( \lambda ^2-\frac 14\right) \eta _0^2\left( \frac \Gamma
{4\phi
_0}\right) ^{1/2}\frac{\ln t}t\right] \,,
\end{equation}
and hence 
\begin{eqnarray}
\phi (t) &\rightarrow &\phi _0\left( 1-\lambda ^2\eta _0^2\sqrt{\frac
\Gamma
{4\phi _0}}t^{-1}\right) \,, \\
a(t) &\rightarrow &\left( \frac{4\Gamma }{\phi _0}\right)
^{1/4}t^{1/2}\left[ 1+\frac 14\left( \lambda ^2-\frac 14\right) \eta
_0^2\left( \frac \Gamma {4\phi _0}\right) ^{1/2}\frac{\ln t}t\right]
\,,
\end{eqnarray}
valid as $t\rightarrow \infty $. At early times the solution becomes 
\begin{eqnarray}
\phi (\eta ) &\rightarrow &2^{2-|\lambda |}\eta _0^{-|\lambda |}\phi
_0\eta
^{|\lambda |}\,, \\
a^2(\eta ) &\rightarrow &\frac{2^{|\lambda |-2}\Gamma \eta
_0^{1+|\lambda |}
}{\phi _0}\eta ^{1-|\lambda |}\,,.
\end{eqnarray}
as $\eta \rightarrow 0$. The behaviour of the $t$-parametrised version
of
this model is given by Eqs.~(\ref{conf11}) -- (\ref{at10}) after
applying
the transformations 
\begin{eqnarray}
\phi _0 &\rightarrow &2^{2-|\lambda |\eta _0^{-|\lambda |}\phi _0}\,,
\\
A &\rightarrow &\Gamma \eta _0\,,
\end{eqnarray}
and remembering $K=1$. Minima are a feature of this model when
$|\lambda |>1$.

{\center \subsection{Radiation solution ($k=-1$)}}

When $k=-1$, Eq.~(\ref{rad1}) becomes 
\begin{equation}  \label{app2}
-\frac{1}{\lambda}\int\frac{du}{u^{\alpha/2}(1-u)} - \ln K = \ln
\left|\frac{ 
(A^2-\Gamma^2)^{1/2} e^{2(\eta+\eta_0)}-\Gamma -A}
{(A^2-\Gamma^2)^{1/2}
e^{2(\eta+\eta_0)}-\Gamma +A}\right|\,.
\end{equation}

{\center \subsubsection{Late-time behaviour}}

Expanding the right-hand side at large $\eta $ and integrating on the
left
as $u\rightarrow 0$ leads to the approximate formula 
\begin{equation}
\frac 1\lambda \frac{u^{1-\alpha /2}}{1-\alpha /2}\simeq \frac{2A}{
(A^2-\Gamma ^2)^{1/2}}e^{-2\eta }\,,
\end{equation}
where we have chosen $K=1$ to ensure $\phi \rightarrow \phi _0$ at
late
times. The right-hand side of this expression tends to zero at late
times
and consistency on the left as $u\rightarrow 0$ requires $1-\alpha
/2>0$, or 
$\alpha <2$. If $\phi _\eta >0$, as it must be when $\phi \rightarrow
\phi_0
$ and $0\leq \phi \leq \phi _0$, we recover the inequality $\lambda
A>0$.
Selecting the $k=-1$ form for $y(\eta )$ from Eq.~(\ref{vacy}) we have
the approximate solutions 
\begin{eqnarray}
\phi (\eta ) &\rightarrow &\phi _0\left[ 1-\left\{ \frac{A\lambda
(2-\alpha )
}{(A^2-\Gamma ^2)^{1/2}}\right\} ^{\frac 2{2-\alpha }}e^{-\frac{4\eta
}{
2-\alpha }}\right] \,,  \label{a16} \\
a^2(\eta ) &\rightarrow &\frac{(A^2-\Gamma ^2)^{1/2}}{4\phi
_0}e^{2\eta
}\left[ 1+\left\{ \frac{A\lambda (2-\alpha )}{(A^2-\Gamma
^2)^{1/2}}\right\}
^{\frac 2{2-\alpha }}e^{-\frac{4\eta }{2-\alpha }}\right] \,,
\end{eqnarray}
as $\eta \rightarrow \infty $. Integrating and inverting
Eq.~(\ref{a16}) to
next-to-lowest-order yields 
\begin{eqnarray}
\eta (t) &\rightarrow &\ln \left[ \frac{2\phi _0^{1/2}t}{(A^2-\Gamma
^2)^{1/4}}\left\{ 1
\begin{array}{c}
\\ 
\\ 
\\ 
\end{array}
\!\!\!\!\!\!\right. \right.   \nonumber  \label{km1th1reta} \\
&&\mbox{}\left. \left. +\frac 12\left( \frac{2+\alpha }{2-\alpha
}\right)
\left[ \frac{A\lambda (2-\alpha )}{(A^2-\Gamma ^2)^{1/2}}\right]
^{\frac
2{2-\alpha }}\left[ \frac{2\phi _0^{1/2}}{(A^2-\Gamma
^2)^{1/4}}\right]
^{-\frac 4{2-\alpha }}t^{-\frac 4{2-\alpha }}
\begin{array}{c}
\\ 
\\ 
\\ 
\end{array}
\!\!\!\!\!\!\right\} \right] \,,
\end{eqnarray}
$t\rightarrow \infty $ as $\eta \rightarrow \infty $ and hence 
\begin{eqnarray}
\phi (t) &\rightarrow &\phi _0\left[ 1-\left\{ \frac{A\lambda
(2-\alpha )}{
4\phi _0}\right\} ^{\frac 2{2-\alpha }}t^{-\frac 4{2-\alpha }}\right]
\,,
\label{km1th1ra} \\
a(t) &\rightarrow &t\left[ 1+\frac 2{(2-\alpha )}\left\{
\frac{A\lambda
(2-\alpha )}{4\phi _0}\right\} ^{\frac 2{2-\alpha }}t^{-\frac
4{2-\alpha
}}\right] \,,
\end{eqnarray}
as $t\rightarrow \infty $. Again, we observe power-law approach to the
GR
Milne solution at large $t$ since $\alpha <2$.

{\center \subsubsection{Early-time behaviour}}

In the early-time limit we find $u\rightarrow 1$ and Eq.~(\ref{app2})
becomes 
\begin{equation}
\ln (1-u)\rightarrow \lambda \ln \left| \frac{(A^2-\Gamma
^2)^{1/2}e^{2(\eta
+\eta _0)}-\Gamma -A}{(A^2-\Gamma ^2)^{1/2}e^{2(\eta +\eta _0)}-\Gamma
+A}
\right| \,,
\end{equation}
and as $u\rightarrow 1$, $\ln (1-u)\rightarrow -\infty $. To simplify
the
analysis we pick $\eta _0$ such that $u\rightarrow 1$ as $\eta
\rightarrow 0$
, which requires 
\begin{equation}
e^{2\eta _0}=\left( \frac{A+\Gamma }{A-\Gamma }\right) ^{1/2}\,.
\end{equation}
Combined with the $k=-1$ version of $y(\eta )$ from Eq.~(\ref{rady}),
this
leads to the limiting forms 
\begin{eqnarray}
\phi (\eta ) &\rightarrow &\phi _0\left( \frac{A+\Gamma }A\right) \eta
^\lambda \,, \\
a^2(\eta ) &\rightarrow &\frac A{\phi _0}\left( \frac{A+\Gamma
}A\right)
^{-\lambda }\eta ^{1-\lambda }\,,
\end{eqnarray}
and the $t$-parametrised evolution will be that of the flat vacuum
model
described by Eqs.~(\ref{conf11}) -- (\ref{at10}) after we apply the
relabelling 
\begin{equation}
K\rightarrow \frac{A+\Gamma }A\,.
\end{equation}

{\center \subsubsection{Minima}}

To complete the study of this class of solutions, we search for points
where
the gradient of $a^2$ vanishes, Eq.~(\ref{mintest}) gives the
condition 
\begin{equation}
\left( 1-\frac{\Gamma ^2}{A^2}\right) ^{1/2}=\frac{\lambda u^{\alpha
/2}}{
\cosh \left[ 2(\eta +\eta _0)\right] }\,,
\end{equation}
for a stationary point to exist. Since $\alpha >0$, $u<1$ and $\cosh
[2(\eta
+\eta _0)]>1$ this expression is equivalent to the inequalities 
\begin{eqnarray}
\left( 1-\frac{\Gamma ^2}{A^2}\right) ^{1/2} &<&\lambda \,,\quad
\lambda
>0\,, \\
\left( 1-\frac{\Gamma ^2}{A^2}\right) ^{1/2} &>&\lambda \,,\quad
\lambda
<0\,.
\end{eqnarray}
Since $A^2>\Gamma ^2$, we know that $0<(1-\Gamma ^2/A^2)^{1/2}<1.$
Hence,
models with $\lambda >1$, in which $a$ begins collapsing from
infinity, 
will
bounce to mimic GR expansion at late times.

{\center \subsection{Radiation solutions ($k=+1$)}}

If we selecting the necessary right-hand side from Eq.~(\ref{radint})
for
the finite, closed-universe models with $k=+1$, then Eq.~(\ref{rad1})
is 
\begin{equation}
-\frac 1\lambda \int \frac{du}{u^{\alpha /2}(1-u)}=\ln \left[ K\left(
\frac{
\Gamma \tan (\eta +\eta _0)+(\Gamma ^2+A^2)^{1/2}-A}{\Gamma \tan (\eta
+\eta
_0)+(\Gamma ^2+A^2)^{1/2}+A}\right) \right] \,.
\end{equation}
As $u\rightarrow 0$, we obtain the following approximate expressions
for the
behaviour of $\phi $ and $a$: 
\begin{eqnarray}
\phi (\eta ) &\rightarrow &\phi _0\left\{ 1-\left[ \lambda \left(
\frac{
\alpha -2}2\right) \right] ^{\frac 2{2-\alpha }}
\begin{array}{c}
\\ 
\\ 
\end{array}
\!\!\!\!\!\!\right.   \nonumber \\
&&\mbox{}\left. \times \ln ^{\frac 2{2-\alpha }}\left[ K\left(
\frac{\Gamma
\tan (\eta +\eta _0)+(\Gamma ^2+A^2)^{1/2}-A}{\Gamma \tan (\eta +\eta
_0)+(\Gamma ^2+A^2)^{1/2}+A}\right) \right] 
\begin{array}{c}
\\ 
\\ 
\end{array}
\!\!\!\!\!\!\right\} \,, \\
a^2(\eta ) &\rightarrow &\frac 1{2\phi _0}\left\{ \Gamma +(\Gamma
^2+A^2)^{1/2}\sin [2(\eta +\eta _0)]\right\} \left\{ 1-\left[ \lambda
\left( 
\frac{\alpha -2}2\right) \right] ^{\frac 2{2-\alpha }}
\begin{array}{c}
\\ 
\\ 
\end{array}
\!\!\!\!\!\!\right.   \nonumber \\
&&\mbox {}\left. \times \ln ^{\frac 2{2-\alpha }}\left[ K\left(
\frac{\Gamma
\tan (\eta +\eta _0)+(\Gamma ^2+A^2)^{1/2}-A}{\Gamma \tan (\eta +\eta
_0)+(\Gamma ^2+A^2)^{1/2}+A}\right) \right] 
\begin{array}{c}
\\ 
\\ 
\end{array}
\!\!\!\!\!\!\right\} ^{-1}\,,
\end{eqnarray}
which are valid as $\phi \rightarrow \phi _0$.

{\center \subsection{Perfect-fluid solutions ($k=0$)}}

We shall analyse the late-time behaviour of Theory 1, specified in
Section 
\ref{coupling}, by the device of using the solution defined by the
choice of 
field evolution 
\begin{equation}
\phi (\xi )=\phi _0\exp \left( A\ln ^B\xi \right) \,,  \label{phi40}
\end{equation}
where $A$ and $B$ are constants. This gives rise to the generating
function 
\begin{equation}
g(\xi )=\frac{\xi ^2}{AB}\ln ^{1-B}\xi \,,
\end{equation}
and the scale-factor 
\begin{equation}
a^{3(2-\gamma )}=\frac{a_0^{3(2-\gamma )}}{AB\phi _0}\xi ^2\ln
^{1-B}\xi
\exp \left( A\ln ^B\xi \right) \,.  \label{a40}
\end{equation}
From Eq.~(\ref{a40}) it is clear that to keep the left-hand side
positive we
require the combination $(\xi -1)^{B-1}AB>0$. This leads to 
\begin{equation}
f(\xi )=\frac{\xi ^2}{AB}\ln ^{1-B}\xi +\frac{4-3\gamma }4\xi ^2-D\,,
\end{equation}
and hence 
\begin{equation}
2\omega (\xi )+3=\frac{4-3\gamma }{3(2-\gamma )^2}\frac{\left[ \frac
2{AB}\xi \ln ^{1-B}\xi +\frac{1-B}{AB}\xi \ln ^{-B}\xi
+\frac{4-3\gamma }
2\xi \right] ^2}{\frac{\xi ^2}{AB}\ln ^{1-B}\xi +\frac{4-3\gamma }4\xi
^2}\,.
\end{equation}
We require $\phi \rightarrow \phi _0$ at late times and this occurs as
$\xi
\rightarrow 1$ when $B>0$ and as $\xi \rightarrow \infty $ when $B<0$.
When $
B=0$, $\phi =\phi _0e^A$ and when $A=0$, $\phi =\phi _0$; in both
cases the
theory is GR at all times.

{\center \subsubsection{Late-time behaviour}}

When $0<B<1$, we find from Eq.~(\ref{timeperf}) 
\begin{equation}
t\propto \ln ^{\frac{1-B}{2-\gamma }}\xi \,,
\end{equation}
as $\xi \rightarrow 1$. The power of the log is always positive for
$0\leq
\gamma <2$ and $t\rightarrow 0$ in this limit. Since we are only
concerned
with limits corresponding to large $t$ we exclude the range $0<B<1$.
When $
B>1$ the coupling function tends to 
\begin{equation}
2\omega (\xi )+3\rightarrow \frac{4-3\gamma }{3(2-\gamma
)^2}\frac{(1-B)^2}{
AB}\ln ^{-(B+1)}\xi \,,  \label{omega1}
\end{equation}
and Eq.~(\ref{timeperf}) becomes 
\begin{equation}
dt=\frac{1-B}{3(2-\gamma )}\frac{a_0^{3(\gamma -1)}}{(AB\phi
_0)^{\frac{ 
\gamma -1}{2-\gamma }}}\left( \frac{4-3\gamma }{AB}\right) ^{1/2}\xi
^{2\left( \frac{\gamma -1}{2-\gamma }\right) }\ln ^{\frac{3\gamma
-4-B\gamma 
}{2(2-\gamma )}}\xi \exp \left[ A\left( \frac{\gamma -1}{2-\gamma
}\right)
\ln ^B\xi \right] d\xi \,.  \label{elem1}
\end{equation}
As $\xi \rightarrow 1$ this tends to 
\begin{equation}
dt\simeq \frac{1-B}{3(2-\gamma )}\frac{a_0^{3(\gamma -1)}}{(AB\phi
_0)^{
\frac{\gamma -1}{2-\gamma }}}\left( \frac{4-3\gamma }{AB}\right)
^{1/2}\left( \xi -1\right) ^{\frac{3\gamma -4-B\gamma }{2(2-\gamma
)}}d\xi
\,,
\end{equation}
and 
\begin{equation}
t(\xi )\simeq F^{-1}\left( \xi -1\right) ^{\frac{\gamma
(1-B)}{2(2-\gamma)} 
}\,,
\end{equation}
where 
\begin{equation}
F=\frac{3\gamma }2\frac{(AB\phi _0)^{\frac{\gamma -1}{2-\gamma }}}{
a_0^{3(\gamma -1)}}\left( \frac{AB}{4-3\gamma }\right) ^{1/2}\,.
\end{equation}
Eventually, this leads to 
\begin{eqnarray}
\xi (t) &\rightarrow &1+F^{\frac{2(2-\gamma )}{\gamma (1-B)}}t^{\frac{
2(2-\gamma )}{\gamma (1-B)}}\,, \\
\phi (t) &\rightarrow &\phi _0\left[ 1+AF^{\frac{2B(2-\gamma )}{\gamma
(1-B)}
}t^{\frac{2B(2-\gamma )}{\gamma (1-B)}}\right] \,, \\
a(t) &\rightarrow &\frac{a_0F^{2/3\gamma }}{(AB\phi _0)^{\frac
1{3(2-\gamma
)}}}t^{2/3\gamma }\left[ 1+\frac{2F^{\frac{2(2-\gamma )}{\gamma
(1-B)}}}{
3(2-\gamma )}t^{\frac{2(2-\gamma )}{\gamma (1-B)}}\right] \,,
\end{eqnarray}
as $t\rightarrow \infty ,$ and there is power-law approach to the GR
solutions in this limit. The coupling, in terms of the field,
approaches 
\begin{equation}
2\omega (\phi )+3\rightarrow \frac{4-3\gamma }{3(2-\gamma
)^2}\frac{(1-B)^2}
BA^{1/B}\ln ^{-\left( \frac{B+1}B\right) }\left( \frac \phi {\phi
_0}\right) 
.  \label{omega3}
\end{equation}
The negativity of $d\xi /dt$ implied by Eq.~(\ref{elem1}) for $B>1$
implies
that $\xi $ approaches unity from above. As we noted earlier (after
Eq.~(\ref
{a40})), when $\xi >1$ positivity of $B$ implies positivity of $A$ and
thus
Eq.~(\ref{phi40}) confirms that $\phi \rightarrow \phi _0$ from above
in
these theories, ie $\phi \in (\phi _0,\;\infty )$. When $B<0$ the
coupling
function, as $\xi \rightarrow \infty $, is 
\begin{equation}
2\omega (\xi )+3\rightarrow \frac{4-3\gamma }{3(2-\gamma )^2}\frac
4{AB}\ln
^{1-B}\xi \,,  \label{omega2}
\end{equation}
and the temporal line element becomes 
\begin{equation}
dt=\frac 2{3(2-\gamma )}\frac{a_0^{3(\gamma -1)}}{(AB\phi
_0)^{\frac{\gamma
-1}{2-\gamma }}}\left( \frac{4-3\gamma }{AB}\right) ^{1/2}\xi ^{\frac{ 
2(\gamma -1)}{2-\gamma }}\ln ^{\frac{\gamma (1-B)}{2(2-\gamma )}}\xi
d\xi \,.
\label{elem2}
\end{equation}
Integrating, this gives 
\begin{equation}
t(\xi )\simeq C^{-1}\xi ^{\frac \gamma {2-\gamma }}\ln ^{\frac{\gamma
(1-B)}{
2(2-\gamma )}}\xi \,,
\end{equation}
where 
\begin{equation}
C=\frac{3\gamma }2\frac{(AB\phi _0)^{\frac{\gamma -1}{2-\gamma }}}{ 
a_0^{3(\gamma -1)}}\left( \frac{AB}{4-3\gamma }\right) ^{1/2}\,.
\end{equation}
Inverting asymptotically in $t$, this gives 
\begin{equation}
\xi (t)\simeq C^{\frac{2-\gamma }\gamma }t^{\frac{2-\gamma }\gamma
}\ln ^{
\frac{B-1}2}t^{\frac{2-\gamma }\gamma }\,,
\end{equation}
and hence there is logarithmic approach to the GR perfect-fluid
solutions 
\begin{eqnarray}
\phi (t) &\rightarrow &\phi _0\left[ 1+A\ln ^Bt^{\frac{2-\gamma
}\gamma
}\right] \,, \\
a(t) &\rightarrow &\frac{a_0C^{2/3\gamma }}{(AB\phi _0)^{\frac
1{3(2-\gamma
)}}}t^{2/3\gamma }\left[ 1+\frac A{3(2-\gamma )}\ln
^Bt^{\frac{2-\gamma }
\gamma }\right] \,,
\end{eqnarray}
as $t\rightarrow \infty $. The late-time behaviour of the coupling as
a function of $\phi $ is given by 
\begin{equation}
2\omega (\phi )+3\rightarrow \frac{4-3\gamma }{3(2-\gamma
)^2}\frac{4A^{-1/B}
}B\ln ^{\frac{1-B}B}\left( \frac \phi {\phi _0}\right) \,. 
\label{omega4}
\end{equation}
At large $t$, $\xi \rightarrow \infty $ and must do so from below to
maintain positivity. Consequentially, $B<0$ implies $A<0$ and from
Eq.~(\ref
{phi40}) $\phi \rightarrow \phi _0$ from below, ie $\phi \in (0,\;\phi
_0)$ in these models.

{\center \subsubsection{Early-time behaviour}}

\label{early1}

When $B>1$, $a(\xi)$ has a zero as $\xi \rightarrow 0$ iff $(-1)^B A
\leq 0$. 
At late times $\xi \rightarrow 1$ from above; however, this is a
manifestation of the fact that $\xi(t)$, for the exact theory defined
by
Eq.~(\ref{phi40}), is not monotonic. That this is correct may be
demonstrated using Eq.~(\ref{timeperf}) and solving $\sqrt{2\omega +
3}=0$
for $\xi$ when $B=2$ then further showing that $\sqrt{2\omega
+3}_{\xi} \neq
0$ at that point. As $\xi \rightarrow 0$, $\ln \xi < 0$ and
$(-1)^{B-1}AB
\geq 0$ so that $a(\xi) \geq 0$. When $B>0$, $(-1)^{B-1}A \geq 0$ and
hence $
(-1)^{B}A \leq 0$, ie the universe is always singular at $\xi=0$ in
$B>1$
theories. In this limit 
\begin{equation}  \label{BDcoup}
2\omega(\xi) +3 \rightarrow \frac{(4-3\gamma)^2}{3(2-\gamma)^2}\,,
\end{equation}
ie a BD theory.

When $B<0$, $a(\xi)$ has a zero as $\xi \rightarrow 1$ iff $A<0$. This
arises since $\sqrt{2\omega+3}>0$ as $\xi \rightarrow 1$ and $\xi$
approaches unity from above ($\ln \xi >0$). As in the preceding
paragraph, $
a(\xi)>0$ implies $AB>0$, $B<0$ enforces $A<0$ and the evolution
begins from
a singularity at $\xi =0$. Again, the form of the coupling function at
early
times is given by Eq.~(\ref{BDcoup}), BD theory.

We will not present explicit solutions showing the approach to BD
theory
since these models are not early-time limits of the theories, defined
in
Section \ref{coupling}, that we are interested in; they are merely
early-time limits of other theories which happen to asymptote (at late
times) to the theories with which we are concerned.

We now highlight some special cases of these models for particular
values of 
$\gamma$.

{\center \subsubsection{Dust models}}

These arise by substituting the choice $\gamma =1$ into the asymptotic
relations already derived. When $B>1$ we find 
\begin{eqnarray}
\phi (t) &\rightarrow &\phi _0\left[ 1+\left( \frac 32\right)
^{\frac{2B}{1-B
}}A^{\frac 1{1-B}}B^{\frac B{1-B}}t^{\frac{2B}{1-B}}\right] \,, \\
a(t) &\rightarrow &\left( \frac 32\right) ^{2/3}\frac{a_0}{\phi
_0^{1/3}} 
t^{2/3}\left[ 1+\left( \frac 32\right) ^{\frac{1+B}{1-B}}\left(
AB\right)
^{\frac 1{1-B}}t^{\frac 2{1-B}}\right] .
\end{eqnarray}
When $B<0$ the solutions tend to the GR solution only logarithmically 
\begin{eqnarray}
\phi (t) &\rightarrow &\phi _0\left[ 1+A\ln ^Bt\right] \,, \\
a(t) &\rightarrow &\left( \frac 32\right) ^{2/3}\frac{a_0}{\phi
_0^{1/3}}
t^{2/3}\left[ 1+\frac A3\ln ^Bt\right] \,.
\end{eqnarray}
All of these expressions are valid as $t\rightarrow \infty $.

{\center \subsubsection{Inflationary models}}

\label{inf1}

Inflationary models driven by a false vacuum equation of state may be
derived from the choice $\gamma =0$. Although there are varieties of
inflationary universe with $-1/3>\gamma >0,$ and these can easily be
found
from the formula for the general $\gamma $ solutions given above, we
shall
confine our attention to the $\gamma =0$ case which is not described
by the
previous formulae. It offers an excellent approximation to many slowly
changing scalar-field potentials. In this case we can view the
scalar-tensor
coupling as providing a second scalar field, thereby offering the
chance for
double inflation to occur. However, it is not sufficient simply to
substitute $\gamma =0$ into the existing expressions since the
qualitative
structure of the solutions is different. When $B>1$, the form of
$2\omega +3$
is as presented in Eq.~(\ref{omega1}) and the temporal line element of
Eq.~(
\ref{elem1}) can be approximated by 
\begin{equation}
dt\simeq \frac{(1-B)\phi _0^{1/2}}{3a_0^3}\ln ^{-1}\xi d\ln \xi \,,
\end{equation}
as $\xi \rightarrow 1$. Integrating this expression we find 
\begin{equation}
t(\xi )\simeq \frac{(1-B)\phi _0^{1/2}}{3a_0^3}\ln \left( \ln \xi
\right) \,,
\end{equation}
which $\rightarrow \infty $ as $\xi \rightarrow 1$. This can be
inverted,
yielding 
\begin{equation}
\xi (t)\simeq \exp \left\{ \exp \left[ \frac{3a_0^3t}{(1-B)\phi
_0^{1/2}}
\right] \right\} \,,
\end{equation}
and hence, 
\begin{eqnarray}
\phi (t) &\rightarrow &\phi _0\left[ 1+A\exp \left\{ \frac{3a_0^3Bt}{ 
(1-B)\phi _0^{1/2}}\right\} \right] \,, \\
a(t) &\rightarrow &\frac{a_0}{\left( AB\phi _0\right) ^{1/6}}\exp
\left(
\frac{a_0^3t}{2(\phi _0)^{1/2}}\right) \left[ 1+\frac 13\exp \left\{
\frac{ 
3a_0^3t}{(1-B)\phi _0^{1/2}}\right\} \right] \,,
\end{eqnarray}
as $t\rightarrow \infty $. Here we see explicitly the possibility of
double inflation arising from the sequential effects of the $\phi $
field and the $%
p=-\rho $ stress. If $B<0$, $t\rightarrow \infty $ as $\xi \rightarrow
\infty $ and $2\omega +3$ is given by Eq.~(\ref{omega2}). The
differential $
\xi -t$ relation, Eq.~(\ref{elem2}), is then well approximated by 
\begin{equation}
dt\simeq \frac{2\phi _0^{1/2}}{3a_0^3}\exp \left( -\frac A2\ln ^B\xi
\right)
d\ln \xi \,.
\end{equation}
Making the substitution $\zeta =\ln ^B\xi $ we can integrate the above
equation to obtain 
\begin{equation}
t(\xi )\simeq \frac{2\phi _0^{1/2}}{3a_0^3}\ln \xi \left[ 1-\frac
A{2(B+1)}\ln ^B\xi \right] \,,
\end{equation}
as $\xi $, and hence $t$, tend to infinity. Asymptotically, we obtain 
\begin{eqnarray}
\xi (t) &\rightarrow &\exp \left( \frac{3a_0^3t}{2\phi
_0^{1/2}}\right) \,,
\label{endth1} \\
\phi (t) &\rightarrow &\phi _0\left[ 1+A\left( \frac{3a_0^3}{2\phi
_0^{1/2}} 
\right) ^Bt^B\right] \,, \\
a(t) &\rightarrow &\frac{a_0}{\left( AB\phi _0\right) ^{1/6}}\left(
\frac{
3a_0^3}{2\phi _0^{1/2}}\right) ^{\frac{1-B}6}t^{\frac{1-B}6}\exp
\left(
\frac{a_0^3t}{2\phi _0^{1/2}}\right) \,,
\end{eqnarray}
as $t\rightarrow \infty $.

{\center \subsubsection{The Connection to the parameters of Theory 1}}

We now use the results derived earlier in this section to model the
late-time behaviour of Theory 1. Consider first universes in which
$\phi
\rightarrow \phi _0$ from above, ie $\phi \in (\phi _0,:\infty )$. The
approach of Theory 1 to the relativistic limit in this direction can
be
accurately modelled using the theory defined by Eq.~(\ref{phi40}) with
$A>0$, $B>1$. From Eq.~(\ref{omega3}), 
\begin{eqnarray}
2\omega (\phi )+3 &\rightarrow &\frac{4-3\gamma }{3(2-\gamma
)^2}\frac{
(1-B)^2}BA^{1/B}\left[ \left( \frac \phi {\phi _0}\right) -1\right]
^{-\left( \frac{B+1}B\right) }\,, \\
&&
\end{eqnarray}
as $\phi \rightarrow \phi _0$. This expression is essentially the same
as
definition of the Theory 1 coupling for $\phi \in (\phi _0,\;\infty )$
introduced in section 4. Explicitly, we may obtain the asymptotic
behaviour
of Theory 1 by making the following identifications between its
parameters
and the parameters of Eq.~(\ref{phi40}): 
\begin{eqnarray}
A &=&\left[ \frac{6B_1\left( 2-\gamma \right) ^2\left( \alpha
-1\right)}{
\left( 4-3\gamma \right) \left( \alpha -2\right) ^2}\right] ^{\frac
1{\alpha
-1}}\,, \\
B &=&\frac 1{\alpha -1}\,.
\end{eqnarray}
The constraint on the $B>0$ models such that they approach GR at late
times,
namely $B>1$, is equivalent to $-(B+1)/B>-2$. This is a very
restrictive
condition because the function $-(B+1)/B$ is naturally bounded above
by the
value $-1$. Thus,  for perfect-fluid universes with $0<\gamma <4/3$
Theory 1
can only be expected to converge to the general relativistic value of
$\phi $
{\em from above} if $1<\alpha <2$. When $\phi $ converges to $\phi _0$
from
below, ie $\phi \in (0,\;\phi _0)$ we can approximate the late-time
behaviour of Theory 1 using the solutions for $A<0$, $B<0$. The
asymptotic
form of the coupling in this case is given by Eq.~(\ref{omega4}) 
\begin{eqnarray}
2\omega (\phi )+3\rightarrow \frac{4-3\gamma }{3(2-\gamma )^2}\frac{
4(-A)^{-1/B}}{(-B)}\left[ 1-\left( \frac \phi {\phi _0}\right) \right]
^{ 
\frac{1-B}B}\,,
\end{eqnarray}
as $\phi \rightarrow \phi _0$. The behaviour of Theory 1 at late times
may
be found by substituting the expressions 
\begin{eqnarray}
A &=&\left[ \frac{3B_1\left( 2-\gamma \right) ^2(-1)^\alpha }{2\left(
4-3\gamma \right) \left( 1-\alpha \right) }\right] ^{\frac 1{\alpha
-1}}\,,
\\
B &=&\frac 1{1-\alpha }\,,
\end{eqnarray}
into the formulae describing the asymptotic evolution of the theory
defined
by Eq.~(\ref{phi40}). The choice $B<0$ leads to the constraint
$\frac{1-B} 
B<-1$, which becomes $\alpha >1$ for Theory 1 as $\phi \rightarrow
\phi_0$ 
from below.

\begin{center}
\section{Theory 2: $2\omega (\phi )+3=B_{2}\left| \ln (\phi /\phi _0)
\right|^{-2\delta}$; $\delta > 1/4$, $B_{2} >0$ constants.}

\label{th2} 
\end{center}

The left-hand side of Eq.~(\ref{vacint}) for this choice of the
coupling
function, when $\phi \in (0,\phi_0)$, is
\begin{equation}  \label{field10}
\int \frac{(2\omega(\phi) +3)^{1/2}}{\phi} d\phi = \left\{ 
\begin{array}{ll}
-\frac{\sqrt{B_{2}}}{1-\delta} \left|\ln \left(\frac{\phi}{\phi_0}
\right)\right|^{1-\delta} - \sqrt{3} \ln K\,, & \delta \neq 1\,, \\ 
&  \\ 
-\sqrt{B_{2}} \ln \left| \ln \left( \frac{\phi}{\phi_0} \right)
\right| - 
\sqrt{3} \ln K\,, & \delta = 1\,,
\end{array}
\right.
\end{equation}
where $K$ is an integration constant. We now investigate vacuum and
radiation solutions for this case, under the assumption that $\phi \in
(0,
\phi_0)$.

{\center \subsection{Vacuum solutions ($k=0$)}}

Selecting the zero-curvature right-hand side from Eq.~(\ref{vacint})
and the
necessary form for $y(\eta )$ from Eq.~(\ref{vacy}),
Eq.(\ref{field10}) for $
\delta \neq 1$ reads 
\begin{equation}
\left[ -\ln \left( \frac \phi {\phi _0}\right) \right] ^{1-\delta
}=-\lambda
_2(1-\delta )\ln \left[ K(\eta +\eta _0)\right] \,,
\end{equation}
where $\lambda _2=\sqrt{3/B_2}$. The left-hand side is positive, for
the
right-hand side to follow suit as $\eta \rightarrow \infty $ requires
$ 
\lambda _2(1-\delta )<0$. As $\eta \rightarrow \infty $, the
right-hand side$ 
\rightarrow \infty $ and we must have $\delta >1$ for this to occur on
the
left as $\phi \rightarrow \phi _0$. This implies $\lambda _2>0$ and we
obtain for the field and the scale-factor, when $\delta \neq 1$, the
exact expressions 
\begin{eqnarray}
\phi (\eta ) &=&\phi _0\exp \left[ -\left\{ \lambda _2(\delta -1)\ln
\left[ 
K(\eta +\eta _0)\right] \right\} ^{\frac 1{1-\delta }}\right] \,,
\label{avacln0} \\
a^2(\eta ) &=&\frac A{\phi _0}(\eta +\eta _0)\exp \left\{ \left[
\lambda
_2(\delta -1)\ln \left[ K(\eta +\eta _0)\right] \right] ^{\frac
1{1-\delta
}}\right\} \,.
\end{eqnarray}
When $\delta =1$, the conformal time-evolution of the field and the
scale-factor are given by 
\begin{eqnarray}
\phi (\eta ) &=&\phi _0\exp \left[ -K(\eta +\eta _0)\right] ^{-\lambda
_2}\,,
\label{avacln01} \\
a^2(\eta ) &=&\frac A{\phi _0}(\eta +\eta _0)\exp \left\{ \left[
K(\eta
+\eta _0)\right] ^{-\lambda _2}\right\} \,.
\end{eqnarray}

{\center \subsubsection{Late-time behaviour}}

As $\eta \rightarrow \infty$ the late-time behaviour can be modelled
by
Eqs.~(\ref{conf10}) -- (\ref{a70}) under the simultaneous
transformations: $
\lambda \rightarrow \lambda_2$, $\alpha \rightarrow 2\delta$. Setting
$ 
\eta_0=0$ for simplicity, we see that the scale-factor tends to zero
as $ 
\eta \rightarrow 0$ when $\delta >1$. When $\delta\neq 1$ we require $
\lambda_2>0$ if $\phi$ is to tend to $\phi_0$ as $\eta \rightarrow
\infty$.
We find 
\begin{equation}
\phi(t) \rightarrow \phi_0\left[1 -
K^{-\lambda_2}\left(\frac{9\phi_0}{4A} 
\right)^{-\lambda_2/3} t^{-2\lambda_2/3}\right]\,,
\end{equation}
which approaches $\phi_0$ as $t\rightarrow \infty$. The asymptotic
form of
the scale-factor is 
\begin{equation}
a(t) \rightarrow \left(\frac{3A}{2\phi_0}\right)^{1/3}t^{1/3}\left[1 +
K^{-\lambda_2}\left(\frac{1-\lambda_2}{3-2\lambda_2}\right)\left(
\frac{
9\phi_0}{4A}\right)^{-\lambda_2/3}t^{-2\lambda_2/3}\right]\,,
\end{equation}
with limiting behaviour $a \propto t^{1/3}$ as $t \rightarrow \infty$.

{\center \subsubsection{Early-time behaviour}}

Examining the form of $a(\eta)$ when $\delta \neq 1$ in
Eq.~(\ref{avacln0}),
we see that the choice $\eta_0=0$ in ensures $a(0) = 0$. As $\eta
\rightarrow 0$, $\ln \eta \rightarrow -\infty$ and the exponential
factor in 
$a$ either tends to zero or a constant, depending upon the sign of
$1-\delta$
. However, the bound $1-\delta <0$ on the power of the logarithm in
the
exponential guarantees that we always obtain $a(\eta) \propto \sqrt{
A/\phi_{0}}\eta^{1/2}$ as $\eta \rightarrow 0$ and hence $\eta
\rightarrow
(9\phi_{0}/4A)^{1/3} t^{2/3}$. This final expression implies $t
\rightarrow
0 $ as $\eta \rightarrow 0$ and the early-time formulae for $\phi(t)$
and $
a(t) $ are identical to Eqs.~(\ref{conf10}) and (\ref{a70}), under the
transformations $\lambda \rightarrow \lambda_2$, $\alpha \rightarrow
2\delta$.

When $\delta =1$, the dominant behaviour in $a(\eta )$ as $\eta
\rightarrow 0
$ is contained in the exponential, from which we may conclude (after
setting $\eta _0=0$) that 
\begin{equation}
t\propto \eta ^{\frac 32+\lambda _2}\exp \left[ \frac 12\left( K\eta
\right)
^{-\lambda _2}\right] \,.  \label{conftlnvac}
\end{equation}
This may be inverted asymptotically in $\eta ^{-1}$. To first-order we
obtain $\eta \sim K^{-1}(2\ln t)^{-1/\lambda _2}$ as $\eta $ and  $t$
tend
to zero (noting the sign of $\lambda _2$ and the necessary monotonicity
of $
\eta (t)$). The next-order corrections at early-times follow by
substituting
this lowest-order result into the weakest dependence in
Eq.~(\ref{avacln01}
), ie the power-law factor, and solving for $\eta $. This yields 
\begin{equation}
\eta \sim K^{-1}\left\{ 2\left[ \ln t+\left( 1+\frac 3{2\lambda
_2}\right)
\ln (\ln t)\right] \right\} ^{-1/\lambda _2}\,,
\end{equation}
and hence 
\begin{eqnarray}
\phi (t) &\rightarrow &\phi _0t^{-2}\ln ^{-\frac 3{\lambda _2}-2}t\,,
\label{a2} \\
a(t) &\propto &t\left( \ln t\right) ^{1+1/\lambda _2}\,,
\end{eqnarray}
as $t\rightarrow 0$.

{\center \subsubsection{Minima}}

Examining the $\delta \neq 1$ vacuum models for the presence of minima
we find 
\begin{equation}
(a^2)_\eta =\frac A{\phi _0}\exp \left\{ -\left[ \lambda _2(\delta
-1)\ln
(K\eta )\right] ^{\frac 1{1-\delta }}\right\} \left[ 1+\left\{
-\frac{\left[
\lambda _2(\delta -1)\right] ^{\frac 1{1-\delta }}}{1-\delta }\ln
^{\frac
\delta {1-\delta }}K\eta \right\} \right] \,.  \label{min20}
\end{equation}
For the exponential to tend to zero we require its argument to tend to
$
-\infty $. the bound on $\delta $, namely $\delta >1$, implies that
$\eta
\rightarrow K^{-1}$ for this to happen. At this point $\ln ^{\frac
\delta
{1-\delta }}(K\eta )\rightarrow \infty $; nevertheless, this
logarithmic
divergence will be insufficient to counter the exponential convergence
of
the pre-factor. We can also expect to see turning points in the
evolution of 
$a$ when the factor in square brackets vanishes. This happens when
$\eta
=\eta _{*}$, where 
\begin{equation}
\eta _{*}=K^{-1}\exp \left[ \frac{(-\lambda _2)^{-1/\delta }}{1-\delta
} 
\right] \,.
\end{equation}
One can show that the conditions for the factor in square brackets in
Eq.~( 
\ref{min20}) to vanish also guarantee that the exponential will be
well
behaved. When $\delta =1$ we have 
\begin{equation}
(a^2)_\eta =\frac A{\phi _0}\left[ 1+\lambda _2K^{-\lambda _2}\eta
^{-\lambda _2}\right] \exp \left\{ -K^{-\lambda _2}\eta ^{-\lambda
_2}\right\} \,.  \label{min30}
\end{equation}
$K\eta >0$ and so the exponential will vanish when $\eta \rightarrow
0$.
When this happens the factor in square brackets diverges, however its
divergence is quashed by the rapid convergence of the exponential,
confirming the presence of a stationary point at $\eta =0$. There will
also
exist a stationary point in the evolution of $a$ when the
square-bracketed
factor itself vanishes in Eq.~(\ref{min30}). This occurs at $\eta
_{*}$, 
where 
\begin{equation}
\eta _{*}=(-\lambda _2)^{1/\lambda _2}K^{-1}\,.
\end{equation}
At this point $-(K\eta )^{-\lambda _2}=\lambda _2^{-1}>0$ and the
exponential is well behaved.

{\center \subsection{Vacuum solutions ($k=-1$)}}

Similarly, we analyse the behaviour of the negatively curved models.
Selecting the $k=-1$ versions of Eqs.~(\ref{vacint}) and (\ref{vacy}),
we
obtain for the field and the metric, when $\delta \neq 1,$ the
expressions 
\begin{eqnarray}
\label{vkm1t1p}
\phi (\eta ) &=&\phi _0\exp \left\{ -\left[ \lambda _2(\delta
-1)\right]
^{\frac 1{1-\delta }}\ln ^{\frac 1{1-\delta }}\left[ K\tanh \left(
\eta
+\eta _0\right) \right] \right\} \,, \\ 
\label{vkm1t1a}
a^2(\eta ) &=&\frac A{2\phi _0}\sinh \left[ 2\left( \eta +\eta
_0\right) 
\right]   \nonumber \\
&&\mbox{}\times \exp \left\{ \left[ \left[ \lambda _2(\delta
-1)\right]
^{\frac 1{1-\delta }}\ln ^{\frac 1{1-\delta }}\left[ K\tanh \left(
\eta
+\eta _0\right) \right] \right] \right\} \,.
\end{eqnarray}
This solution will approach the Milne model at late times under the
conditions $0<\delta <1$, $K=1$. The approach of $\phi $ to $\phi _0$
from
below requires $\lambda _2<0$. When $\delta =1$, setting $\eta _0=0$,
we
have 
\begin{eqnarray}
\phi (\eta ) &=&\phi _0\exp \left\{ -K^{-\lambda _2}\tanh ^{-\lambda
_2}\eta
\right\} \,, \\
a^2(\eta ) &=&\frac A{2\phi _0}\sinh (2\eta )\exp \left\{ K^{-\lambda
_2}\tanh ^{-\lambda _2}\eta \right\} \,.
\end{eqnarray}
Because there is no choice of $K$ which allows $\phi $ to tend to
$\phi_0$
at late times we do not pursue this model any further.

{\center \subsubsection{Late-time behaviour}}

The late-time behaviour is given by Eqs.~(\ref{etakm1}) --
(\ref{akm1})
under the substitutions $\lambda=\lambda_2$, $\alpha=2\delta$.

{\center \subsubsection{Early-time behaviour}}

Picking $\eta _0=0$, $K=1$ in Eqs.~(\ref{vkm1t1p}) and
(\ref{vkm1t1a}), we
have as $\eta \rightarrow 0,$ 
\begin{eqnarray}
\phi (\eta ) &\rightarrow &\phi _0\exp \left\{ -\left[ \lambda
_2(\delta
-1)\right] ^{\frac 1{1-\delta }}\ln ^{\frac 1{1-\delta }}\eta \right\}
\,, \\
a^2(\eta ) &\rightarrow &\frac A{\phi _0}\eta \exp \left\{ \left[
\lambda
_2(\delta -1)\right] ^{\frac 1{1-\delta }}\ln ^{\frac 1{1-\delta
}}\eta
\right\} \,.
\end{eqnarray}
We know $0<\delta <1$ and hence $1/(1-\delta )>1$ and $a^2(\eta )$
will be
dominated by the exponential factor as $\eta \rightarrow 0$, and hence
$
a\rightarrow 0$. Integrating and inverting $a(\eta )$, we obtain 
\begin{equation}
\eta (t)\simeq \exp \left\{ \frac{2^{1-\delta }\ln ^{1-\delta
}(-t)}{\lambda
_2(\delta -1)}\left[ 1+3\left( \frac{2^{-\delta }}{\lambda _2}\right)
\ln
^{-\delta }(-t)\right] \right\} \,,
\end{equation}
where $t\rightarrow -\infty $ as $\eta \rightarrow 0$. Using this
relation
the early-time behaviour is 
\begin{eqnarray}
\phi (t) &\rightarrow &\phi _0t^{-2}\exp \left\{ -3\left(
\frac{2^{1-\delta }
}{\lambda _2(1-\delta )}\right) \ln ^{1-\delta }(-t)\right\} \,, \\
a^2(t) &\rightarrow &\sqrt{\frac A{\phi _0}}t\exp \left\{
\frac{2^{1-\delta
}\ln ^{1-\delta }(-t)}{\lambda _2(1-\delta )}\right\} \,,
\end{eqnarray}
as $t\rightarrow -\infty $.

{\center \subsubsection{Minima}}

Differentiating Eq.~(\ref{vkm1t1a}) yields 
\begin{eqnarray}
(a^2)_\eta  &=&\frac A{\phi _0}\left\{ \cosh (2\eta )+\left( -\lambda
_2\right) ^{\frac 1{1-\delta }}(1-\delta )^{\frac \delta {1-\delta
}}\ln
^{\frac \delta {1-\delta }}(\tanh \eta )\right\}   \nonumber \\
&&\mbox{}\times \exp \left\{ \left[ \lambda _2(1-\delta )\right]
^{\frac
1{1-\delta }}\ln ^{\frac 1{1-\delta }}\left[ K\tanh (\eta +\eta
_0)\right]
\right\} \,.
\end{eqnarray}
The exponential cannot tend to zero, since its argument is always
positive.
If we examine the pre-factor we find stationary points exist at
$\eta=$ $
\eta _{*}$, where 
\begin{equation}
\cosh (2\eta _{*})=-(-\lambda _2)^{\frac 1{1-\delta }}(1-\delta
)^{\frac
\delta {1-\delta }}\left[ \ln \left( \tanh \eta _{*}\right) \right]
^{\frac
\delta {1-\delta }}\,.
\end{equation}
This is not soluble analytically, although we may gain a bound on its
value
by demanding that the cosh function always be greater than unity. We
obtain 
\begin{equation}
\eta _{*}<{\rm arctanh}\left\{ \exp \left( -\frac{\lambda
_2^{-1/\delta }}{
1-\delta }\right) \right\} \,.
\end{equation}

{\center \subsection{Vacuum solutions ($k=+1$)}}

When $\phi$ lies in the range $0<\phi<\phi_0$, and $\delta \neq 1$ we
have
the exact solution 
\begin{eqnarray}
\phi(\eta) & = & \phi_0 \exp \left\{ -
\left[\lambda_2(\delta-1)\right]^{
\frac{1}{1-\delta}}\ln^{\frac{1}{1-\delta}}\left[K\tan(\eta+\eta_0)
\right]\right\}\,, \\
a^2(\eta) & = & \frac{A}{2\phi_0}\sin\left[2(\eta + \eta_0)\right]
\exp
\left\{\left[\lambda_2(\delta-1)\right]^{\frac{1}{1-\delta}}\ln
^{\frac{1}{ 
1-\delta}}\left[K\tan(\eta + \eta_0)\right]\right\}\,.
\end{eqnarray}
In the particular case $\delta=1$, we have instead 
\begin{eqnarray}
\phi(\eta) & = & \phi_0\exp\left\{-K^{-\lambda_2}\tan^{-\lambda_2}
(\eta + 
\eta_0)\right\}\,, \\
a^2(\eta) & = & \frac{A}{2\phi_0}\sin\left[2(\eta + \eta_0)\right]\exp
\left\{-K^{-\lambda_2}\tan^{-\lambda_2}(\eta + \eta_0)\right\}\,.
\end{eqnarray}

{\center \subsection{Radiation solutions ($k=0$)}}

Fixing the origin of conformal time in Eqs.~(\ref{radint}) and
(\ref{rady})
such that $2\Gamma \eta_{0} = A$, we obtain the exact results, for
$\delta
\neq 1$ 
\begin{eqnarray}  
\label{phi6}
\phi(\eta) & = & \phi_{0} \exp \left\{-\left[\lambda_2(\delta-1)
\right]^{
\frac{1}{1-\delta}} \ln^{\frac{1}{1-\delta}} \left[ K \left|
\frac{\eta}{
\eta + 2\eta_{0}}\right| \right] \right\}\,, \\
\label{a6}
a^{2}(\eta) & = & \frac{\Gamma}{\phi_{0}} \eta (\eta + 2\eta_{0})\exp
\left\{ \left[\lambda_2(\delta-1)\right]^{\frac{1}{1-\delta}}
\ln^{\frac{1}{
1-\delta}} \left[ K \left| \frac{\eta}{\eta + 2\eta_{0}}
\right|\right]
\right\}\,,
\end{eqnarray}
In the special case of $\delta = 1$, we have the exact relations 
\begin{eqnarray}  \label{phid1t2r}
\phi(\eta) & = & \phi_0 \exp \left[ -\left|\frac{K \eta}{\eta +
2\eta_0}
\right|^{-\lambda_2}\right]\,, \\
a^2(\eta) & = & \frac{\Gamma}{\phi_0} \eta (\eta + 2\eta_0)\exp \left[
\left|
\frac{K \eta}{\eta + 2\eta_0}\right|^{-\lambda_2} \right]\,.
\end{eqnarray}
To ensure $a^2>0$, we require $\eta(\eta+2\eta_0)>0$ and the modulus
signs
in the above expressions can be dropped.

{\center \subsubsection{Late-time behaviour}}

The asymptotic behaviour may be obtained from Eqs.~(\ref{etarkzt1}) --
(\ref
{arkzt1}), by applying the transformations $\alpha \rightarrow 2\delta
$, $
\lambda \rightarrow \lambda _2$. When $\delta \neq 1$ Eq.~(\ref{phi6})
serves to bound the allowed parameter values in order that $\phi
\rightarrow
\phi _0$ as $\eta \rightarrow \infty $. We can see that we require
$\delta
\leq 1$ and $K=1$, delimiting the allowed range of $\delta $ to
$1/4<\delta
<1$. We take $\lambda _2\eta _0>0$ to ensure both sides of
Eq.~(\ref{radint}
) exist on the correct domains. Examining Eq.~(\ref{phid1t2r}), we see
that
when $\delta =1$ there is no non-trivial choice of $K$ which permits
$\phi
\rightarrow \phi _0$ at late times. Therefore we exclude it from
further analysis.

{\center \subsubsection{Early-time behaviour}}

At early times when $\delta \neq 1$ the behaviour is harder to
ascertain and
we need to make use of logarithmic approximations. From
Eq.~(\ref{a6}), by
demanding that $\phi \in (0,\phi _0)$, we see that the lower limit of
$\eta $
occurs as 
\begin{eqnarray}
{\rm (i)}\;\eta  &\rightarrow &0\,,\;\lambda _2>0,\,\eta _0>0,\,{\rm
or}\,,
\\
{\rm (ii)}\;\eta  &\rightarrow &-2\eta _0\,,\;\lambda _2<0,\,\eta
_0<0\,. \\
&&
\end{eqnarray}
Case (i) gives 
\begin{equation}
a(\eta )\simeq \eta ^{1/2}\exp \left\{ \frac 12\left[ \lambda
_2(\delta
-1)\right] ^{\frac 1{1-\delta }}\ln ^{\frac 1{1-\delta }}\left( \frac
\eta
{2\eta _0}\right) \right\} \,,
\end{equation}
which, for the range of values of $\delta $ to which we are confined,
is
dominated by the exponential at small $\eta $. Integrating this
expression and inverting approximately then leads to 
\begin{equation}
\eta (t)=2\eta _0\exp \left\{ -\frac{2^{-\delta }}{\lambda _2(1-\delta
)}
\left[ \ln (-t)+3\frac{2^{-\delta }}{\lambda _2(1-\delta )}\ln
^{1-\delta
}(-t)\right] ^{1-\delta }\right\} \,,  \label{t2kzret}
\end{equation}
and so $t\rightarrow -\infty $ as $\eta \rightarrow 0$. The field and
the
metric are then given by  
\begin{eqnarray}
\phi (t) &\simeq &t^{-2}\exp \left\{ 3\frac{2^{1-\delta }}{\lambda
_2(\delta
-1)}\ln ^{1-\delta }(-t)\right\} \,,  \label{t2kzrpt} \\
a(t) &\simeq &t\exp \left\{ \frac{2^{1-\delta }}{\lambda _2(1-\delta
)}\ln
^{1-\delta }(-t)\right\} \,,
\end{eqnarray}
as $t\rightarrow -\infty $. Case (ii) can be modelled by applying the
transformations: $\eta _0\rightarrow -\eta _0$, $\eta \rightarrow \eta
+2\eta _0$, $\lambda _2\rightarrow -\lambda _2$, in this order, to
Eqs.~(\ref {t2kzret}) -- (\ref{t2kzrpt}).

{\center \subsubsection{Minima}}

Since the scale-factor is infinite at early-times, we know there must
exist
at least one minimum in its evolution, in order that we obtain general
relativistic expansion at late times. Differentiating Eq.~(\ref{a6}),
we find 
\begin{eqnarray}
(a^2)_\eta  &=&\frac{2\Gamma }{\phi _0}\left[ \eta +\eta _0+\eta
_0\left(
-\lambda _2\right) ^{\frac 1{1-\delta }}(1-\delta )^{\frac \delta
{1-\delta
}}\ln ^{\frac \delta {1-\delta }}\left( \frac \eta {\eta +2\eta
_0}\right)
\right]   \nonumber \\
&&\mbox{}\times \exp \left\{ \left[ \lambda _2(\delta -1)\right]
^{\frac
1{1-\delta }}\ln ^{\frac 1{1-\delta }}\left( \frac \eta {\eta +2\eta
_0}\right) \right\} \,.
\end{eqnarray}
For the parameter choices we are confined to, the exponential factor
in the
above expression is a monotonic function, existing in the range
$(1,\,\infty
)$. Thus we search for zeros of the pre-factor, finding them to exist
at $
\eta _{*}$, where 
\begin{equation}
\eta _{*}+\eta _0+\eta _0(-\lambda _2)^{\frac 1{1-\delta }}(1-\delta
)^{\frac \delta {1-\delta }}\ln ^{\frac \delta {1-\delta }}\left(
\frac{\eta
_{*}}{\eta _{*}+2\eta _0}\right) =0\,,
\end{equation}
which is non-analytic and must be solved numerically for particular
$\delta
,\lambda _2$ and $\eta _0$.

{\center \subsection{Radiation solutions ($k=-1$)}}

For the negatively-curved models with $\delta \neq 1$ we have exactly
\begin{eqnarray}  
\label{pkm1th2r}
\phi(\eta) & = & \phi_0\exp \left\{\left[
\begin{array}{c}
\\ 
\\ 
\end{array}
\!\!\!\!\!\! \lambda_2(\delta-1)\right]^{\frac{1}{1-\delta}}\right. 
\nonumber \\
& & \mbox{} \left. \times \ln^{\frac{1}{1-\delta}}\left[K\left|\frac{
\Gamma
\tanh(\eta+\eta_0) +(\Gamma^2 + A^2)^{1/2} -A}{\Gamma \tanh
(\eta+\eta_0)
+(\Gamma^2 + A^2)^{1/2} +A}\right|\right]\right\}\,, \\
\label{akm1th2r}
a^2(\eta) & = & \frac{1}{2\phi_0}\left[-\Gamma + (\Gamma^2 +
A^2)^{1/2}
\sinh\left[2(\eta + \eta_0)\right]\right] \exp\left\{ 
\begin{array}{c}
\\ 
\\ 
\end{array}
\!\!\!\!\!\! -\left[\lambda_2
(\delta-1)\right]^{\frac{1}{1-\delta}}\right. 
\nonumber \\
& & \mbox{} \left. \times 
\begin{array}{c}
\\ 
\\ 
\end{array}
\!\!\!\!\!\! \ln^{\frac{1}{1-\delta}}\left[K \left|\frac{\Gamma \tanh 
(\eta+\eta_0) +(\Gamma^2 + A^2)^{1/2} -A} {\Gamma \tanh (\eta+\eta_0)
+(\Gamma^2 + A^2)^{1/2} +A}\right|\right] 
\begin{array}{c}
\\ 
\\ 
\end{array}
\!\!\!\!\!\!\right\}\,.
\end{eqnarray}
When $\delta =1$ these become 
\begin{eqnarray}
\phi(\eta) & = & \phi_0 \exp \left\{-K^{-\lambda_2}\left|\frac{\Gamma
\tanh
(\eta+\eta_0) +(\Gamma^2 + A^2)^{1/2} -A} {\Gamma \tanh (\eta+\eta_0)
+(\Gamma^2 + A^2)^{1/2} +A}\right|^{-\lambda_2} \right\}\,, \\
a^2(\eta) & = & \frac{1}{2\phi_0}\left\{-\Gamma + (A^2 -
\Gamma^2)^{1/2}
e^{2(\eta+ \eta_0)}\right\}  \nonumber \\
& & \mbox{} \exp \left\{K^{-\lambda_2}\left|\frac{\Gamma \tanh
(\eta+\eta_0)
+(\Gamma^2 + A^2)^{1/2} -A} {\Gamma \tanh (\eta+\eta_0) +(\Gamma^2 +
A^2)^{1/2} +A}\right|^{-\lambda_2} \right\}\,.
\end{eqnarray}

{\center \subsubsection{Late-time behaviour}}

The late-time behaviour may be derived from the Theory 1 solutions, by
the
simultaneous transformations $\lambda \rightarrow \lambda _2$, $\alpha
\rightarrow 2\delta $. When $\delta \neq 1$, Eqs.~(\ref{km1th1reta})
-- (\ref
{km1th1ra}) require $A\lambda _2>0$ and $\delta <1$ to ensure
late-time
approach to GR. When $\delta =1$, there exists no value of $K$ that
will
admit a tendency of $\phi $ to $\phi _0$ at large $\eta $, and for
this reason we pursue these models no further.

{\center \subsubsection{Early-time behaviour}}

We make the simplifying choice for the origin of $\eta $-time, 
\begin{equation}
e^{2\eta _0}=\left( \frac{A+\Gamma }{A-\Gamma }\right) ^{1/2}\,.
\end{equation}
Eq.~(\ref{pkm1th2r}) for the $\delta \neq 1$ solution then becomes 
\begin{equation}
\phi (\eta )\simeq \phi _0\exp \left\{ -\left[ \lambda _2(\delta
-1)\ln
\left[ \left( \frac{\Gamma +A}A\right) \eta \right] \right] ^{\frac
1{1-\delta }}\right\} \,,
\end{equation}
as $\eta \rightarrow 0$. When $\lambda _2>0$, $\phi \rightarrow 0$ as
$\eta
\rightarrow 0$. When $\lambda _2<0$ however, there only exist
solutions when 
$(-1)^{\frac 1{1-\delta }}$ is real. For the limiting form of the
solution,
we find 
\begin{eqnarray}
\eta (t) &\propto &t^{2/3}\,, \\
\phi (t) &\propto &\exp \left\{ -\left[ \frac{2\lambda _2}3(\delta
-1)\ln
t\right] ^{\frac 1{1-\delta }}\right\} \,, \\
a(t) &\propto &t^{1/3}\,,
\end{eqnarray}
as $\eta $, and hence $t$, tend to zero.

{\center \subsubsection{Minima}}

We define 
\begin{equation}
\Psi \equiv \frac{e^{2\eta }-1}{e^{2\eta }-e^{-4\eta _0}}\,,
\end{equation}
such that 
\begin{eqnarray}
(a^2)_\eta  &=&\frac 1{\phi _0}\left\{ 
\begin{array}{c}
\\ 
\\ 
\end{array}
\!\!\!\!\!\!(A^2-\Gamma ^2)^{1/2}\cosh [2(\eta +\eta _0)]\right.  
\nonumber 
\label{min31} \\
&&\ \mbox{}\left. +\left[ -\Gamma +(A^2-\Gamma ^2)^{1/2}\sinh \left[
2(\eta
+\eta _0)\right] \right] \right.   \nonumber \\
&&\ \mbox{}\left. \times (-\lambda _2)^{\frac 1{1-\delta }}(1-\delta
)^{\frac \delta {1-\delta }}\left[ \frac{e^{2\eta }(1-e^{-4\eta
_0})}{\left(
e^{2\eta }-e^{-4\eta _0}\right) ^2}\right] \ln ^{\frac \delta
{1-\delta
}}\Psi 
\begin{array}{c}
\\ 
\\ 
\end{array}
\!\!\!\!\!\!\right\}   \nonumber \\
&&\ \mbox{}\times \exp \left\{ \left[ \lambda _2(\delta -1)\right]
^{\frac
1{1-\delta }}\ln ^{\frac 1{1-\delta }}\Psi \right\} \,.
\end{eqnarray}
At a stationary point in the evolution of the scale-factor we require
the
expression on the right-hand side of Eq.~(\ref{min31}) to vanish.
Since a
zero of the exponential would require $\phi \rightarrow \infty $,
which is
outside our range of consideration, we search for zeros of the
pre-factor.
Minima exist at $\eta _{*}$, given by the implicit formula 
\begin{eqnarray}
(A^2-\Gamma ^2)^{1/2}\cosh \left[ 2(\eta _{*}+\eta _0)\right] +\left[
-\Gamma +(A^2-\Gamma ^2)^{1/2}\sinh \left[ 2(\eta _{*}+\eta _0)\right]
\right]  &&  \nonumber \\
\mbox{}\times \left( -\lambda _2\right) ^{\frac 1{1-\delta }}\left(
1-\delta
\right) ^{\frac \delta {1-\delta }}\left[ \frac{e^{2\eta
_{*}}(1-e^{-4\eta
_0})}{(e^{2\eta _{*}}-e^{-4\eta _0})^2}\right] \ln ^{\frac \delta
{1-\delta
}}\left( \frac{e^{2\eta _{*}}-1}{e^{2\eta _{*}}-e^{-4\eta _0}}\right) 
&=&0\,.
\end{eqnarray}

{\center \subsection{Radiation models ($k=+1$)}}

For the closed models with $\delta \neq 1$ we have 
\begin{eqnarray}
\phi(\eta) & = & \phi_0 \exp \left\{ 
\begin{array}{c}
\\ 
\\ 
\end{array}
\!\!\!\!\!\! -\left[\lambda_2 (\delta-1) \right]^{\frac{1}{1-\delta}}
\right.
\nonumber \\
& & \mbox{} \left. \times \ln^{\frac{1}{1-\delta}} \left[ K\left|
\frac{ 
\Gamma \tan (\eta + \eta_0 ) + (\Gamma^2 + A^2)^{1/2} -A}{\Gamma \tan
(\eta
+ \eta_0 ) + (\Gamma^2 + A^2)^{1/2} +A}\right| \right] 
\begin{array}{c}
\\ 
\\ 
\end{array}
\!\!\!\!\!\! \right\}, \\
a^2(\eta) & = & \frac{1}{2\phi_0}\left( \Gamma + (\Gamma^2 +A^2)^{1/2}
\sin[ 
2(\eta + \eta_0)]\right) \exp \left\{ 
\begin{array}{c}
\\ 
\\ 
\end{array}
\!\!\!\!\!\! \left[\lambda_2 (\delta-1)\right]^{\frac{1}{1-\delta}}
\right. 
\nonumber \\
& & \mbox{} \left. \times \ln^{\frac{1}{1-\delta}} \left[ K\left|
\frac{ 
\Gamma \tan (\eta + \eta_0 ) + (\Gamma^2 + A^2)^{1/2} -A}{\Gamma \tan
(\eta
+ \eta_0 ) + (\Gamma^2 + A^2)^{1/2} +A}\right| \right] 
\begin{array}{c}
\\ 
\\ 
\end{array}
\!\!\!\!\!\! \right\}\,,
\end{eqnarray}
and when $\delta = 1$ these become 
\begin{eqnarray}
\phi(\eta) & = & \phi_0 \exp \left\{ -\exp \left( -\lambda_2 
\begin{array}{c}
\\ 
\\ 
\end{array}
\!\!\!\!\!\! \right. \right.  \nonumber \\
& & \mbox{} \left. \left. \times \ln \left[ K \left|\frac{\Gamma \tan
(\eta
+ \eta_0 ) + (\Gamma^2 + A^2)^{1/2} -A}{\Gamma \tan (\eta + \eta_0 ) +
(\Gamma^2 + A^2)^{1/2} +A}\right| \right] \right) \right\} \!\!\! \,,
\\
a^2(\eta) & = & \frac{1}{2\phi_0}\left( \Gamma + (\Gamma^2 +A^2)^{1/2}
\sin[
2(\eta + \eta_0)]\right)\exp \left\{ \exp \left( -\lambda_2 
\begin{array}{c}
\\ 
\\ 
\end{array}
\!\!\!\!\!\! \right. \right.  \nonumber \\
& & \mbox{} \left. \left. \times\ln\left[ K \left|\frac{\Gamma \tan
(\eta +
\eta_0 ) + (\Gamma^2 + A^2)^{1/2}-A}{\Gamma \tan (\eta + \eta_0 ) +
(\Gamma^2 + A^2) ^{1/2} +A}\right|\right] \right) \right\} \,.
\end{eqnarray}

{\center \subsection{Perfect-fluid solutions ($k=0$)}}

When $\phi \rightarrow \phi_0$ from above, ie $\phi \in
(\phi_0,\:\infty)$
the approach of Theory 2 to the general relativistic limit can be
accurately
modelled using the theory defined by Eq.~(\ref{phi40}) with $A>0$,
$B>1$.
Recall Eq.~(\ref{omega3}), 
\begin{eqnarray}
2\omega(\phi) + 3 & \rightarrow & \frac{4-3\gamma}{3(2-\gamma)^2}\frac
{(1-B)^2}{B}A^{1/B} \ln^{-\left(\frac{B+1}{B}\right)}\left(\frac
{\phi}{\phi_0}\right)\,,
\end{eqnarray}
as $\phi \rightarrow \phi_0$. Thus we can deduce the late-time
evolution of
Theory 2 by enforcing the relations on the perfect-fluid solutions
presented
in Section \ref{th1} 
\begin{eqnarray}
A & = & \left[\frac{3B_2\left(2-\gamma\right)^2 \left(2\delta-1
\right)}{
4\left(4-3\gamma\right)\left(\delta-1\right)^2}\right]
^{\frac{1}{2\delta -1}
}\,, \\
B & = & \frac{1}{2\delta-1}\,.
\end{eqnarray}
when $\phi \rightarrow \phi_0$ from above. Theory 2 can only tend to
GR at
late times with $\phi$ going to $\phi_0$ from above if $1/2 < \delta <
1$,
as follows from the constraint $B>0$. When $\phi \rightarrow \phi_0$
from
below the behaviour of the coupling is given by Eq.~(\ref{omega4}), 
\begin{eqnarray}
2\omega(\phi) + 3 & \rightarrow & \frac{4-3\gamma}{3(2-\gamma)^2}
\frac{ 
4A^{-1/B}}{B}\ln^{\frac{1-B}{B}}\left(\frac{\phi}{\phi_0} \right)\,,
\\
\end{eqnarray}
The late-time behaviour of Theory 2 when $\phi$ is in the range $\phi
\in
(0,\;\phi_0)$ is then given by Eqs.~(\ref{phi40}) -- (\ref{endth1})
under
the substitutions 
\begin{eqnarray}
A & = & \left[\frac{3\left(2-\gamma\right)^2 B_2}{4\left(4-3\gamma
\right)\left(1-2\delta\right)}\right]^{\frac{1}{2\delta-1}}\,, \\
B & = & \frac{1}{1-2\delta}\,.
\end{eqnarray}
The bound $B<0$ leads to the constraint $\delta>1/2$, hence there
exists a
wide spectrum of models with $0<\gamma<4/3$ in which $\phi \rightarrow
\phi_0 $ from below.

\begin{center}
\section{Theory 3: $2\omega (\phi )+3=B_{3}\left| 1-(\phi /\phi _0)
^\beta\right| ^{-1}$, $\beta >0$, $B_{3}>0$ constants.}

\label{th3} 
\end{center}

For this choice of the coupling function, we find 
\begin{equation}
\int \frac{\left(2\omega+3\right)^{1/2}}{\phi}d\,\phi = -
\frac{\sqrt{B_3}}{
\beta} \ln \left|\frac{1+\sqrt{u_2}}{1-\sqrt{u_2}} \right|
-\sqrt{3}\ln K\,,
\end{equation}
where $0<\phi<\phi_0$ and 
\begin{equation}  \label{u2def}
u_2 = 1-\left(\frac{\phi}{\phi_0}\right)^{\beta}\,.
\end{equation}

{\center \subsection{Vacuum solutions ($k=0$)}}

Starting with the flat $k=0$ models with $\phi \in (0, \phi_0)$, we
obtain 
\begin{equation}
\frac{1-\sqrt{u_2}}{1+\sqrt{u_2}} = \left( K \eta
\right)^{\lambda_2\beta}\,,
\end{equation}
with $\lambda_2 = \sqrt{3/B_3}$ and fixing $\eta_0 = 0$. Using
Eq.~(\ref
{vacy}) we can then deduce the evolution of the field and the metric
to be 
\begin{eqnarray}  \label{phi20}
\phi(\eta) & = & \frac{4^{1/\beta} \phi_0 \left(K \eta
\right)^{\lambda_2}}
{\left[1 + \left(K \eta\right)^{\lambda_2\beta}\right]^{2/\beta}}\,,
\\
a^2(\eta) & = & \frac{AK^{-\lambda_2}}{4^{1/\beta}\phi_0}\eta^
{1-\lambda_2}\left[1 + \left(K \eta\right)^{\lambda_2\beta}\right]
^{2/\beta}\,.
\end{eqnarray}

{\center \subsubsection{Late-time behaviour}}

Examining the form of Eq.~(\ref{phi20}) we see that there is no choice
of $K$
for which \linebreak $\phi \rightarrow $constant at large $\eta $,
precluding any
possible approach to GR at late times. In spite of this, we find for
the
asymptotic behaviour 
\begin{eqnarray}
\phi (\eta ) &\rightarrow &4^{1/\beta }\phi _0K^{-|\lambda _2|}\eta
^{-|\lambda _2|}\,, \\
a^2(\eta ) &\rightarrow &\frac{AK^{|\lambda _2|}}{4^{1/\beta }\phi
_0}\eta
^{1+|\lambda _2|}\,.
\end{eqnarray}
At late times we find 
\begin{eqnarray}
\eta (t) &\rightarrow &\left[ \left( \frac{3+|\lambda _2|}2\right)
\left( 
\frac{4^{1/\beta }\phi _0}{AK^{|\lambda _2|}}\right) ^{1/2}\right]
^{\frac
2{3+|\lambda _2|}}t^{\frac 2{3+|\lambda _2|}}\,, \\
\phi (t) &\rightarrow &\left( \frac{4^{1/\beta }\phi _0}{K^{|\lambda
_2|}}
\right) ^{\frac 3{3+|\lambda _2|}}\left( \frac{3+|\lambda
_2|}{2A^{1/2}} 
\right) ^{-\frac{2|\lambda _2|}{3+|\lambda |}}t^{-\frac{2|\lambda
_2|}{ 
3+|\lambda _2|}}\,, \\
a(t) &\rightarrow &\left( \frac{AK^{|\lambda _2|}}{4^{1/\beta }\phi
_0}
\right) ^{\frac 1{3+|\lambda _2|}}\left( \frac{3+|\lambda _2|}2\right)
^{
\frac{1+|\lambda _2|}{3+|\lambda _2|}}t^{\frac{1+|\lambda
_2|}{3+|\lambda_2| 
}}\,.
\end{eqnarray}
Since there is no late-time approach to GR, we do not pursue the
early-time
behaviour or probe for the existence of minima in these models.

{\center \subsection{Vacuum solutions ($k=-1$)}}

Similarly, for the $k=-1$ cases we have 
\begin{equation}
\frac{1-\sqrt{u_2}}{1+\sqrt{u_2}} = K^{\lambda_2\beta}
\tanh^{\lambda_2\beta} \eta \,,
\end{equation}
and convergence to GR as $\eta \rightarrow \infty$ demands $K=1$. By
direct
comparison with the solution for flat models we obtain the exact
results for
the evolution of $\phi$ and $a$ 
\begin{eqnarray}  
\label{phi21}
\phi(\eta) & = & \frac{4^{1/\beta} \phi_0 \tanh^{\lambda_2}
\eta}{\left(1 +
\tanh^{\lambda_2\beta} \eta \right)^{2/\beta}}\,, \\
\label{a21}
a^2(\eta) & = & \frac{A}{2^{\frac{\beta+2}{\beta}}\phi_0}\sinh(2\eta)
\frac{
\left(1 + \tanh^{\lambda_2\beta} \eta \right)^{2/\beta}}
{\tanh^{\lambda_2}
\eta}\,.
\end{eqnarray}

{\center \subsubsection{Late-time behaviour}}

Asymptotically, these relationships tend to the pair 
\begin{eqnarray}
\phi (\eta ) &\rightarrow &\phi _0\left[ 1-\lambda _2^2\beta e^{-4\eta
}\right] \,,  \label{a90} \\
a(\eta ) &\rightarrow &\frac 12\sqrt{\frac A{\phi _0}}e^\eta \left[
1+\frac{ 
\lambda _2^2\beta }2e^{-4\eta }\right] \,,
\end{eqnarray}
as $\eta \rightarrow \infty $, and we find 
\begin{equation}
\eta (t)\simeq \ln \left[ 2\sqrt{\frac{\phi _0}A}t\left(
1+\frac{A^2\lambda
_2^2\beta }{96\phi _0^2}t^{-4}\right) \right] \,,
\end{equation}
leading to 
\begin{eqnarray}
\phi (t) &\rightarrow &\phi _0\left[ 1-\frac{A^2\lambda _2^2\beta
}{16\phi
_0^2}t^{-4}\right] \,, \\
a(t) &\rightarrow &t\left[ 1+\frac{A^2\lambda _2^2\beta }{24\phi _0^2}
t^{-4}\right] \,,
\end{eqnarray}
as $t\rightarrow \infty ,$ and so there is power-law approach to the
Milne
universe of GR at late times.

{\center \subsubsection{Early-time behaviour}}

Setting $\eta _0=0$, we can expand Eqs.~(\ref{phi21}) and (\ref{a21})
about $
\eta =0$. We find 
\begin{eqnarray}
\phi (\eta ) &\simeq &4^{1/\beta }\phi _0\eta ^{|\lambda _2|}\,, 
\label{a24}
\\
a^2(\eta ) &\simeq &4^{-1/\beta }\frac A{\phi _0}\eta ^{1-|\lambda
_2|}\,.
\end{eqnarray}
The latter of these leads to the early-time $\eta (t)$ relation, 
\begin{equation}
\eta (t)=2^{\frac{2(1-\beta )}{\beta (3-|\lambda _2|)}}\left(
\frac{\phi_0} 
A\right) ^{\frac 1{3-|\lambda _2|}}\left( 3-|\lambda _2|\right)
^{\frac
2{3-|\lambda _2|}}t^{\frac 2{3-|\lambda _2|}}\,.  \label{eta10}
\end{equation}
Studying the form of the exponent in Eq.~(\ref{eta10}) reveals that
when $
|\lambda _2|<3$, $t\rightarrow 0$ as $\eta \rightarrow 0$ and when
$|\lambda 
_2|\geq 3$, $t\rightarrow -\infty $ as $\eta \rightarrow 0$. The
early-time
evolution of the universe as a function of $t$ is then given by 
\begin{eqnarray}
\phi (t) &\rightarrow &2^{\frac{2(3-|\lambda _2|\beta )}{\beta
(3-|\lambda
_2|)}}\phi _0^{\frac 3{3-|\lambda _2|}}A^{\frac{|\lambda _2|}
{|\lambda
_2|-3} 
}\left( 3-|\lambda _2|\right) ^{\frac{2|\lambda _2|}{3-|\lambda
_2|}}t^{
\frac{2|\lambda _2|}{3-|\lambda _2|}}\,,  \label{a23} \\
a(t) &\rightarrow &2^{\frac{\beta (|\lambda _2|-1)-2}{\beta
(3-|\lambda _2|)}
}\left( \frac A{\phi _0}\right) ^{\frac 1{3-|\lambda _2|}}\left(
3-|\lambda
_2|\right) ^{\frac{1-|\lambda _2|}{3-|\lambda _2|}}t^{\frac{1-|\lambda
_2|}{ 
3-|\lambda _2|}}\,,
\end{eqnarray}
as $t\rightarrow 0$, $-\infty $ accordingly. When $|\lambda _2|=3$,
the $t$
-dependent evolution is given by 
\begin{eqnarray}
\eta (t) &\rightarrow &\exp \left[ 2^{1/\beta }\sqrt{\frac{\phi _0}A}
t\right] \,,  \label{a91} \\
\phi (t) &\rightarrow &4^{1/\beta }\phi _0\exp \left[ 3.2^{1/\beta
}\sqrt{ 
\frac{\phi _0}A}t\right] \,, \\
a(t) &\rightarrow &2^{-1/\beta }\sqrt{\frac A{\phi _0}}\exp \left[
-2^{1/\beta }\sqrt{\frac{\phi _0}A}t\right] \,,
\end{eqnarray}
as $t\rightarrow -\infty $.

{\center \subsubsection{Minima}}

Differentiating Eq.~(\ref{a21}), we find 
\begin{equation}
\left( a^2\right) _\eta =\frac{a^2}{\sinh (2\eta )}\left[ \cosh (2\eta
)+
\frac{2\lambda _2}{\tanh ^{-\lambda _2\beta }\eta +1}-\lambda
_2\right] \,.
\label{t3km1vm}
\end{equation}
At early times, ie as $\eta \rightarrow 0$, $a^2\rightarrow 0$ if
$|\lambda
_2|<1$. Since $\sinh (2\eta )\simeq 2\eta $ as $\eta \rightarrow 0$,
Eq.~(
\ref{a24}) implies 
\begin{equation}
\left( a^2\right) _\eta \propto \eta ^{-|\lambda _2|}\left[ \dots
\right] \,,
\end{equation}
as $\eta \rightarrow 0$, and thus the power-law pre-factor in $\left(
a^2\right) _\eta $ can never vanish at early times. The field $\phi $
evolves monotonically, by Eq.~(\ref{trace1}), and tends to a constant
at
late times, which guarantees that $\phi _\eta $ cannot diverge during
the
subsequent evolution. Eq.~(\ref{trace1}), in conjunction with the
bound $
0<\phi <\phi _0$, then ensures that there can be no further zeros of
$a$ as
the universe evolves. Eq.~(\ref{a90}) ensures $\left( a^2\right) _\eta
$ is
non-zero as $\eta \rightarrow \infty $. In general, stationary points
in the
evolution of $a$ arise at $\eta _{*}$, given by 
\begin{equation}
\cosh \left( 2\eta _{*}\right) +\frac{2\lambda _2}{\tanh ^{-\lambda
_2\beta
}\eta _{*}+1}-\lambda _2=0\,,
\end{equation}
obtained from the square-bracketed factor in Eq.~(\ref{t3km1vm}).
Remembering the positivity of both $\cosh \left( 2\eta _{*}\right) -1$
and $
\eta _{*}$, we can derive the interesting result that when $|\lambda
_2|>1$ 
the location of the minimum is constrained by 
\begin{equation}
\eta _{*}<{\rm arctanh}\left( \frac{\lambda _2-1}{\lambda _2+1}\right)
^{1/\lambda _2\beta }\,,
\end{equation}
and when $|\lambda _2|\leq 1$ stationary points do not exist.

{\center \subsection{Vacuum solutions ($k=+1$)}}

Finally, we note the existence of $k=+1$ closed-universe solutions,
governed by 
\begin{equation}
\frac{1-\sqrt{u_2}}{1+\sqrt{u_2}} = K^{\lambda_2\beta}
\tan^{\lambda_2\beta}\eta\,.
\end{equation}
This leads to the exact solution, in conformal time, 
\begin{eqnarray}
\phi(\eta) & = & \frac{4^{1/\beta}\phi_0 K^{\lambda_2}
\tan^{\lambda_2}\eta}{ 
\left(1+K^{\lambda_2\beta}
\tan^{\lambda_2\beta}\eta\right)^{2/\beta}}\,, \\
a^2(\eta) & = & \frac{A}{2^{\frac{\beta+2}{\beta}}\phi_0}\sin
(2\eta)\frac{ 
\left(1+K^{\lambda_2\beta}\tan^{\lambda_2\beta}\eta
\right)^{2/\beta}}{ 
K^{\lambda_2}\tan^{\lambda_2}\eta}\,.
\end{eqnarray}

{\center \subsection{Radiation solutions ($k=0$)}}

The behaviour of the $k=0$ radiation models for this choice of the
coupling
function is determined by 
\begin{equation}
\ln \left( \frac{1-\sqrt{u_2}}{1+\sqrt{u_2}} \right) = \lambda_2
\beta\ln\left[\frac{K\eta}{\eta + 2\eta_0}\right]\,,
\end{equation}
where we have exploited our freedom in $\eta_0$ to set $A=2\Gamma
\eta_0$
and $u_2$ is as defined by Eq.~(\ref{u2def}). We also require K=1, so
that $
\phi \rightarrow \phi_0$ as $\eta \rightarrow \infty$. Using
Eq.~(\ref{rady}
) we can calculate the exact evolution of these models 
\begin{eqnarray}  \label{a92}
\phi(\eta) & = & 4^{1/\beta} \phi_0 \left(\frac{\eta}{\eta + 2\eta_0}
\right)^{\lambda_2}\left[1 + \left( \frac{\eta}{\eta+2\eta_0} \right)
^{\lambda_2\beta}\right]^{-2/\beta}\,, \\
a^2(\eta) & = & \frac{\Gamma}{4^{1/\beta}\phi_0} \eta^{1-\lambda_2}
\left(\eta+2\eta_0\right)^{1+\lambda_2}\left[1+\left(\frac{\eta}{\eta
+2\eta_0}\right)^{\lambda_2\beta}\right]^{2/\beta}\,.
\end{eqnarray}

{\center \subsubsection{Late-time behaviour}}

At late times these equations may be approximated by 
\begin{eqnarray}
\phi (\eta ) &\rightarrow &\phi _0\left[ 1-\lambda _2^2\beta
\frac{\eta_0^2 
}{\eta ^2}\right] \,,  \label{a25} \\
a(\eta ) &\rightarrow &\sqrt{\frac \Gamma {\phi _0}}\eta \left[
1+\frac{\eta
_0}\eta +\frac 12\left( \lambda _2^2\beta -1\right) \frac{\eta
_0^2}{\eta^2}
\right] \,.
\end{eqnarray}
It is necessary to extend the computation of $a(\eta )$ to
second-order
since the first-order contributions will later vanish. Eq.~(\ref{a25})
allows asymptotic calculation of conformal time as a function of
cosmic
time, 
\begin{equation}
\eta (t)\rightarrow \sqrt{2}\left( \frac{\phi _0}\Gamma \right)
^{1/4}t^{1/2}\left[ 1-\frac{\eta _0}{\sqrt{2}}\left( \frac \Gamma
{\phi
_0}\right) ^{1/4}t^{-1/2}-\frac{\eta _0^2}8\sqrt{\frac \Gamma {\phi
_0}}
\left( \lambda _2^2\beta -1\right) \frac{\ln t}t\right] \,,
\end{equation}
and hence there is power-law approach to the GR solution 
\begin{eqnarray}
\phi (t) &\rightarrow &\phi _0\left[ 1-\frac{\lambda _2^2\beta }2\eta
_0^2
\sqrt{\frac \Gamma {\phi _0}}t^{-1}\right] \,, \\
a(t) &\rightarrow &\sqrt{2}\left( \frac \Gamma {\phi _0}\right)
^{1/4}t^{1/2}\left[ 1-\frac{\eta _0^2}8\sqrt{\frac \Gamma {\phi
_0}}\left(
\lambda _2^2\beta -1\right) \frac{\ln t}t\right] \,,
\end{eqnarray}
as $t\rightarrow \infty $.

{\center \subsubsection{Early-time behaviour}}

We now probe the early-time behaviour of these models, finding two
distinct
cases. There exist zeros of $\phi$ as $\eta \rightarrow 0$ and $\eta
\rightarrow -2\eta_0$. Since we require $\phi>0$ during the portion of
the
evolution in which we are interested, and ultimately at late times, we
shall
take our early-time limit to be the most recent of these zeros. We
know from
Eq.~(\ref{trace1}) that 
\begin{equation}  \label{pevt3kzr}
\phi_\eta a^2 = \lambda_2 A\left|1-\left(\frac{\phi}{ \phi_0}
\right)^{\beta}\right|^{1/2}\,,
\end{equation}
and since $\phi \in (0,\phi_0)$ and $\phi \rightarrow \phi_0$ at late
times
we deduce that $\phi_{\eta} >0$. From Eq.~(\ref{pevt3kzr}), we derive
$
\lambda_2 A >0$. We have fixed $A=2\Gamma \eta_0$, with $\Gamma>0$, so
we
may also say $\lambda_2\eta_0 >0$, ie when $\lambda_2 >0$, $\eta
\rightarrow
0$ is the more recent zero of $\phi$ and when $\lambda_2 <0$, $\eta
\rightarrow -2\eta_0$ is most recent. When $\lambda_2>0$ we find 
\begin{eqnarray}
\phi(\eta) & \rightarrow & 4^{1/\beta}\phi_0\left(\frac{
\eta}{2\eta_0} 
\right)^{\lambda_2}\,, \\
a^2(\eta) & \rightarrow & \frac{A}{4^{1/\beta}\phi_0}
\left(2\eta_0\right)^{ 
\lambda_2}\eta^{1-\lambda_2}\,,
\end{eqnarray}
as $\eta \rightarrow 0$. We can model the $t$-parametrised early-time
behaviour by Eqs.~(\ref{eta10}) -- (\ref{a91}), for the $k=-1$ vacuum
models, via the transformation 
\begin{equation}
\phi_0 \rightarrow \frac{\phi_0}{(2\eta_0)^{\lambda_2}}\,.
\end{equation}
When $\lambda_2<0$ 
\begin{eqnarray}
\phi(\eta) & \rightarrow & 4^{1/\beta}\phi_0\left( \frac{\eta +
2\eta_0}{ 
-2\eta_0}\right)^{-\lambda_2}\,, \\
a^2(\eta) & \rightarrow & -\frac{A}{4^{1/\beta}\phi_0} \left(-2\eta_0 
\right)^{-\lambda_2}\left(\eta + 2\eta_0 \right)^{1+\lambda_2}\,,
\end{eqnarray}
as $\eta \rightarrow -2\eta_0$. Note that both $A$ and $\eta_0$ are
negative
here. The behaviour as a function of $t$ can be gleaned from
Eqs.~(\ref
{eta10}) -- (\ref{a91}), by the redefinitions 
\begin{eqnarray}
\phi_0 & \rightarrow & \phi_0 \left(-2\eta_0\right) ^{\lambda-2}\,, \\
\eta & \rightarrow & \eta + 2\eta_0 \,, \\
A & \rightarrow & -A\,.
\end{eqnarray}

{\center \subsubsection{Minima}}

Differentiating Eq.~(\ref{a92}) yields 
\begin{equation}  \label{t3kzrm}
\left(a^2\right)_{\eta} = \frac{a^2}{\eta\left( \eta + 2\eta_0
\right)} 
\left[2\eta_0 \left(1+\lambda_2\right) + 2\eta - \frac{4\lambda_2
\eta_0}{1
+ \left(\frac{\eta} {\eta + 2\eta_0}\right)^{\lambda_2
\beta}}\right]\,.
\end{equation}
If the early-time limit is given by $X \rightarrow 0$ where $X$ is
either $
\eta$ or $\eta +2\eta_0$, then 
\begin{equation}
\left(a^2\right)_{\eta} \propto X^{-|\lambda_2|}\left[\dots \right]\,,
\end{equation}
and thus never vanishes at early times. From Eq.~(\ref{pevt3kzr}) we
deduce
that when $0<\phi<\phi_0$, $\phi$ is bound to evolve monotonically.
The
necessity that it tend to a constant at late times then occludes the
possibility of it diverging after the early-time regime. The non-zero
right-hand side of Eq.~(\ref{pevt3kzr}) on the interval
$0<\phi<\phi_0$ then
guarantees that $a$ is non-zero. Consequently, the stationary points
in the
evolution are obtained by setting the square-bracketed factor in
Eq.~(\ref
{t3kzrm}) to zero. They lie at $\eta_*$ where 
\begin{equation}
\frac{2\lambda_2 \eta_0}{\left(\frac{\eta_*}{\eta_*+2\eta_0}
\right)^{\lambda_2 \beta} +1} - \eta_* = \eta_0\left(1 +
\lambda_2\right) \,,
\end{equation}
which must be solved numerically for the particular values of $\beta$,
$\lambda_2$, and $\eta_0$.

{\center \subsection{Radiation solutions ($k=-1$)}}

When the curvature is negative, the evolution is determined by the
equation 
\begin{equation}  \label{radnegt1}
\frac{1-\sqrt{u_2}}{1+\sqrt{u_2}} = K^{\lambda_2 \beta}\left|\frac{ 
\Gamma\tanh(\eta+\eta_0) + \left(A^2 - \Gamma^2\right)^{1/2} -A}{
\Gamma\tanh(\eta+\eta_0) + \left( A^2 - \Gamma^2\right)^{1/2} + A} 
\right|^{\lambda_2 \beta} \,, A^2 \geq \Gamma^2\,,
\end{equation}
with $u_2$ and $\lambda_2$ as defined earlier. We make the simplifying
choice of $\eta_0$: 
\begin{equation}  \label{etkm1rt3}
\tanh\left(2\eta_0\right) = \frac{\Gamma}{A}\,,
\end{equation}
which ensures that the early-time behaviour (ie, $\phi \rightarrow 0$)
occurs at $\eta=0$. To recover GR at late times, we fix 
\begin{equation}  \label{kdef20}
K= \frac{\Gamma + \left(A^2 - \Gamma^2\right)^{1/2} +A}{\Gamma +
\left(A^2 -
\Gamma^2\right)^{1/2} -A}\,.
\end{equation}
Substituting Eq.~(\ref{etkm1rt3}) into the $k=-1$ right-hand side of
Eq.~(
\ref{rady}) we find 
\begin{equation}
y(\eta) = \frac{\Gamma}{2}\left[\cosh(2\eta)-1\right] + \frac{A}{2}
\sinh(2\eta)\,.
\end{equation}
Lastly, we note that with $\eta_0$ given by Eq.~(\ref{etkm1rt3}), the
expression within the moduli on the right-hand side of
Eq.~(\ref{radnegt1})
is monotonically increasing from zero and thus positive. Dropping the
moduli, we find 
\begin{eqnarray}
\phi(\eta) & = & \frac{4^{1/\beta}\phi_0 K^{\lambda_2}\left( \frac{
\Gamma\tanh(\eta + \eta_0) + \left(A^2 - \Gamma^2 \right)^{1/2} - A}{ 
\Gamma\tanh(\eta+\eta_0) + \left( A^2 - \Gamma^2\right)^{1/2} + A} 
\right)^{\lambda_2}}{\left[1 + K^{ \lambda_2\beta}\left(\frac{
\Gamma\tanh(\eta + \eta_0) + \left(A^2 - \Gamma^2 \right)^{1/2} - A}{ 
\Gamma\tanh(\eta +\eta_0) + \left( A^2 -\Gamma^2\right)^{1/2} +
A}\right)
^{\lambda_2\beta}\right]^{2/\beta}}\,, \\
a^2(\eta) & = & \frac{4^{-1/\beta} K^{-\lambda_2}}{2\phi_0}
\left[\Gamma\left[\cosh(2\eta) -1\right] + A\sinh(2\eta)\right] 
\nonumber \\
& & \mbox{}\times\frac{\left[1+ K^{ \lambda_2\beta}\left(\frac{
\Gamma\tanh(\eta + \eta_0) + \left(A^2 - \Gamma^2 \right)^{1/2} - A}{
\Gamma\tanh(\eta +\eta_0) + \left( A^2 -\Gamma^2\right)^{1/2} +
A}\right)
^{\lambda_2\beta}\right]^{2/\beta}}{\left(\frac{\Gamma\tanh( \eta +
\eta_0)
+ \left(A^2 - \Gamma^2\right)^{1/2} - A}{ \Gamma\tanh(\eta+\eta_0) +
\left(A^2 - \Gamma^2\right)^{1/2} +A}\right)^{\lambda_2}}\,.
\end{eqnarray}

{\center \subsubsection{Late-time behaviour}}

As $\eta \rightarrow \infty $ we find 
\begin{eqnarray}
\phi (\eta ) &\rightarrow &\phi _0\left[ 1-\frac{\lambda
_2^2A^2\beta}{
A^2-\Gamma ^2}\left( \frac{A-\Gamma -\left( A^2-\Gamma ^2\right)
^{1/2}}{ 
A+\Gamma -\left( A^2-\Gamma ^2\right) ^{1/2}}\right) ^2e^{-4\eta
}\right] \,,
\\
a^2(\eta ) &\rightarrow &\frac{\Gamma +A}{4\phi _0}e^{2\eta }\left[
1-\frac{ 
2\Gamma }{\Gamma +A}e^{-2\eta }\right] \,,
\end{eqnarray}
as $\eta \rightarrow \infty $. These lead to the $t$-dependent
behaviours: 
\begin{eqnarray}
\eta (t) &\rightarrow &\ln \left[ 2\sqrt{\frac{\phi _0}{\Gamma
+A}}t\left(
1-\frac \Gamma {4\phi _0}t^{-2}\right) \right] \,, \\
\phi (t) &\rightarrow &\phi _0\left[ 1-\frac{\lambda _2^2A^2\beta
}{16\phi
_0^2}\left( \frac{A+\Gamma }{A-\Gamma }\right) \left(
\frac{A-\Gamma-\left(
A^2-\Gamma ^2\right) ^{1/2}}{A+\Gamma -\left( A^2-\Gamma
^2\right)^{1/2}}
\right) ^2t^{-4}\right] \,, \\
a(t) &\rightarrow &t\left[ 1-\frac \Gamma {4\phi _0}t^{-2}\right]\,,
\end{eqnarray}
as $t\rightarrow \infty $. Again, we see the approach of this model to
the
Milne universe as $t$ increases.

{\center \subsubsection{Early-time behaviour}}

Analysing the behaviour in the neighbourhood of the last zero of
$\phi$ (ie,
around $\eta = 0$) we obtain a description of the early-time
behaviour. We find 
\begin{eqnarray}
\phi(\eta) & \rightarrow & 4^{1/\beta} \phi_0 K^{\lambda_2}
\left[\frac{ 
\Gamma}{A}\sqrt{\frac{A^2}{\Gamma^2}-1}\left( \frac{A}{\Gamma}-
\sqrt{\frac{ 
A^2}{\Gamma^2} -1}\right) \right]^{\lambda_2}\eta^{\lambda_2}\,, \\
a^2(\eta) & \rightarrow & \frac{4^{-1/\beta}AK^
{-\lambda_2}}{\phi_0}\left[ 
\frac{\Gamma}{A} \sqrt{\frac{A^2}{\Gamma^2}-1}\left(\frac{A}{\Gamma}-
\sqrt{
\frac{A^2}{\Gamma^2} -1}\right)\right]^{-\lambda_2}
\eta^{1-\lambda_2}\,,
\end{eqnarray}
with $K$ as given by Eq.~(\ref{kdef20}). The reality condition $A^2 >
\Gamma^2$ and the inequality $\lambda_2A > 0$ (from
Eq.~(\ref{pevt3kzr}),
using $\phi_{\eta}>0$) ensure all of the pre-factors in the above
relations
are real and positive. We obtain the $t$-parametrised behaviour at
early
times by applying the substitutions 
\begin{eqnarray}
|\lambda_2| & \rightarrow & \lambda_2\,, \\
\phi_0 & \rightarrow & \phi_0 K^{\lambda_2}\left[
\frac{\Gamma}{A}\sqrt{ 
\frac{A^2}{\Gamma^2}-1}\left( \frac{A}{\Gamma}-
\sqrt{\frac{A^2}{\Gamma^2}-1}
\right) \right]^{\lambda_2}\,,
\end{eqnarray}
to Eqs.~(\ref{eta10}) -- (\ref{a91}).

{\center \subsubsection{Minima}}

If $\lambda_2>1$ then the scale-factor diverges at early times. For
such
models to look like GR as $t \rightarrow \infty$ requires the presence
of a
minimum. Using Eq.~(\ref{mintest}) we find 
\begin{equation}
\cosh\left[2(\eta_*+\eta_0)\right] = \lambda_2 \cosh(2\eta_0)
\left\{\frac{ 
1- K^{\lambda_2 \beta} \left[\frac{\Gamma \tanh (\eta_* + \eta_0) +
\left(A^2-\Gamma^2\right)^{1/2} -A}{ \Gamma\tanh(\eta_*+\eta_0)
+\left(A^2-\Gamma^2\right)^{1/2} + A}\right]^{\lambda_2\beta}}{1+
K^{\lambda_2\beta}\left[
\frac{\Gamma\tanh(\eta_*+\eta_0)+\left(A^2-\Gamma^2 
\right)^{1/2}
-A}{\Gamma\tanh(\eta_*+\eta_0)+\left(A^2-\Gamma^2\right)^{1/2}
+A}\right]^{\lambda_2\beta}}\right\}\,,
\end{equation}
defines $\eta_*$, where the minimum is situated. Although this
expression is
not soluble analytically, we may obtain bounds on the value of
$\eta_*$.
Eq.~(\ref{mintest}), for Theory 3, may be written 
\begin{equation}
\Gamma \sinh\left(2\eta_*\right) + A\cosh\left(2\eta_* \right)=
\lambda_2 A
u_2^{1/2}\,.
\end{equation}
The bounds on $u_2$, namely $u_2 \in (0, 1)$ then imply 
\begin{equation}
0 < \cosh\left[2\left(\eta_* + \eta_0\right)\right] < \lambda_2
\cosh\left(2
\eta_0\right)\,,
\end{equation}
after using Eq.~(\ref{etkm1rt3}). The $\cosh$ function is greater than
unity
and monotonically increasing when its argument is positive, allowing
us to
strengthen the above inequality to 
\begin{equation}
0<\eta_*<\frac{1}{2}{\rm arccosh}\left[\lambda_2\cosh \left(
2\eta_0\right)\right] -\eta_0\,,
\end{equation}
where the lower limit follows from the positivity 0f $\eta$.

{\center \subsection{Radiation solutions ($k=+1$)}}

For the positively-curved models ($k=+1$) we obtain 
\begin{equation}
\frac{1-\sqrt{u_2}}{1+\sqrt{u_2}} = \left[K\left|\frac{\Gamma
\tan(\eta+\eta_0)+(\Gamma^2+A^2)^{1/2}-A}{\Gamma
\tan(\eta+\eta_0)+(\Gamma^2+A^2)^{1/2} +A} \right|
\right]^{\lambda_2\beta}
= \sigma(\eta)\,.
\end{equation}
This leads to the conformal time-parametrised set of equations 
\begin{eqnarray}
\phi(\eta) & = & \phi_0\left[\frac{4\sigma(\eta)}{\left(1+\sigma(\eta)
\right)^{2}}\right]^{1/\beta}\,, \\
a^2(\eta) & = & \frac{1}{2\phi_0}\left[\frac{\left(1+\sigma(\eta)
\right)^{2} 
}{4\sigma(\eta)}\right]^{1/\beta}\left[\Gamma + \left(\Gamma^2 +
A^2\right)^{1/2} \sin\left[2(\eta+\eta_0)\right] \right]\,.
\end{eqnarray}

{\center \subsection{Perfect-fluid solutions ($k=0$)}}

We examine the late-time form of the solutions resulting from a
scalar-tensor cosmology driven by the third type of coupling function
using
the theory first studied by Barrow and Mimoso and defined by 
\begin{equation}  \label{phi41}
\phi(\xi) = \phi_0 \exp\left(P\xi^Q \right)\,,
\end{equation}
where $P$ and $Q$ are constants. This choice of $\phi(\xi)$ arises
from the
generating function 
\begin{equation}
g(\xi) = \frac{1}{PQ}\xi^{2-Q}\,,
\end{equation}
and results in the scale-factor 
\begin{equation}  \label{a41}
a^{3(2-\gamma)}(\xi) = \frac{a_0^{3(2-\gamma)}}{PQ\phi_0}\xi^{2-Q}
\exp\left(-P\xi^{Q}\right)\,.
\end{equation}
Positivity of the left-hand side of Eq.~(\ref{a41}) requires that
$\xi^{J}
PQ>0$. Hence, we may derive 
\begin{equation}
f(\xi) = \frac{1}{PQ}\xi^{2-Q} + \frac{4-3\gamma}{4}\xi^2 -D\,,
\end{equation}
and 
\begin{equation}  \label{omega20}
2\omega(\xi) + 3 = \frac{4-3\gamma}{3(2-\gamma)^2}\frac{\left[\frac
{2-Q}{PQ}\xi^{1-Q} +
\frac{4-3\gamma}{2}\xi\right]^2}{\left[\frac{1}{PQ}
\xi^{2-Q} + \frac{4-3\gamma}{4}\xi^2\right]}\,.
\end{equation}
Again, at late times we demand $\phi \rightarrow \phi_0$. This can be
realised as $\xi \rightarrow 0$ when $Q>0$ or as $\xi \rightarrow
\infty$
when $Q<0$. When $P=0$ and/or $Q=0$ we have GR.

{\center \subsubsection{Late-time behaviour}}

The limiting forms of the coupling function as $\xi \rightarrow 0$
when $Q>0$%
, and as $\xi \rightarrow \infty $ when $Q<0$, are identical: 
\begin{equation}
2\omega (\xi )+3\rightarrow \frac{4-3\gamma }{3(2-\gamma
)^2}\frac{(2-Q)^2}{ 
PQ}\xi ^{-Q}\,.  \label{omega21}
\end{equation}
In this case, Eq.~(\ref{timeperf}) is 
\begin{equation}
dt\simeq \left( \frac{2-Q}{2-\gamma }\right) \frac{a_0^{3(\gamma
-1)}}{ 
3(PQ\phi _0)^{\frac{\gamma -1}{2-\gamma }}}\left( \frac{4-3\gamma
}{PQ} 
\right) ^{1/2}\xi ^{\frac{4\gamma -4-Q\gamma }{2(2-\gamma )}}\exp
\left[ 
-P\left( \frac{\gamma -1}{2-\gamma }\right) \xi ^Q\right] d\xi \,,
\label{elem3}
\end{equation}
as $\xi ^Q\rightarrow 0$. This expression may be integrated
approximately
for small $\xi ^Q$, giving 
\begin{equation}
t(\xi )\simeq S^{-1}\xi ^{\frac{\gamma (2-Q)}{2(2-\gamma )}}\left[
1-P\left( 
\frac{\gamma -1}{2-\gamma }\right) \xi ^Q\right] \,,  \label{t16}
\end{equation}
where $S$ is defined by 
\begin{equation}
S\equiv \frac{3\gamma }2\frac{(PQ\phi _0)^{\frac{\gamma -1}{2-\gamma
}}}{
a_0^{3(\gamma -1)}}\left( \frac{PQ}{4-3\gamma }\right) ^{1/2}\,.
\end{equation}
Ignoring the square brackets in Eq.~(\ref{t16}), we obtain the
lowest-order
inversion of this expression 
\begin{equation}
\xi (t)\simeq S^{\frac{2(2-\gamma )}{\gamma (2-Q)}}t^{\frac{2(2-\gamma
)}{
\gamma (2-Q)}}\,.
\end{equation}
Introducing the corrections associated with the terms in the square
bracket
leads, at the next-order, to 
\begin{equation}
\xi (t)\simeq S^{\frac{2(2-\gamma )}{\gamma (2-Q)}}t^{\frac{2(2-\gamma
)}{
\gamma (2-Q)}}\left[ 1+\frac{2P}\gamma \left( \frac{\gamma
-1}{2-Q}\right)
S^{\frac{2Q(2-\gamma )}{\gamma (2-Q)}}t^{\frac{2Q(2-\gamma )}{\gamma
(2-Q)}
}\right] \,.
\end{equation}
From this we can see that $\xi \rightarrow 0$ as $t\rightarrow \infty
$ iff $%
Q>2$. As $\xi \rightarrow \infty $, $t\rightarrow \infty $ iff $Q<2$;
the
condition $\phi \rightarrow \phi _0$ as $\xi \rightarrow \infty $
requires $
Q<0$, we thus deduce that the range of theories demarcated by $0<Q<2$
will
not approach GR as $t\rightarrow \infty $. Substituting the above $\xi
(t)$
into Eqs.~(\ref{phi41}) and (\ref{a41}) we obtain the $t$-dependent
evolution 
\begin{eqnarray}
\label{phi42}
\phi (t) &\rightarrow &\phi _0\left[ 1-PS^{\frac{2Q(2-\gamma )}{\gamma
(2-Q)}
}t^{\frac{2Q(2-\gamma )}{\gamma (2-Q)}}\right] \,, \\ 
\label{a42} 
a(t) &\rightarrow &\frac{a_0S^{2/3\gamma }}{(PQ\phi _0)^{\frac
1{3(2-\gamma
)}}}t^{2/3\gamma }\left[ 1-\frac 13PS^{\frac{2Q(2-\gamma )}{\gamma
(2-Q)}}t^{
\frac{2Q(2-\gamma )}{\gamma (2-Q)}}\right] \,,
\end{eqnarray}
as $t\rightarrow \infty $. The corresponding late-time evolution of
the
coupling as a function of $\phi $ is given by 
\begin{equation}
2\omega (\phi )+3\rightarrow \frac{4-3\gamma }{3(2-\gamma
)^2}\frac{(2-Q)^2} 
Q\frac 1{\ln \left( \phi /\phi _0\right) }\,.
\end{equation}
When $Q>0$ the requirement that $a$ be positive ensures $P\xi ^Q>0$.
From
Eq.~(\ref{timeperf}) we see that for $Q>2$ $d\xi /dt<0$, ie $\xi
\rightarrow
0$ from above and hence $P>0$. Eq.~(\ref{phi41}) confirms that $\phi
\rightarrow \phi _0$ from above when $P$ and $Q$ lie in these domains.
Conversely, when $Q<0$ the requirement $a>0$ implies $P\xi ^Q<0$. At
late
times $\xi \rightarrow \infty $ in these models and so $P<0$ and $\phi
$
approaches the GR value, $\phi _0,$ from below.

We remarked earlier that the $Q=0$ case gives pure GR at all times.
When $ 
Q=2 $ Eq.~(\ref{omega20}) for $2\omega +3$ does not approach the limit
given
in Eq.~(\ref{omega21}) as $\xi \rightarrow 0$, instead we have 
\begin{equation}
2\omega(\xi) + 3 \rightarrow \frac{2(4-3\gamma)^3 P}{12(2-\gamma)^2}
\xi^2\,,
\end{equation}
and $2\omega +3$ decays to zero as $\xi \rightarrow 0$. The range of
$Q$
incompatible with current observations, that GR is a good
approximation to
the time description of gravitation today, is then $0<Q\leq 2$.

{\center \subsubsection{Early-time behaviour}}

When $Q>2$, $P>0$ the scale-factor $a$ approaches zero as $\xi
\rightarrow 
\pm\infty$. 
\begin{equation}  \label{omega22}
2\omega + 3 \rightarrow \frac{(4-3\gamma)^2}{3(2-\gamma)^2}\,,
\end{equation}
ie, BD theory. The square root of this limit is positive and so it
follows
from Eq.~(\ref{timeperf}) that $d\xi/dt$ must also be positive,
demanding
that $\xi$ approach $-\infty$. The positivity of $a$, $Q$ and $P$
imply $
(-1)^Q >0$ and hence $\phi \rightarrow \infty$ as $\xi \rightarrow
-\infty$, 
$a \rightarrow 0$.

When $Q<0$, a necessary condition for the scale-factor to approach
zero is $
\xi \rightarrow 0$ and in this limit we recover BD theory again, the
coupling given by Eq.~(\ref{omega22}). As before $\sqrt{2\omega
+3}>0$, $
d\xi /dt>0$ and $\xi \rightarrow 0$ from above. Positivity of $a$ thus
requires $PQ>0$, which is guaranteed from the late-time behaviour.
However,
for $a$ to converge to zero as $\xi \rightarrow 0$ requires $P>0$ and
hence $
Q>0$. This direct contradiction ensures that $a(\xi )$ will be
non-singular
when $Q<0$. Explicitly, $a(\xi )$ possesses a minimum at 
\begin{equation}
\xi _{*}=\left( \frac{2-Q}{PQ}\right) ^{1/Q}\,.
\end{equation}
For reasons given in Section \ref{early1} we refrain from presenting
the
explicit form of the solution here.

{\center \subsubsection{Dust models}}

The late-time evolution of the $\gamma=1$ models is, from
Eqs.~(\ref{phi42})
and (\ref{a42}), that 
\begin{eqnarray}
\phi(t) & \rightarrow & \phi_0\left[1 - P\left(\frac{3}{2(PQ)^{1/2}}
\right)^{\frac{2Q}{2-Q}} t^{\frac{2Q}{2-Q}}\right]\,, \\
a(t) & \rightarrow & a_0 \left(\frac{9}{4\phi_0(PQ)^2}\right)^{1/3}
t^{2/3}\left[1 - \frac{P}{3}\left(\frac{3}{2(PQ)^{1/2}}\right)
^{\frac{2Q}{ 
2-Q}}t^{\frac{2Q}{2-Q}}\right]\,,
\end{eqnarray}
as $t \rightarrow \infty$.

{\center \subsubsection{Inflationary models}}

As was the case in Section \ref{inf1}, the solutions when $\gamma=0$
are
qualitatively different to their more general counterparts,
Eqs.~(\ref{phi42} 
) and (\ref{a42}). In this case Eq.~(\ref{elem3}) is 
\begin{equation}
dt \simeq \frac{(2-Q)\phi_0^{1/2}}{3a_0^3}\xi^{-1} \exp\left[
\frac{P}{2}
\xi^{Q}\right]d\xi\,,
\end{equation}
as $\xi^Q \rightarrow 0$, which integrates approximately in this limit
to give 
\begin{equation}  \label{t10}
t(\xi) \simeq \frac{(2-Q)\phi_0^{1/2}}{3a_0^3}\ln \xi \left[1 +
\frac{P\xi^Q
}{2Q \ln \xi}\right]\,.
\end{equation}
The first-order inversion of this at large $t$ gives 
\begin{equation}
\xi(t) \simeq \exp \left(\frac{3a_0^3 t}{(2-Q)\phi_0^{1/2}}\right)\,,
\end{equation}
which arises by neglecting the square brackets on the right of
Eq.~(\ref{t10}
). The next-order correction to this result is 
\begin{equation}
\xi(t) \simeq \left(\frac{3a_0^3 t}{(2-Q)\phi_0^{1/2}}\right)\left[1 -
\frac{
P}{2Q}\exp \left(\frac{3Qa_0^3 t}{(2-Q)\phi_0^{1/2}}\right) \right]\,,
\end{equation}
as $t \rightarrow \infty$. Substituting this into Eqs.~(\ref{phi41})
and (
\ref{a41}) yields 
\begin{eqnarray}
\phi(t) & \rightarrow & \phi_0\left[1 + P\exp \left(\frac{3Qa_0^3 t}
{(2-Q)\phi_0^{1/2}}\right)\right]\,, \\
a(t) & \rightarrow & \frac{a_0}{(PQ\phi_0)^{1/6}}\exp \left(\frac
{a_0^3
t}{2\phi_0^{1/2}}\right)\left[1 - \frac{P(2+Q)}{12Q}\exp
\left(\frac{3Qa_0^3
t}{(2-Q)\phi_0^{1/2}}\right)\right]\,.
\end{eqnarray}

{\center \subsubsection{Connection to the parameters of Theory 3  }}

At late times we have for all of the models considered in this section 
\begin{eqnarray}  
\label{forma}
2\omega(\phi) + 3 & \rightarrow & \frac{4-3\gamma}{3(2-\gamma)^2}
\frac{
(2-Q)^2 E}{Q}\left[\left(\frac{\phi}{\phi_0}\right)^{E}-1
\right]^{-1}\,,\;
{\rm as}\;\phi \rightarrow \phi_0\,, \\
\label{formb}
& \rightarrow & \frac{4-3\gamma}{3(2-\gamma)^2}\frac{(2-Q)^2 E}{(-Q)}
\left[1 - \left(\frac{\phi}{\phi_0}\right)^{E}\right]^{-1}\,,
\end{eqnarray}
where $E$ is a constant. When $\phi>\phi_0$ and $Q>0$ and we can model
the
form of the coupling as a function $\phi$ for Theory 3 by
Eq.~(\ref{forma}).
When $\phi<\phi_0$ and $Q<0$ we can use Eq.~(\ref{formb}) for the
coupling
as a function of $\phi$. This leads to the consistency relations
\begin{eqnarray}
\frac{(2-Q)^2}{{\rm sgn}(Q)} & = & \frac{3(2-\gamma)^2}{4-3\gamma}
\frac{B_3
}{\beta}\,, \\
E & = & \beta \,,
\end{eqnarray}
connecting Theory 3 as defined in Section \ref{coupling} with the
solutions presented here.

{\center \subsubsection{Exact solution}}

We note the existence of an exact solution, defined by the choice 
\begin{equation}  \label{phi43}
\phi(\xi) = \phi_0\left(1-\xi^{2-\lambda}\right)\,,
\end{equation}
where $\lambda$ is a constant. This leads, by Eq.~(\ref{gdef}), to 
\begin{equation}
g(\xi) = \frac{1}{\lambda -2}\xi^{\lambda}\left(1 - \xi^{2-\lambda}
\right)\,,
\end{equation}
and by Eq.~(\ref{aofg}) to 
\begin{equation}
a^{3(2-\gamma)} = \frac{a_0^{3(2-\gamma})}{(\lambda-2)\phi_0}
\xi^{\lambda}\,.
\end{equation}
The exact form of coupling driving this behaviour is given by
Eq.~(\ref
{omphid}) as 
\begin{equation}
2\omega(\xi) + 3 = \frac{4-3\gamma}{3(2-\gamma)^2 (\lambda-2)} \frac{ 
\left[\lambda \xi^{\lambda-2} -2J\right]^2}{\xi^{\lambda -2} -J}\,,
\end{equation}
or, using Eq.~(\ref{phi43}), 
\begin{equation}
2\omega(\phi) + 3 = \frac{4-3\gamma}{3(2-\gamma)^2 (\lambda-2)} \frac{ 
\left[\lambda\left(1-\phi/\phi_0\right)^{-1} - 2J\right]}{%
\left(1-\phi/\phi_0\right)^{-1} - J}\,,
\end{equation}
where 
\begin{equation}
J = 1 + \frac{(3\gamma - 4)(\lambda-2)}{4}\,.
\end{equation}

\begin{center}
\section{Discussion}
\label{disc}
\end{center}

In this paper we have supplied a comprehensive study of isotropic
cosmological models in scalar-tensor theories, extending the earlier
work of
refs.\cite{BAR1} and \cite{BAR3}. We have explored the behaviour of
isotropic cosmological models in these theories using a combination of
two
basic mathematical techniques introduced in section \ref{methods}. In
the
case where the trace of the energy-momentum tensor of matter vanishes
(ie
vacuum or radiation) exact solutions can be found directly for all
curvatures if the requisite integrals can be performed. asymptotic
forms are
easily derived in all cases. This technique exploits the conformal
relationship between scalar-tensor theories and general relativity
that
exists when the trace of the energy-momentum tensor vanishes. However,
when
the energy-momentum tensor is not trace-free the conformal equivalence
disappears and the indirect method of Barrow and Mimoso must be used
to find
exact solutions. This only works for zero curvature cosmological
models but
includes the important cases of all $p=0$ universes and inflationary
universes with $-\rho \leq p\leq -\frac 13\rho $. It also permits a
simple
means of comparing the behaviour of cosmological solutions in any
scalar-tensor theory with those of Brans-Dicke theory at early and
late
times. Since this procedure does not commence with the specification
of $
\omega (\phi )$, but with a choice of generating function that
produces the
entire solution by non-linear transformations, it is necessary to
build up
intuition by a thorough exploration of the results of employing
particular
classes of generating function. In particular, we were able to find
generating functions which gave rise to dust universes in
scalar-tensor
theories for which the exact radiation and vacuum solutions can be
found
exactly by the direct method. This provides us with descriptions of
scalar-tensor cosmologies throughout the entire radiation and dust and
vacuum dominated eras. We were also able to find a wide range of new
inflationary universe solutions with $p=-\rho $ in these theories.

In section \ref{coupling} we introduced three classes of scalar-tensor
theory which permit asymptotic approach to general relativity at late
times
when $\phi \rightarrow \phi _0$. The parameters defining the
functional form
of $\omega (\phi )$ which specify these gravity theories can be
restricted
further if we require the theory to approach general relativity in the
weak-field limit ($\omega \rightarrow \infty $ and $\omega ^{\prime
}\omega
^{-3}\rightarrow 0$), and describe expanding universes. For each of
these
general classes of theory we have determined the behaviour of flat,
open,
and closed universes by a combination of exact solutions and
asymptotic
studies of the early and late-time behaviours.

The catalogue of solutions and asymptotes that we have found will
enable
scalar-tensor theories to be constrained in new ways because they
enable
complete cosmological histories to be constructed through initial
vacuum,
radiation, dust and final vacuum-dominated eras. The standard sequence
of
physical processes responsible for events like monopole production,
inflation, baryosynthesis, primordial black hole formation,
electroweak
unification, the quark-hadron phase transition, and nucleosynthesis
can be
explored in the cosmological environment provided by scalar-tensor
gravity
theories. The constraints derived from these considerations can be
compared
directly with those imposed by weak-field tests in the solar system
and
observations of astrophysical objects like white dwarfs and the binary
pulsar. The ubiquity of scalar fields in current string theories of
high-energy physics has led to continued interest in the detailed
behaviour
of scalar-tensor gravity theories and their associated cosmologies. In
this
paper we have displayed some of the diversity that these cosmologies
possess
together with a collection of methods for solving other specific
theories
that may be motivated by future developments in high-energy physics.

{\center \section*{Acknowledgements}}

The authors are supported by the PPARC. They would like to
thank Jos\'e Mimoso for discussions and Starlink for
computing facilities.

{\center \section*{Appendix -- Conformal equivalence}}

The action for trace-free matter is conformally invariant and we may
exploit
this fact to study the behaviour of vacuum and radiation models with $
\phi>\phi_0$. Under a conformal transformation to a new metric
$\tilde{{\bf g
}}$, with components 
\begin{equation}
g_{{\rm ab}} = \phi^{-2} \tilde{g}_{{\rm ab}}\,,
\end{equation}
and a re-definition of the field 
\begin{equation}
\tilde{\phi} = \tilde{\phi}_0\left(\frac{\phi_0}{\phi}\right)\,,
\end{equation}
the action becomes 
\begin{equation}
S_{G} = \int d^4 x \sqrt{-\tilde{g}} \left[-\tilde{\phi}\tilde{{\cal
{R}}} + 
\frac{\omega(\phi_0\tilde{\phi}_0/\tilde{\phi})}{\tilde
{\phi}}\tilde{g}^{
{\rm ab}}\partial_{{\rm a}}\tilde{\phi} \partial_{{\rm b}}\tilde{\phi}
\right]\,,
\end{equation}
neglecting overall constant factors. When $0<\phi<\phi_0$ the coupling
function for Theory 2, as defined in section \ref{coupling}, is 
\begin{equation}
2\omega(\phi) + 3 = B_2\left[-\ln\left(\frac{\phi}{\phi_0} \right)
\right]^{-2\delta}\,,
\end{equation}
and so 
\begin{equation}  \label{omegatrans}
2\omega(\phi_0\tilde{\phi}_0/\tilde{\phi}) + 3 = B_2\left[-\ln
\left(\frac{
\tilde{\phi}_0}{\tilde{\phi}}\right) \right]^{-2\delta}\,,
\end{equation}
with $\tilde{\phi}>\tilde{\phi}_0$ when $0<\phi<\phi_0$. Eq.~(\ref
{omegatrans}) is exactly the form of the coupling for Theory 2 with a
gravitational scalar field $\tilde{\phi}>\tilde{\phi}_0$. Thus we may
obtain
solutions when $\phi>\phi_0$ in Theory 2 by applying the
transformations  
\begin{eqnarray}
a & \rightarrow & \phi a\,, \\
\phi & \rightarrow & \frac{1}{\phi}\,,
\end{eqnarray}
in this order, to the solutions presented in Section \ref{th2}. These
are
transformations which render the form of $y$ in Eq.~(\ref{ydef})
invariant.

Lastly, we remark that the asymptotic behaviours of Theories 1 and 3
when $
\phi >\phi _0$ may also be examined in this way, since the coupling
functions for both these theories may be approximated by logarithms as
$\phi \rightarrow \phi _0$.\\


\end{document}